\pdfoutput=1
\documentclass[aip,jcp,12pt,floatfix]{revtex4-1}

\usepackage{amsmath}
\usepackage{amssymb}
\usepackage{amscd}
\usepackage[pdftex]{graphicx}
\usepackage{nicefrac}
\usepackage[np,autolanguage]{numprint}
\usepackage{longtable}
\usepackage{bm}
\usepackage[svgnames]{xcolor}
\usepackage{hyperref}
\hypersetup{
    colorlinks,
    citecolor=blue,
    linkcolor=blue,
    urlcolor=red
}
\usepackage{array}
\usepackage{multirow}
\usepackage{framed}

\newcommand*\openquote{\makebox(25,-22){\scalebox{5}{``}}}
\newcommand*\closequote{\makebox(25,-22){\scalebox{5}{''}}}
\colorlet{shadecolor}{Azure}
\makeatletter
\newif\if@right
\def\shadequote{\@righttrue\shadequote@i}
\def\shadequote@i{\begin{snugshade}\begin{quote}\openquote}
\def\endshadequote{%
  \if@right\hfill\fi\closequote\end{quote}\end{snugshade}}
\@namedef{shadequote*}{\@rightfalse\shadequote@i}
\@namedef{endshadequote*}{\endshadequote}
\makeatother

\newcommand{\erf}{\mathop{\mathrm{erf}}}

\begin{document}

\makeatletter
  \def\tagform@#1{\maketag@@@{[#1]\@@italiccorr}}
\makeatother

\npthousandsep{}

\title{Noncovalent Interactions in Density-Functional Theory}
\thanks{This chapter will appear in volume 29 of Reviews in Computational Chemistry.}
\author{Gino A. DiLabio}
\affiliation{Department of Chemistry, University of British Columbia,
Okanagan, 3333 University Way, Kelowna, British Columbia V1V 1V7, Canada}
\affiliation{National Institute for Nanotechnology, National Research
Council of Canada, 11421 Saskatchewan Drive, Edmonton, Alberta T6G
2M9, Canada}
\author{Alberto Otero-de-la-Roza}
\affiliation{National Institute for Nanotechnology, National Research
Council of Canada, 11421 Saskatchewan Drive, Edmonton, Alberta T6G
2M9, Canada}

\date{\today}


\maketitle
\tableofcontents

\section*{Introduction}
\label{s:intro}
Density-functional
theory\cite{hohenberg1964,kohn1965,parrbook,dreizlerbook,kochbook,burkebook,stefaanbook,fiolhaisbook,shollbook,martinbook}
(DFT) is arguably the most successful approach to the calculation of
the electronic structure of matter. The success of the theory is
largely based on the fact that many DFT approximations can predict
properties such as thermochemistry, kinetics parameters, spectroscopic
constants, and a large range of properties with an accuracy rivaling those obtained by high-level {\it{ab initio}} wavefunction
theory methods in terms of agreement with experimental quantities. The
computational cost of DFT scales formally as $N^3$, where $N$ is the
number of electrons in the system, as compared to the $N^{5}-N^{7}$
scaling (or even higher) of correlated wavefunction methods,
indicating that DFT can be applied to much larger systems than
wavefunction methods, and to the same systems at a much lower
computational cost. Furthermore, DFT can be applied to molecular
systems using atom-centered basis sets and to molecular and solid
state systems through periodic, plane wave approaches, thus allowing
for the prediction of the properties of molecular and condensed matter
systems on the same theoretical footing.    

Despite their broad success in predicting many chemical and physical
properties, conventional\cite{footnote1} 
density-functional approximations have well-known
shortcomings.\cite{cohen2008,cohen2011,burke2012} In recent years, a
great deal of attention has been paid to the inability of conventional
DFT methods to predict dispersion interactions accurately. This
particular failing of DFT was first illustrated in the
1990s.\cite{lacks1993,kristyan1994,hobza1995,seponer1996,perezjorda1995,perezjorda1999,couronne1999} An
early work by Kristy\'{a}n and Pulay\cite{kristyan1994} demonstrated
that the local density approximation of DFT significantly overbinds
the noble-gas dimers He$_2$, Ne$_2$, and Ar$_2$, while ``improved''
DFT methods based on generalized gradient approximations significantly
underbind or predict their interactions to be completely
repulsive. This work serves as one of the early descriptions of the
``dispersion problem'' of DFT that underpinned two decades of effort
to understand and correct DFT in this capacity. 

\begin{figure}
\includegraphics[width=0.70\textwidth]{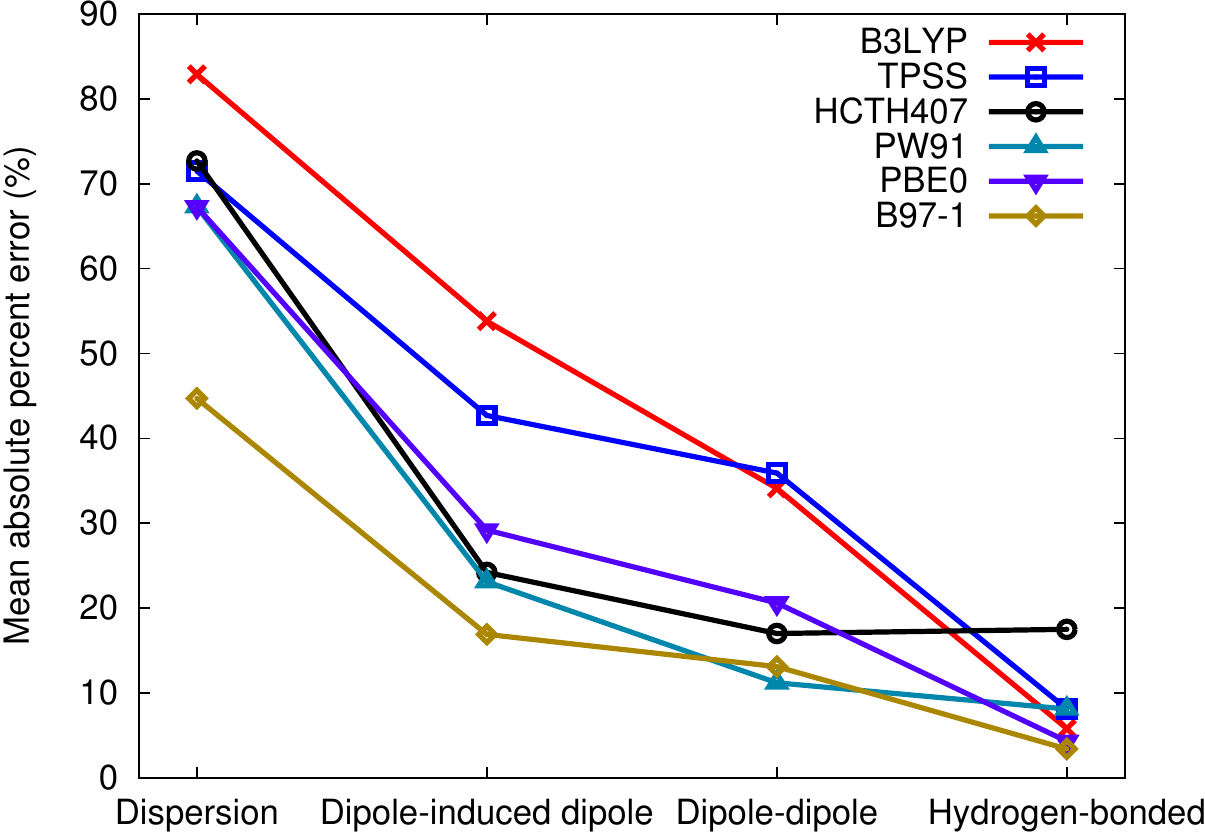}
\caption{Mean absolute percent errors in the predicted binding energies
of noncovalently-interacting dimers using various DFT methods for
different types on interactions (adapted from
ref.~\citenum{johnson2006}). \label{fig:interaction_types}} 
\end{figure}

The absence of explicit dispersion physics in common approximations to
DFT naturally focussed the attention of researchers on this
problem. Current understanding amongst some members of the DFT
community is that, of the van der Waals forces in general, only
dispersion is poorly treated. The prevailing opinion is that DFT can treat
electrostatics and other effects
accurately.\cite{distasio2012,kochbook} Considering the percent errors
in the binding energies of noncovalently-interacting dimers predicted
by various DFT methods, as shown in
Figure~\ref{fig:interaction_types}, this view may
seem justified. The figure indicates that DFT methods tend to offer poor predictions of binding energies in predominantly
dispersion-bound systems but work well for hydrogen-bonded
systems. But is this true?  

Applying the very popular B3LYP method to a set of 23
predominantly dispersion-bound dimers yields a mean error of 5.1
kcal/mol, which supports the notion that approximate DFT methods
underbinds in the case of dispersion. However, for a set of 23 dimers
in which hydrogen-bonding is the dominant interaction, B3LYP
underbinds by an average of 1.7 kcal/mol, an error that is large
enough to contradict the notion that hydrogen-bonding is well-treated
by DFT methods. Some of the 1.7 kcal/mol error in binding may come
from the absence of dispersion in B3LYP but this, as we shall see, is
not likely the only deficiency.  

The shortcomings of B3LYP are not unique and there is evidence in the literature that
other DFT methods are likewise deficient with respect to predicting the
strength of hydrogen bonding interactions. For
instance, Xu and Goddard studied a range of conventional DFT methods
for their ability to reproduce a number of properties in water
dimer.~\cite{xu2004} Based on their design principles, there is no
{\it a priori} reason to believe that the DFT methods used in their
study have any particular deficiencies or advantages when it comes to
modeling hydrogen bonding. And yet, Xu and Goddard found that
different DFT methods gave errors in binding energies ranging 
from overbinding by 0.41 kcal/mol (PWPW functional) to underbinding by
1.42 kcal/mol (BPW91 functional), with the latter result being worse
than uncorrelated wavefunction theory. If all of these functionals are
missing dispersion to a similar extent, the broad range of error in
the DFT-predicted binding energies offers evidence that DFT-based
methods do not accurately reproduce electrostatic interactions in
general, and point to broader difficulties in predicting noncovalent
interactions.  

On the basis of the foregoing discussion, we consider the ``dispersion
problem'' a ``noncovalent interaction problem''. This distinction is
important because the nomenclature that is used directs one's
thinking, and the term ``dispersion-corrected DFT'' leads to the
notion that the shortcomings in common DFT approximations are related
to dispersion alone. We are not advocating a change in
the nomenclature of ``dispersion''-corrected DFT
methods at this point because it has been in wide-spread use for
nearly a decade; however, it is important to understand the breadth of
the problems DFT methods have in modeling noncovalent
interactions. Part of the motivation for this chapter is to
underscore, where appropriate, the limitations of dispersion-corrected
DFT methods. The flurry of activity associated with the development of
new dispersion-correcting methods may obfuscate that
these methods cannot correct all of the underlying deficiencies of
the functional to which they are applied.  

We begin this chapter by providing an overview of the different
categories of noncovalent interactions, as generally described in
chemistry.  This is accompanied by some recent examples of
noncovalent interactions that focus on dispersion. We then provide some background on general density-functional theory, which
is structured so that a reader who is familiar with this material can
skip to the remainder
of the chapter without loss of continuity. We then introduce some modern DFT methods that are
capable of treating dispersion and other noncovalent
interactions. The
chapter closes with a comparison of methods using standard benchmark
data sets and some perspectives on the general applicability of the
methods and outlook.

\subsection*{Overview of Non-Covalent Interactions}
\label{s:noncovalentinteractions}
Dispersion is the weakest of the van der Waals forces that arises from
instantaneous charge fluctuations (e.g. induced dipoles) that occur in
otherwise non-polar systems. In chemistry, dispersion forces (or
interactions) are often called ``London'' forces. In the physics
community, Casimir forces~\cite{casimir1948} are described as arising
from quantum fluctuations in a quantized field that polarizes nearby
systems to induce the formation of dipoles. Both have their origins in
the same physical phenomenon.\cite{grimme2011}

Dispersion forces play a critical role at the molecular scale. As a
simple example, dispersion (and other van der Waals) interactions are
responsible for the deviation from the ideal gas behavior of most
real gases. Friction and wetting phenomena are also influenced by
dispersion forces. Dispersion can lead to the attraction of molecules
to a surface, often referred to as ``physisorption''. The measurement
of the physisorption of gases on solids is used to determine, among
other properties, the porosity and surface area of
materials,~\cite{sing1985} and it may precede important chemical
events like catalytic steps of chemical reactions, or surface
modifications.

\begin{figure}
\includegraphics[width=0.60\textwidth]{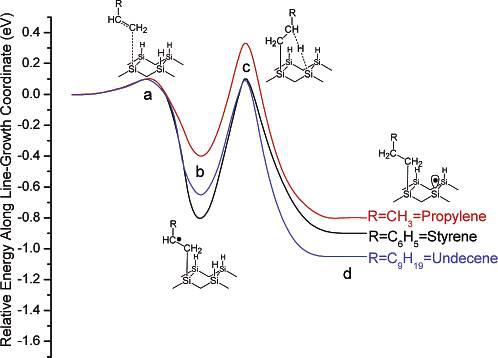}\\[\baselineskip]
\includegraphics[width=0.60\textwidth]{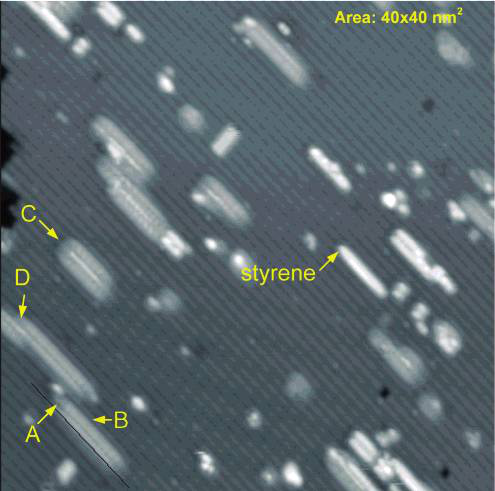}
\caption{Top panel: Potential energy
curves showing the relative energetics associated with one-dimensional
organic nanostructure formation on hydrogen-terminated silicon
surfaces. Following the chemisorption of a molecule another molecule
may add at a neighbouring silicon surface site that holds a radical
and this continues the line growth process. Bottom panel: Scanning
tunneling microscope image showing 1-undecene (labeled ''B'', ''C'',
and ''D'') and styrene-derived lines on the silicon surface (taken
with permission from
reference~\citenum{dilabio2004}). \label{fig:Xdilabio2004}}
\end{figure}

An interesting demonstration of surface physisorption is provided by
the formation of one-dimensional organic nanostructures on silicon
surfaces. Under ultrahigh vacuum conditions, it was demonstrated that
styrene (C$_6$H$_5$CHCH$_2$) is capable of undergoing a
radical-mediated line growth process by reacting with rare silicon
surface dangling bonds on an otherwise hydrogen-terminated silicon
surface. The reaction produces lines through a successive
addition-abstraction reaction mechanism that connects individual
molecules to the silicon surface such that they are juxtaposed. While
styrene can undergo line-growth, efforts to grow lines derived from
propylene (H$_3$CCHCH$_2$) failed.\cite{lopinski2000} The rationale
at the time was that styrene could undergo the line-growth process
because when its alkene moiety added to the silicon surface dangling
bond, the resulting carbon-centered radical was significantly
stabilized by radical delocalization of the unpaired electron into the
phenyl moiety. The propylene addition product does not benefit from
delocalization stabilization and so it undergoes desorption rather
than line growth and it was speculated that all linear alkenes could
not be made to undergo line growth for this reason. However, it was
later hypothesized that dispersion interaction between a longer chain
alkene, like 1-undecene (C$_9$H$_{19}$CHCH$_2$), could stabilize the
addition intermediate long enough to enable the growth of molecular
lines on the silicon surface. This hypothesis was verified by scanning
tunneling microscopy studies, which showed ``caterpillar''-like
molecular structures derived from 1-undecene with styrene-derived
lines nearby (see Figure~\ref{fig:Xdilabio2004}).~\cite{dilabio2004}

The macroscale action of dispersion was nicely demonstrated by the
work by Autumn et al.~\cite{autumn2002} They showed that geckos use
dispersion as the primary means of adhesion between their feet
(specifically small structures on their feet called setae) and
hydrophobic surfaces. This example illustrates that, although
dispersion tends to be the weakest of the noncovalent interactions, it
can result in significant interaction strengths when integrated over
large areas and/or over many atoms.

The macroscopic nature of the dispersion force is dependent upon the
media separating objects. From the Casimir force perspective, the
electromagnetic vacuum fluctuations give rise to polarization in
nearby atoms or molecules almost instantaneously at small
(nanometer) distances. However, the finite speed of light results
in a retardation in polarization when objects are far apart. Munday et
al.~\cite{munday2009} demonstrated experimentally that this
retardation can lead to repulsive interactions between objects
immersed in a solvent when the materials have particular relative
dielectric functions. In other words, the dispersion/Casimir force can
be exploited at large distances to levitate objects!

When two molecules interact, the noncovalent attraction between them
contains other forces along with dispersion. Dipole-induced dipole
forces are somewhat stronger than dispersion forces when compared on a
per atom basis. This force is created when the permanent electric
dipole in one molecule induces an electric dipole in an otherwise
non-polar molecule. The dipole arises from the redistribution of
electrons between bonded atoms having different
electronegativities. The strength of the interaction that results
depends on the magnitude of the permanent dipole moment and the
polarizability of the molecule with which the dipole interacts.

The dipole-dipole force tends to be stronger than the dipole-induced
dipole force and arises through the interaction between two (or more)
permanent electric dipoles. The most energetically favorable alignment
between dipoles is such that the positive ``head'' of one dipole is
arranged in space to be as close as possible to the negative ``tail''
of a second dipole. Dipoles arranged in this fashion and oriented in a
line interact most strongly. ``Head-to-tail'' dipole arrangements
where the dipoles reside in a plane interact less strongly.

Hydrogen bonding, which tends to be much stronger than the other
noncovalent interactions on a per atom basis, is often considered to
be a special case of dipole-dipole interaction, as is suggested by the
IUPAC definition.~\cite{muller1994} This definition may be due to the importance
that hydrogen bonding has in the determination of the structure of
biological molecules such as proteins~\cite{pauling1952} and
deoxyribonucleic acid (DNA),~\cite{watsonandcrick} and the central
role that water plays in life. Hydrogen bonding is also important in
materials chemistry and provides a valuable desgin motif for the
production of industrially relevant polymers like nylon and kevlar.

\begin{figure}
\includegraphics[width=0.70\textwidth]{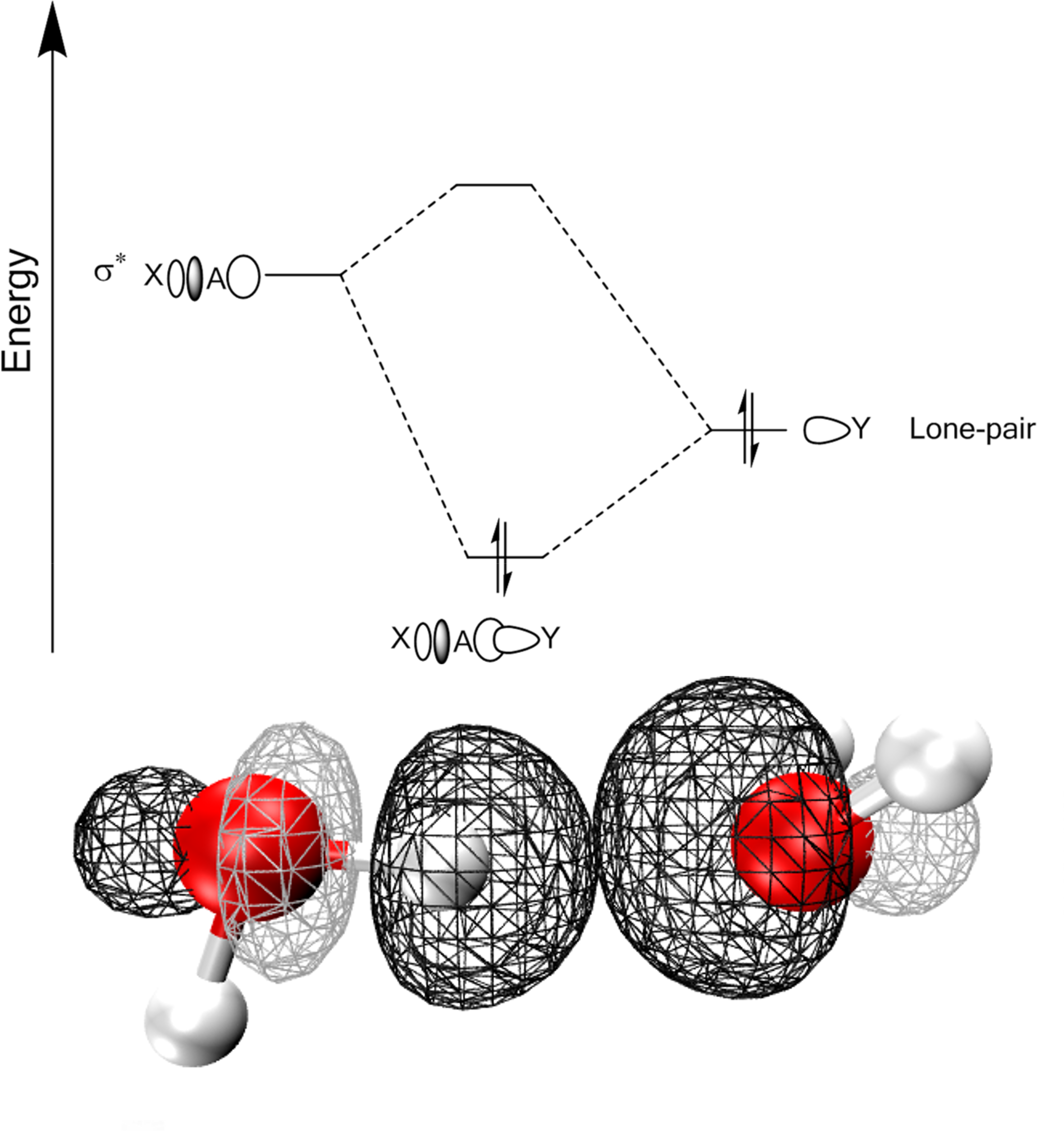}
\caption{Diagram illustrating the
stabilization that occurs between the $\sigma$ antibonding orbital
associated with an X-A bond and a lone-pair orbital of Y in
hydrogen-bonded water dimer. Opposite molecular orbital phases are
shown in different shades.\label{fig:OrbitalPicture}}
\end{figure}

While the nature of the atoms involved in dipole-dipole forces are
generally not specified, hydrogen bonds are usually described as
occurring between a donor species having a bond of the type X-H, where
X is an electronegative atom like oxygen or nitrogen, and an acceptor
atom that has a lone-pair of electrons. In this case, the donor has a
dipole resulting from the electronegativity difference in the X-H bond
and the acceptor has a dipole that exists between the center of the
negative charge distribution of the lone-pair of electrons and the
positive nucleus of the atom to which it belongs. There is also a
secondary orbital interaction associated with hydrogen bonding. The
donor lone-pair orbital overlaps to some extent with the antibonding
$\sigma_{XH}^{*}$ orbital of the X-H moiety, which contributes to the
overall stability of the hydrogen bond. A diagram illustrating this is
given in Figure~\ref{fig:OrbitalPicture}, where the atom A represents a hydrogen atom. The molecular orbitals
involved in the hydrogen bonding in the water dimer are shown in the
bottom panel of Figure \ref{fig:OrbitalPicture}. The stability comes
at the expense of a weakening of the X-H bond because of the extra
electron density in its antibonding orbital as a result of the overlap
with a lone-pair. The weakening is manifested by the lengthening of
the X-H bond and is often, but not always,~\cite{litwinienko2009}
accompanied by the reduction of the frequency of the X-H
vibration.~\cite{badger1937} There are many excellent books and
reviews on the different facets and the impacts of hydrogen bonding to
which the interested reader is
directed.~\cite{jeffrey1997,scheiner1997,etter1990}

In some cases, C-H groups can also be donors to strong hydrogen
bonds. Experimental and theoretical work by Salamone and collaborators
showed that the benzyloxyl radical (BnO, C$_6$H$_5$CH$_2$O) is capable
of forming strongly bound pre-reaction complexes with
amines.~\cite{salamone2011,johnsonradicals2013} Calculations indicate
that the strength of binding in dimers of this kind in vacuum is ca. 7
kcal/mol, which is stronger than that of the water dimer. The large
binding energy is attributed to the strong electron withdrawing
effects of the oxygen-centered radical that is $\alpha$ to the C-H
hydrogen bond donor group. The experimental outcome is that the
reactivity of BnO with amine substrates is more than a factor of 3300
larger than the cumyloxyl radical (C$_6$H$_5$C(CH$_3$)$_2$O), which
cannot engage in hydrogen bonding.

In general, the interactions of higher-order multipoles with each
other or with non-polar species are not allocated a separate category
of noncovalent interaction, despite the fact that their effects can sometimes be
significant. For example, the complementary
quadrupole-quadrupole interactions that arise between benzene and
hexafluorobenzene, which are both liquids at room temperature,
results in the formation of a crystal.~\cite{patrick1960} In its
unperturbed state, benzene has no dipole moment but it does have a
substantial quadrupole moment arising from the electron density above
and below the ring plane which is composed of the relatively positive
carbon nuclei. In most cases, however, as the order of the multipole
increases the forces that arise from them diminish in magnitude.

A number of new nomenclatures associated with noncovalent interactions
have emerged over the last few decades that contrast with the
preceding classification. For instance, the interactions of dipoles
with the $\pi$ face of aromatic molecules, like benzene, may be
considered a special case of dipole-induced interactions.

Cation-$\pi$ interactions,~\cite{dougherty1996} which describe the
attraction between a cation and the negative electron cloud above the
plane of an aromatic system such as benzene, may be considered a
monopole-induced dipole interaction.  The charge associated with the
cationic center polarizes the ``fluffy'' cloud of the aromatic moiety,
resulting in moderately strong interactions. Work by
Dougherty~\cite{dougherty1996} and others has characterized the
strengths of interaction in various cation-$\pi$ systems using various
quantum mechanical tools. It is generally believed that these
interactions are operative in biological systems where salt cations,
such as Na$^{+}$, interact with aromatic moieties in proteins
(e.g. tryptophan, phenylalanine, and tyrosine). Cation-$\pi$
interactions, however, will be strongly attenuated, or eliminated, by
solvation effects.

Related to cation-$\pi$ interactions, but weaker and perhaps less
well-known, are anion-$\pi$ interactions.~\cite{quinonero2002} It may
be counterintuitive at first that a negatively charged atom could
interact attractively with the $\pi$ cloud of an aromatic
system; indeed, beyond a critical distance the interactions between an
anion and the $\pi$-face are repulsive. However, when the two
components are within a certain distance, charge-induced polarization
provides attraction between the two moieties. Anion-$\pi$ interactions
have been shown to exist in inorganic crystals~\cite{schottel2008} and
may be utilized for anion sensing applications.~\cite{rosokha2004}

Halogen-bonding (XB) is another type of noncovalent interaction that
has gathered a great deal of attention recently, although some have
argued that XB effects were reported more than 100 years
ago.~\cite{metrangelo2005} This interaction occurs between a halogen
bond acceptor in the form of X-X or X-Y (X = halogen; Y = H, C, N, O)
and a donor, which is usually a Lewis base (e.g. acetonitrile and
formaldehyde).  For a set of 51 small neutral species used for
theoretical methods benchmarking purposes, gas-phase interaction
strengths were predicted to be as large as ca. 34 kcal/mol, depending
on the nature of the donor and acceptor species.~\cite{kozuch2013}
Interestingly, there does not appear to be a need to incorporate
dispersion corrections into DFT methods in order to accurately predict
the halogen bonding interaction strengths. In fact, dispersion
corrections were found to be detrimental for this purpose in most
cases. The consensus amongst those that study halogen bonding is that
it arises largely from the interaction between the lone-pair of
electrons on the Lewis base with the so-called $\sigma$-hole on the
halogen bond acceptor. The $\sigma$-hole is the area of relatively
positive charge on the otherwise electronegative halogen on the
acceptor that arises from the antibonding orbital associated with the
X-halide $\sigma$-bond. Halogen bonds, like hydrogen bonds, have a
directionality that corresponds to maximum overlap between the Y-X
$\sigma^{*}$ and the acceptor lone-pair, which is 180$^{\circ}$, and
are derived from electrostatic and orbital overlap
effects~\cite{pinter2013} (see Figure~\ref{fig:OrbitalPicture}) with A
the halogen atom.

Finally, we mention here pnictogen and chalcogen bonding, which are
similar in nature to hydrogen and halogen bonding in that they derive
from electrostatic and orbital overlap effects. Pnictogen bonding
involves the lone-pair orbitals of group 15 (N to Bi) donor atoms and
chalcogen bonding involves the lone-pair orbitals of group 16 atoms (O
to Po). In this respect the pnictogen/chalcogen atoms behave as a Lewis
base, while the acceptors can be any empty anti-bonding orbital, and
can be depicted using the same orbital diagram of
Figure \ref{fig:OrbitalPicture} that was used to illustrate hydrogen
and halogen bonding.  From this perspective, halogen, pnictogen and
chalcogen bonding may all be considered variations of Lewis acid-base
interactions and one may question the need to provide them with
different names. Indeed, hyperconjugation, which is the name given to
the overlap between the bonding (i.e. doubly-occupied) orbital on one
center and an anti-bonding (i.e. empty) orbital on an adjacent center
covalently bonded to the first,~\cite{mulliken1941} strongly parallels
the concepts behind pnictogen, chalcogen, halogen and hydrogen
bonding. It seems reasonable to focus on the notion of secondary
orbital interactions as being broadly operative in all of these
noncovalent interactions.

\section*{Theory Background}
\label{s:theory}

\subsection*{Density-Functional Theory}
\label{s:dft}
Density-functional theory\cite{hohenberg1964,kohn1965,%
parrbook,dreizlerbook,kochbook,burkebook,stefaanbook,fiolhaisbook,%
shollbook,martinbook} (DFT) is at present the most popular method to
study the electronic structure of chemical systems. DFT
approximates the solution of the time-independent Schr\"odinger
equation under the Born-Oppenheimer approximation, which is the
foundation of most of quantum chemistry and materials physics:
\begin{equation}
\hat{H}\Psi = E\Psi
\end{equation}
The non-relativistic many-electron Hamiltonian (atomic units are used
throughout this chapter) that describes the problem is:
\begin{equation}
\label{eq:bigh}
\hat{H} = -\frac{1}{2}\sum_i\nabla^2_i - \sum_{i,A}\frac{Z_A}{R_{iA}}
+ \sum_{i>j}\frac{1}{r_{ij}}
\end{equation} 
where $i$ runs over electrons, $A$ runs over atoms, $Z_A$ are the
atomic numbers, $r_{ij}$ are electron-electron distances, and $R_{iA}$
are electron-nucleus distances. The many-electron wavefunction
($\Psi$) contains all the information about the system. Because of the
electron-electron interaction (the last term in the Hamiltonian), the
electronic structure of a system with more than one electron is a
complicated many-body problem, and impossible to solve analytically.

Traditional approaches in quantum chemistry are based on the orbital
approximation. Under it, a single Slater determinant is proposed as an
ansatz for the eigenfunctions of the many-body Hamiltonian. The
determinant is composed of one-electron functions (orbitals), which
are in turn expressed as linear combinations of basis
functions. Application of the variational principle, which states that
the correct ground-state wavefunction minimizes the expectation value
of the Hamiltonian, leads to a set of one-electron equations that must
be solved iteratively in what is called the self-consistent field
(SCF) method. This procedure is the Hartree-Fock (HF) method, and the
difference between the exact ground-state and the HF energy is called
the correlation energy (not to be confused with the DFT correlation
energy, see below).

Many approaches have been developed to improve the accuracy of the HF
results and calculate the missing correlation energy by working with
the HF solution. We collectively refer to these as wavefunction
methods, which have been described extensively
elsewhere.\cite{szabobook,helgakerbook} In the wavefunction approach,
the exact many-electron wavefunction is written as a linear
combination of Slater determinants, that correspond to excitations of
one or more reference configurations. If enough computing power were
available, the exact solution of the Schr\"{o}dinger equation could be
found by employing enough determinants in the wavefunction
expansion. Hence, there is a systematic recipe to improve the
calculation level in wavefunction theory, but the scaling of the
computational cost prevents the application of higher-level
wavefunction theory in all but the simplest systems. Two popular
wavefunction methods are M{\o}ller-Plesset perturbation theory
(specifically, to second-order, MP2) and coupled cluster (CC). The
coupled-cluster singles doubles and perturbative triples (CCSD(T))
variant of the latter is the preferred method in the literature to
obtain accurate reference data for the binding energies of small
noncovalently-bound dimers (see the section below titled ``Description of Non-covalent Interaction Benchmarks'').

In contrast to wavefunction theory, the essential quantity in DFT is
not the wavefunction but the electron density ($\rho(\bm{r})$), a
three-dimensional scalar function that describes the probability of
finding electrons in real space. DFT, in general, reduces the
computational cost compared to wavefunction theory, but there is no
systematic recipe to approach the exact solution for a given
system. The idea of using the electron density instead of the
many-electron wavefunction as the central quantity first appeared in
the Thomas-Fermi theory,\cite{thomas1927,fermi1927,parrbook} an early
version of DFT from the computerless days when even the simplest
wavefunction calculation was impossible to carry out. However,
Thomas-Fermi theory predicts no molecular
binding\cite{teller1962,lieb1977,parrbook} rendering the method
useless in practical applications.

The foundation of modern-day DFT was laid by two theorems proven by
Hohenberg and Kohn\cite{hohenberg1964} (HK) in 1964. The first HK
theorem establishes that there is a one-to-one correspondence between
the external potential (the electrostatic potential created by the
nuclei at the chosen molecular geometry) and the ground-state electron
density. Since the external potential is the only non-universal
(i.e. system-dependent) part of equation~\ref{eq:bigh}, the first
theorem establishes a one-to-one correspondence between density and
the many-electron wavefunction. Hence, any observable can be obtained
as a functional of the electron density, including the energy
$E[\rho]$. The second HK theorem establishes that the ground state
electron density is a minimum of the exact energy functional. This
theorem enables the use of the variational principle, which is a very
powerful tool in the search for the ground-state electron density.

The energy functional $E[\rho]$ is unknown and the HK theorems provide
no indication as to how to obtain it. However, large contributions to
the energy, like the classical electron-electron repulsion ($J[\rho]$)
or the electron-nuclei attraction ($E_{ne}[\rho]$), depend directly on
the density and can be calculated in a straightforward manner. The
energy functional can be expressed as a sum of component functionals
and written as:
\begin{align}
\label{eq:efunctional} E[\rho] & = T[\rho] + J[\rho] + E_{ne}[\rho] + E_{xc}[\rho] \\
J[\rho] & = \int\frac{\rho(\bm{r})\rho(\bm{r}^{'})}{|\bm{r}-\bm{r}^{'}|}d\bm{r}d\bm{r}^{'} \\
E_{ne}[\rho] & = - \sum_A\int\frac{Z_A\rho(\bm{r})}{|\bm{R_A}-\bm{r}|}d\bm{r}
\end{align} 
where $T[\rho]$ is the kinetic energy functional and
$E_{xc}[\rho]$ is the exchange-correlation functional, which
encapsulates the missing energy contributions not contained in the
other functionals. The exchange-correlation functional is usually
partitioned into a exchange part and a correlation functional:
\begin{equation}
E_{xc} = E_x + E_c
\end{equation}
The exchange functional is defined as the difference between the
classical electron-electron repulsion and the expectation value of the
many-body electron-electron energy term:
\begin{equation}
E_x = \langle\sum_{i>j}r_{ij}^{-1}\rangle - J[\rho]
\end{equation}
For a one-electron system, the exchange term would cancel exactly the
spurious self-interaction of the electron with itself coming from
$J[\rho]$, a role that is fulfilled by the exchange term in HF
theory. Hence, $E_x$ contains the energetic contribution coming from
the antisymmetry requirement imposed on the many-body wavefunction and
corrects for double-counting of electrons in $J[\rho]$. The
correlation energy ($E_c$) is defined as the missing energy necessary
to make $E_{\text{xc}}$ exact. Note that in
equation~\ref{eq:efunctional}, only the $E_{ne}$ term depends on the
geometry of the system. The rest is a \emph{universal} functional,
that is, it is the same regardless of the details of the system under
calculation.

The second seminal paper in DFT was published by Kohn and
Sham\cite{kohn1965} a year after the HK theorems were proposed. The
Kohn-Sham (KS) formulation of DFT gives a practical recipe to the
calculation of the ground-state energy and electron density, and
uses much of the same technology (programs, algorithms) as
does HF theory. In KS-DFT, one assumes there is a collection
of non-interacting quasi-particles, similar to electrons and equal in
number, that has the same particle density as the actual electron
density for the system of interest. By doing so, the electron density
of a system has the same expression as if it were derived from a
Slater determinant:
\begin{equation}
\rho(\bm{r}) = \sum_i |\psi_i(\bm{r})|^2
\end{equation}
where the $\psi_i$ are the occupied orbitals (called the
\emph{Kohn-Sham orbitals}). The KS scheme provides a
simple kinetic energy functional expression:
\begin{equation}
T[\rho] \approx T_{KS} = -\frac{1}{2}\sum_i |\nabla\psi_i(\bm{r})|^2
\end{equation}
$T_{KS}$ is the Kohn-Sham kinetic energy, which is only an
approximation to the true kinetic energy functional. In the KS scheme,
the difference between the exact kinetic energy and $T_{KS}$ is
incorporated into the correlation energy $E_{c}$. It is important to
note that the theory is formally exact even for systems that
traditionally can not be treated accurately with a single Slater
determinant (e.g. low-energy excited states, bond breaking,
biradicals, etc.). 

Minimization of the energy functional within the KS scheme with
respect to variations in the electron density leads to the
one-electron Kohn-Sham operator:
\begin{equation}
\label{eq:hks1}
H_{\text{KS}} = T + V_H + V_{\text{ext}} + V_{\text{xc}} \\
\end{equation}
with:
\begin{equation}
T = -\frac{1}{2}\nabla^2 \quad ; \quad
V_H = \int\frac{\rho(\bm{r}^{'})}{|\bm{r}-\bm{r}^{'}|}d\bm{r}^{'} 
\quad ; \quad
V_{\text{ext}} = - \sum_A\frac{Z_A}{|\bm{R}_A-\bm{r}|} 
\end{equation}
and the exchange-correlation potential being defined as the functional
derivative of $E_{\text{xc}}$ with respect to the electron density:
\begin{equation}
V_{\text{xc}} = \frac{\delta E_{\text{xc}}}{\delta\rho(\bm{r})}
\end{equation}
Equation~\ref{eq:hks1}, when combined with orbitals expressed as
linear combinations of basis functions, yields matrix equations 
similar to those in HF theory, which simplified (and still
does) the implementation of DFT in preexisting quantum chemistry
software. 

The advantage of DFT with respect to traditional wavefunction methods
is that, at a computational cost similar to or even less than HF, it is possible to obtain electronic properties that in many cases
rival correlated wavefunction approaches in accuracy. The downside is, in contrast to wavefunction theory where increasingly complex
methods yield better results, there is no systematic approach to
improve the approximations to the exact exchange-correlation
functional $E_{\text{xc}}$, which, recall, is unknown in the formalism. The
design of exchange-correlation functionals is, consequently, the
cornerstone of development in DFT and users should be aware of the
strong and weak points of the functionals being used.

The earliest and simplest method to approximate the exchange
correlation functional is the local-density
approximation\cite{hohenberg1964,kohn1965} (LDA). In LDA, the $E_{\text{xc}}$ is
calculated by assuming the system behaves locally as a uniform
electron gas. That is:
\begin{equation}
E_{xc} = \int \rho(\bm{r})\varepsilon_{xc}^{LDA}(\rho(\bm{r})) d\bm{r}
\end{equation}
where $\varepsilon_{xc}^{LDA}(\rho)$ is the exchange-correlation energy
density per electron of a uniform electron gas with density
$\rho$. The exchange contribution to $\varepsilon_{xc}^{LDA}$ is analytical
($\varepsilon_x^{LDA} = -3/4(3/\pi)^{1/3}\rho^{1/3}$), while the correlation
energy was obtained from accurate quantum Monte Carlo
calculations\cite{ceperley1980} and is parametrized.\cite{vosko1980,perdew1981}

The performance of LDA in actual calculations is surprisingly good for
such a crude model. Unlike Thomas-Fermi theory, LDA binds molecules
and, while there are hundreds of more modern functionals, it is still
occasionally used in the materials science community. However, gross
overestimation of bond energies and poor thermochemistry have ruled
out its use to solve problems of interest in chemistry.

The most basic class of functionals that improve upon LDA rely on the
generalized gradient approximation (GGA). Here which the
exchange-correlation functional depends on both the value and the
gradient of the electron density:
\begin{equation}
E_{xc} = \int \rho(\bm{r})\varepsilon_{xc}^{\text{GGA}}(\rho(\bm{r}),\nabla\rho(\bm{r})) d\bm{r}
\end{equation}
By making the energy density depend on the density gradient, it is
possible to account for local inhomogeneity in the electron density.
Unlike LDA, there is not a single GGA, that is, the expression for
$\varepsilon_{xc}^{\text{GGA}}$ is not unique. The existing GGA
functionals (there are tens of them) vary in the exact constraints
that they fulfill, as well as in the amount of empiricism in their
construction and in the number of adjustable parameters they
contain. Popular exchange GGA functionals include the
Perdew-Burke-Erzenhof\cite{pbe} (PBE) and subsequently revised
versions (revPBE,\cite{revpbe} PBEsol,\cite{pbesol}), Perdew-Wang 1986
(PW86),\cite{pw86} Becke 1986b,\cite{b86b} (B86b), and Becke
1988\cite{b88} (B88). Standalone gradient-corrected correlation
functionals include the popular Lee-Yang-Parr functional\cite{lyp}
(LYP) as well as the correlation part of the PBE
functional.\cite{pbe} Exchange and correlation functionals are usually
combined to give composite functionals, such as PW86PBE and B88LYP
(often simply BLYP). PBE is the most popular functional in solid-state
calculations, and it is non-empirical (its parameters are not
determined by resorting to fits to reference data). B86b and B88 and
the correlation functional LYP contain fitted parameters, but their
performance in the calculation of thermochemical quantities is notably
better than PBE. In general, GGA functionals provide much better
results for the calculation of most properties, although not enough
to be useful in the calculation of chemical reaction energies. GGA
functionals are also very popular in the solid-state field because
they yield accurate geometries, elastic properties of periodic solids
and qualitatively correct electronic band structures. However, they
severely underestimate the electronic band gaps.

Meta-GGA functionals increase the flexibility in the functional
definition by using, in addition to the density and its derivatives,
the Kohn-Sham kinetic energy density ($\tau_{KS}$):
\begin{equation}
E_{xc} = \int \rho(\bm{r})\varepsilon_{xc}^{\text{GGA}}(\rho(\bm{r}),\nabla\rho(\bm{r}),\nabla^2\rho(\bm{r}),\tau_{KS}(\bm{r})) d\bm{r}
\end{equation}
where:
\begin{equation}
\tau_{KS}(\bm{r}) = -\frac{1}{2}\sum_i|\nabla_i(\bm{r})|
\end{equation}
The development of accurate meta-GGAs is still an active area of
research.\cite{tpss,revtpss,sun2013,perdew2013,mn12l,n12,n12sx}
Popular meta-GGA approximations to exchange include the
Tao-Perdew-Staroverov-Scuseria\cite{tpss} functional (TPSS, there is
also a meta-GGA correlation functional proposed in the same work), its
revised version revTPSS\cite{revtpss} and the Minnesota functionals
reviewed later in the section titled ``Minnesota Functionals''. With
an increased degree of freedom, meta-GGAs usually improve upon GGAs in
the accuracy of calculated properties.

LDA, GGAs, and meta-GGAs are \emph{semilocal} or \emph{pure}
functionals, for which the exchange-correlation energy density at a
point depends solely on the properties at that point. In a seminal
article,\cite{becke1993} Becke showed that the calculation of
molecular thermochemistry (particularly, atomization energies,
ionization potentials and electron affinities) can be greatly improved
by using an admixture of a GGA and a fraction of exact exchange, which
is calculated as the exchange energy in HF theory but obtained using
the KS orbitals. The use of exact exchange in a functional is
justified by invoking the adiabatic connection
formula.\cite{harris1974,gunnarsson1976,langreth1977}

The adiabatic connection is a rigorous formula for the calculation of
the exact exchange-correlation functional. It says:
\begin{equation}
\label{eq:acf}
E_{xc}[\rho] = \int_0^1 \left(\langle\sum_{i>j}r_{ij}^{-1}\rangle_{\lambda} - J[\rho^{\lambda}] \right) d\lambda 
= \int_0^1 U_{\text{xc}}^{\lambda} d\lambda
\end{equation}
where $\lambda$ is a parameter that turns on the electron-electron
interaction (the $r_{ij}^{-1}$ term in Eq.~\ref{eq:bigh}). $\lambda=0$
is the non-interacting Kohn-Sham system and $\lambda=1$ is the
fully-interacting real system. The integrand ($U_{\text{xc}}$) is
defined as in equation~\ref{eq:acf}, and is called the potential
exchange-correlation energy. Equation~\ref{eq:acf} represents an
interpolation with the $\lambda=0$ endpoint being the exact exchange
energy calculated using the Kohn-Sham orbitals:
\begin{equation}
U_{xc}^0 = -\frac{1}{2}\sum_{ij}^{\text{occ}}\int
\frac{\psi_i^{*}(\bm{r}_1)\psi_j^{*}(\bm{r}_2)\psi_j(\bm{r}_1)\psi_i(\bm{r}_2)}
{r_{12}} d\bm{r}_1 d\bm{r}_2
\end{equation}
where the sum runs over all pairs of occupied Kohn-Sham states. Hence,
it makes sense to define the exchange-correlation functional
approximation as an interpolation between the known $\lambda=0$ limit
(exact exchange) and $\lambda=1$, represented by the semilocal
functional:\cite{becke1993}
\begin{equation}
E_{xc}[\rho] = a_x U_{xc}^0 + (1-a_x) E_{xc}^{\text{semilocal}}[\rho]
\end{equation}
where $a_x$ is the parameter controlling the amount of exact
exchange in the approximate functional.

The functionals that use a fraction of exact exchange in their
definition are called \emph{hybrids} and, of those, the most popular
by far is B3LYP, a combination of Becke's 1993 exchange
hybrid\cite{b3} and LYP correlation.\cite{lyp} In B3LYP, 20\%\ exact
exchange is used, a number that was obtained by fitting to a set of
reference thermochemical values (atomization energies, ionization
potentials, and proton affinities) and total energies. Subsequently, a
25\%\ fraction of exact exchange was justified on theoretical grounds
by Perdew et al.,\cite{perdew1996} resulting in the definition of
PBE0,\cite{pbe0} the non-empirical hybrid extension of PBE. Another
popular hybrid is B3P86 (same exchange as B3LYP but using Perdew 1986
correlation\cite{p86}).

By including exact exchange, hybrid functionals are no longer
semilocal: the exact exchange energy involves a double integration
over real-space. Thus, they are computationally more expensive than
semilocal functionals. This is particularly true in periodic solids
with plane wave basis sets, for which they are feasible only in very
simple systems. For this reason, and also because of unphysical
features in the HF description of metals,\cite{monkhorst1979} hybrids
are not much used in materials studies,\cite{janesko2009} but they are
very popular in quantum chemistry, where B3LYP is the most used
functional by number of citations. The improved thermochemistry with
respect to GGAs enable accurate studies of reaction energetics,
justifying their continued popularity.

Even though hybrids provide improved accuracy in many
chemically-relevant properties, they still face problems. One of these
is ``self-interaction'' error. Because the antisymmetry of the
wavefunction is not enforced as in HF, there can be overcounting (or
undercounting) of electron-electron interactions, which results in
electrons interacting with themselves. The simplest instance of
self-interaction error happens in the hydrogen atom, for which most
functionals fail to find the correct ground state energy ($-1/2$
Hartree) because the $E_{xc}[\rho]$ does not cancel $J[\rho]$ exactly.

A popular approach to deal with this problem is to use range-separated
or (also called long-range corrected)
functionals.\cite{savin1995,savin1996,gill1996} Similar to hybrids,
range-separated hybrids combine exact exchange with a semilocal
functional, but they do so by partitioning the electron-electron
interaction kernel ($1/r_{ij}$) into long-range ($\erf(\omega
r_{ij})/r_{ij}$) and short-range parts ($(1-\erf(\omega
r_{ij}))/r_{ij}$), where $\erf$ is the standard error function. The
range-separation parameter ($\omega$) controls the relative extent of
the short-range and long-range electron interactions.  The idea behind
range-separated hybrid functionals is to recover the correct
long-range behavior of the exchange-correlation potential. For
semilocal functionals, $V_{xc}$ decays exponentially when moving away
from the system, but the correct tail goes as $-1/r$. This behavior is
recovered by using exact exchange as the limit when $r\to\infty$. This
does \emph{not} mean, however, that long-range corrected functionals
model dispersion, but it does mean that the treatment of
non-dispersive intermolecular electron-electron interactions are, in
general, improved.

In most range-separated hybrids, the short-range part corresponds to
the semilocal functional, while exact exchange is the long-range
part. Common functionals in this category are
LC-$\omega$PBE,\cite{vydrov2006,lcwpbe2} CAM-B3LYP,\cite{camb3lyp} and
$\omega$B97,\cite{wb97} (and also its reparametrized $\omega$B97X
version\cite{wb97}). Range-separated functionals give improved charge
transfer excitation energies, reaction barriers and, in general,
minimize self-interaction error. Their behavior for thermochemistry is
good, outperforming, in general, their hybrid counterparts.  Some
range-separated functionals, most notably the Heyd-Scuseria-Ernzerhof
(HSE) functional,\cite{hse03,hse06} use a short-range exact exchange
and a long-range semilocal functional. The reason is that these
functionals are designed to recover some of the good properties of the
hybrids in periodic solid-state calculations. At a cost of 2--4 times
over semilocal functionals, HSE delivers increased accuracy in the
calculation of geometries and bulk moduli (by about 50\%), and,
particularly, band gaps (errors from 1.3 eV to 0.2 eV on
average).\cite{heyd2004}

A major application of range-separated functionals, and a very active
area of research is time-dependent density functional
theory\cite{marquesbook} (TDDFT). TDDFT is based on the extension of
the Hohenberg-Kohn theorems to time-dependent electron densities put
forward by Runge and Gross.\cite{runge1984} It is mostly used in the
calculation of excited-state transition energies and probabilities
(optical spectra), as well as properties of the excited states, and
ground-state properties related to the excitations
(e.g. polarizabilities and hyperpolarizabilities). Range-separated
functionals are essential in alleviating some of the problems in
TDDFT, including the modelling of excitations involving long-range
charge transfer.

\begin{table}
\caption{Comparative Assessment
of Several Functionals in Standard Thermochemical Tests. The Side
Column Labels the Type of Functional Approximation (mGGA$=$meta-GGA,
RS hybrids$=$range-separated hybrids). The Calculations were Run Using
aug-cc-pVTZ Basis Sets. The Entries are Mean Absolute Deviations (MAD)
in kcal/mol.\label{tab:thermo}}
\begin{tabular}{c|crrrrrrr}
\hline\hline
& Functional     & 
\multicolumn{1}{c}{G3/99} &  
\multicolumn{1}{c}{BDE}  & 
\multicolumn{1}{c}{Isod} & 
\multicolumn{1}{c}{Isom} & 
\multicolumn{1}{c}{BH}   & 
\multicolumn{1}{c}{TM}   &
\multicolumn{1}{c}{CT}   \\
\hline
\parbox[t]{3ex}{\multirow{1}{*}{\rotatebox[origin=c]{90}{LDA}}} & 
\raisebox{-2ex}{\rule{0pt}{5.5ex}}
LDA\cite{hohenberg1964,kohn1965,vosko1980,slaterbook} 
                                       &  117.5  &   11.9  &     0.2  &     2.5  &   18.1  &    34.2 &      6.4 \\
\hline
\parbox[t]{3ex}{\multirow{3}{*}{\rotatebox[origin=c]{90}{GGA}}} & 
PBE\cite{pbe}                          &   18.9  &    4.9  &     3.4  &     1.9  &    9.6  &    10.4 &      2.6 \\
& PW86PBE\cite{pw86,pbe}               &    9.4  &    7.2  &     3.8  &     2.3  &    8.0  &     7.8 &      2.5 \\
& BLYP\cite{b88,lyp}                   &   11.4  &    7.6  &     4.8  &     3.3  &    7.9  &     5.8 &      1.4 \\
\hline
\parbox[t]{3ex}{\multirow{2}{*}{\rotatebox[origin=c]{90}{mGGA}}} & 
TPSS\cite{tpss}                        &    4.7  &    5.9  &     4.9  &     2.5  &    8.1  &    10.0 &      1.9 \\
& M06-L\cite{m06l}                     &    5.1  &    3.9  &     3.2  &     2.0  &    4.6  &    31.8 &      1.7 \\
\hline
\parbox[t]{3ex}{\multirow{5}{*}{\rotatebox[origin=c]{90}{Hybrids}}} & 
B3LYP\cite{b3,lyp}                     &    7.8  &    5.7  &     4.4  &     2.3  &    4.6  &     4.5 &      0.5 \\
& BHandHLYP\cite{b3,lyp}               &   32.3  &    7.2  &     4.1  &     1.6  &    2.4  &    18.2 &      0.7 \\
& PBE0\cite{pbe0}                      &    5.5  &    4.6  &     3.5  &     2.0  &    4.6  &     2.8 &      0.8 \\
& B97-1\cite{b971}                     &    6.1  &    3.8  &     3.9  &     1.5  &    4.6  &     3.2 &      0.9 \\
& B3P86\cite{b3,p86}                   &   23.4  &    2.9  &     3.9  &     1.8  &    6.0  &     3.3 &      0.8 \\
\hline
\parbox[t]{3ex}{\multirow{3}{*}{\rotatebox[origin=c]{90}{RS hybrids}}} & 
CAM-B3LYP\cite{camb3lyp}                  &    4.2  &    4.0  &     3.4  &     1.7  &    3.4  &     4.2 &      0.3 \\
& LC-$\omega$PBE\cite{vydrov2006,lcwpbe2} &    5.1  &    4.4  &     3.4  &     2.4  &    1.3  &     3.0 &      1.2 \\
& HSE06\cite{hse03,hse06}                 &    5.1  &    5.0  &     3.5  &     1.8  &    4.6  &     2.9 &      1.0 \\
\hline\hline
\end{tabular}
\end{table}

Table~\ref{tab:thermo} shows the comparative performance of several
functionals from different approximations. The benchmark sets chosen
are the same as in ref.~\citenum{oterodelaroza2013b}: the G3/99 set
comprising 222 atomization energies,\cite{curtiss2000} the bond
dissociation energy database of Johnson et al.\cite{johnson2003}
(BDE), the hydrogen-transfer reaction set by Lynch and
Truhlar\cite{lynch2001} (BH), the set of linear alkane isodesmic
reactions (Isod) used by Wodrich et al.\cite{wodrich2006} (with the
geometries from the G3X set\cite{curtiss2001}), the isomerization of
organic molecules set by Grimme et al.\cite{grimme2007} (Isom), the
charge-transfer complex set of Zhao and Truhlar\cite{zhao2005b} (CT)
and the database of mean ligand-removal enthalpies in transition-metal
complexes (TM) by Johnson and Becke.\cite{johnson2009b} The results
can be used as an estimate of the performance of different functionals
for those common chemical problems. 

As mentioned earlier, functionals, in general, perform better in the
order: range-separated $>$ hybrids $>$ meta-GGAs $>$ GGAs. Range-separated
hybrids partially address the problem with self-interaction error,
that is particularly relevant in the barrier height set
(BH). LC-$\omega$PBE achieves an excellent result, and so does
BHandHLYP at the hybrid level. However, the good performance of
BHandHLYP for self-interaction error problems comes at a cost.
It fails spectacularly for atomization energies (G3) and
ligand-removal energies in transition metal complexes. The latter
failure is caused by the multideterminant character of these systems,
whose correlation is roughly approximated by semilocal density
functionals.

\subsection*{Failure of Conventional DFT for Non-Covalent Interactions}
\label{s:dftfail}
The last twenty years have been a bright story of success for
density-functional theory.\cite{burke2012} DFT can, at a relatively
modest computational cost, give a reliable picture of such diverse
properties as structures of molecules and solids, excitation energies,
spectroscopic properties, reaction energies, and so on. DFT is nowadays used widely
in the physics and chemistry communities, with the most popular
density functionals (B3LYP in gas-phase chemistry and PBE in condensed
matter) having more than three thousand citations every year and
growing.\cite{burke2012} Despite its popularity, current
density-functional approximations have well-known
shortcomings,\cite{cohen2008,cohen2011,burke2012} including the
inability to calculate noncovalent interactions accurately.

The first studies of the applicability of common density-functionals
to noncovalent interactions were carried out in the
1990s.\cite{lacks1993,kristyan1994,hobza1995,%
seponer1996,perezjorda1995,perezjorda1999,couronne1999} Among the
first works were the articles by Lacks and Gordon\cite{lacks1993} and
Kristy\'{a}n and Pulay,\cite{kristyan1994} which serve as an
illustration of the state of the DFT field at the time as well as of
some of the problems dispersion functionals face today. Lacks and
Gordon\cite{lacks1993} showed that common exchange functionals
reproduce the exact exchange energy of noble gases to within 1\%\. 
However, these variations in the exchange contributions stand out against the very small binding energies in the noble gas dimers. This results in exchange contributions to the
binding energies that can range from 0 to more than 100 \%\ of the exact exchange.\cite{hepburn1975} Kristy\'{a}n and Pulay\cite{kristyan1994} tried to reproduce
the binding energy curves of the noble-gas dimers He$_2$, Ne$_2$, and
Ar$_2$, only to find that all GGAs and B3LYP are repulsive, to a
varying extent, whereas LDA overbinds these systems significantly.

\begin{figure}
\includegraphics[width=0.48\textwidth]{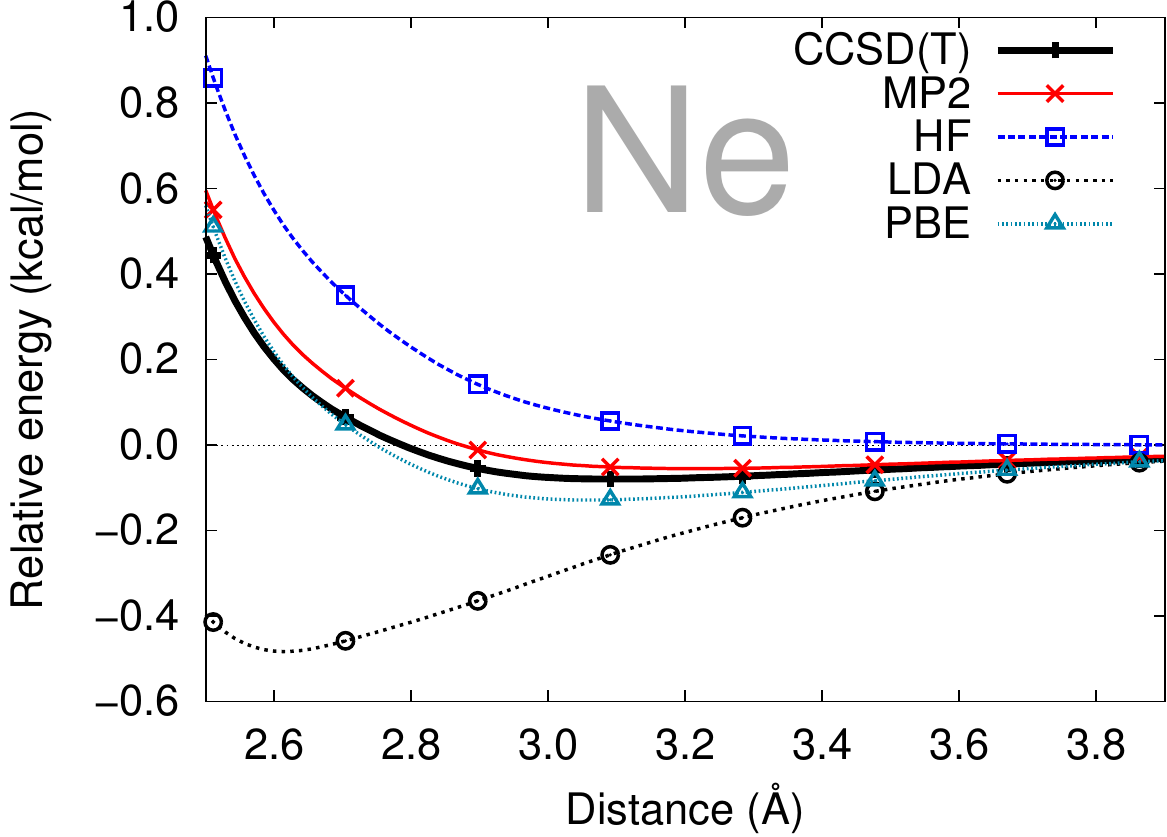}
\includegraphics[width=0.48\textwidth]{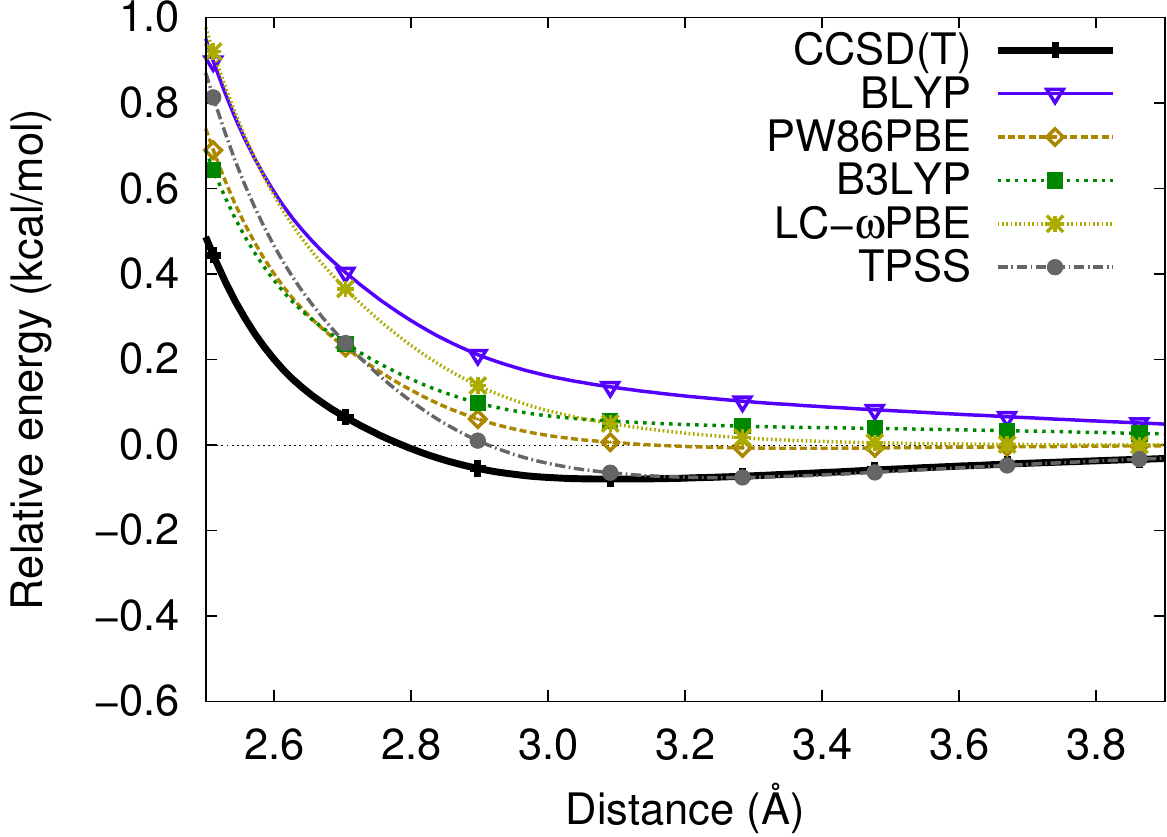}\\
\includegraphics[width=0.48\textwidth]{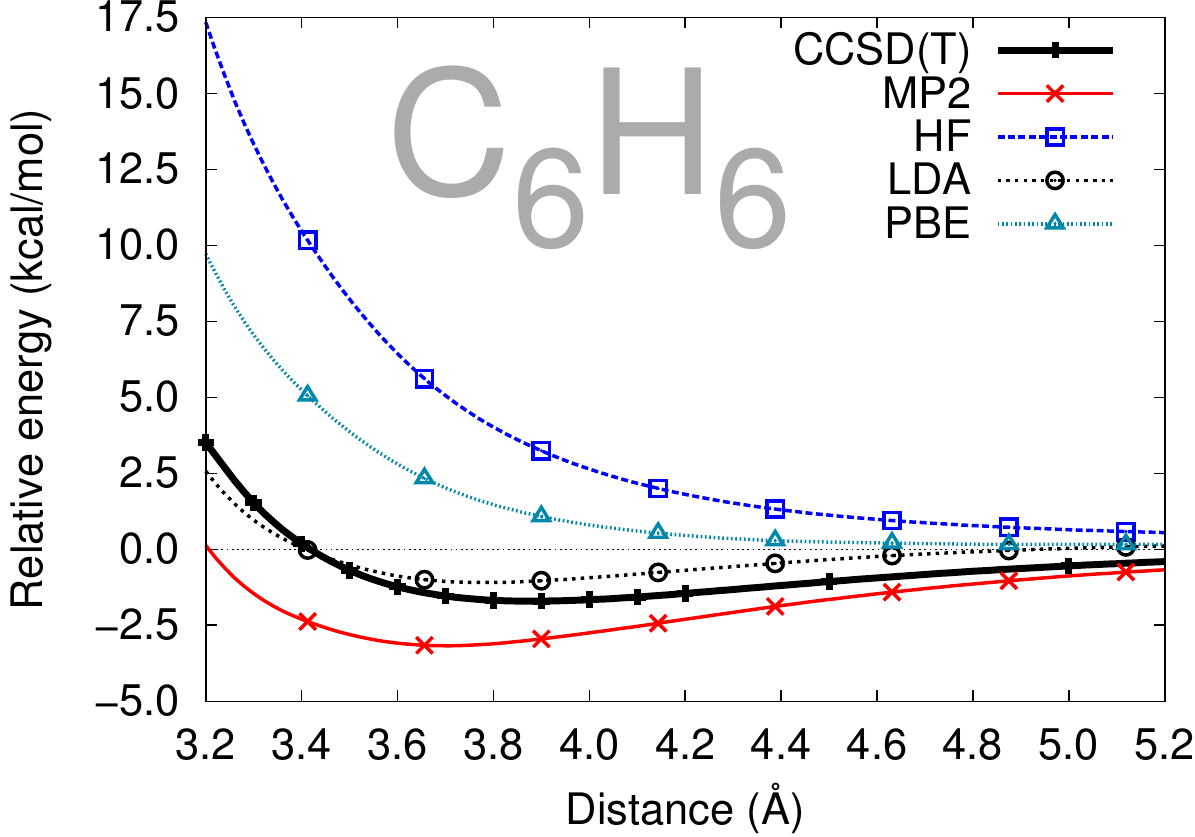}
\includegraphics[width=0.48\textwidth]{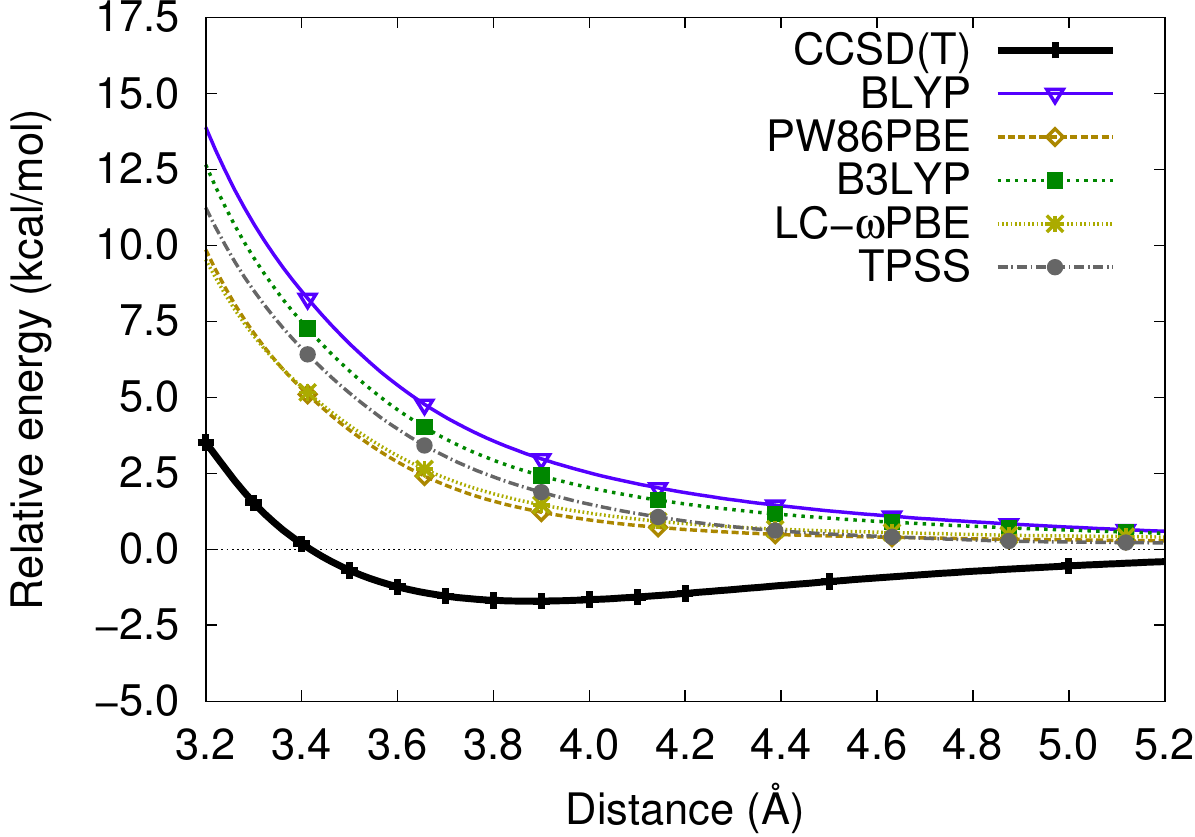}\\
\includegraphics[width=0.48\textwidth]{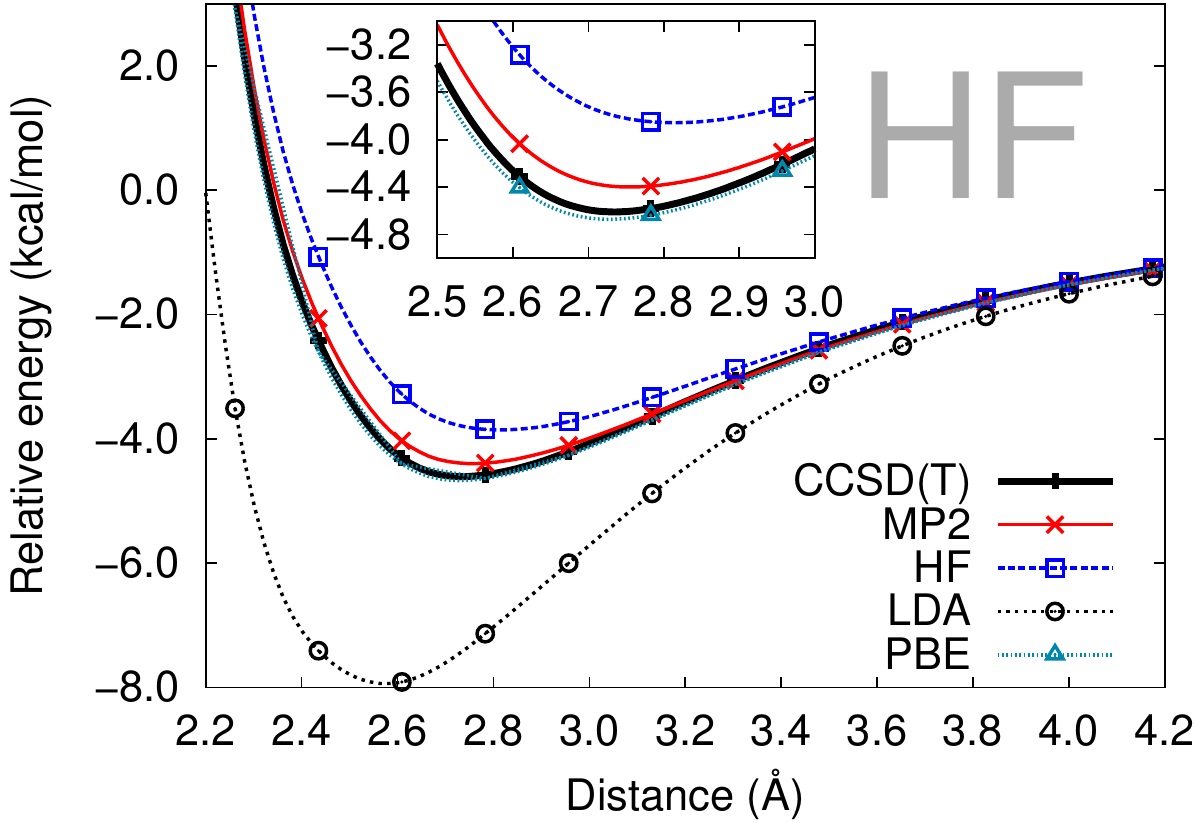}
\includegraphics[width=0.48\textwidth]{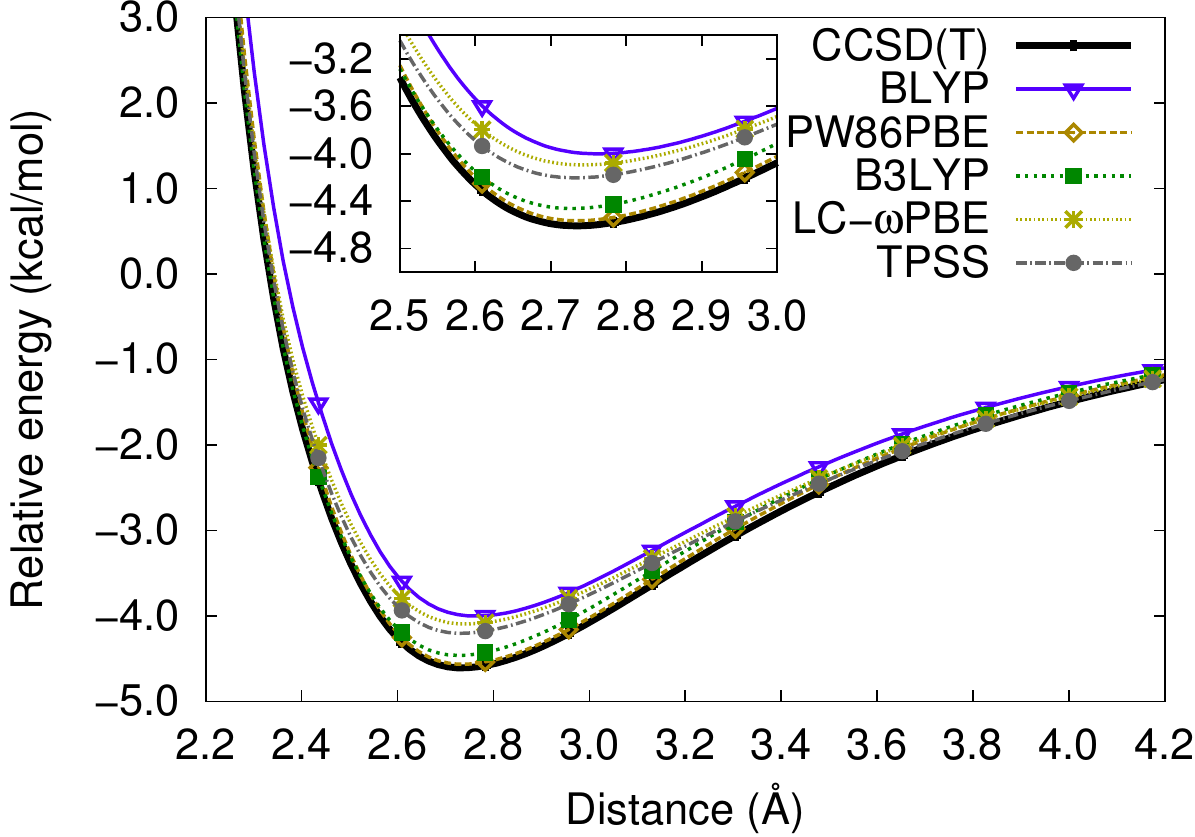}
\caption{Binding energy curves of neon (top), benzene (middle), and
hydrogen fluoride (bottom) dimers, calculated using wavefunction
theory methods, LDA, and PBE (left), and other common density
functionals (right). Note the different energy scales in the three
dimers. For benzene dimers, all DFT calculations were run at
aug-cc-pVTZ level, MP2 calculations used counterpoise-corrected
aug-cc-pVQZ, and the CCSD(T) results come from
ref.~\citenum{sherrill2009}. For the HF dimer, DFT and HF calculations
used aug-cc-pVTZ, MP2 and CCSD(T) used counterpoise-corrected
aug-cc-pVQZ. Ne dimer was run using aug-cc-pV5Z in all cases
(counterpoise-corrected in MP2 and CCSD(T)).
\label{fig:1c:repulsive}} 
\end{figure}

The picture is equally dismal for other types of noncovalent
interactions as well. Figure~\ref{fig:1c:repulsive} shows the
performance of several common density-functional approximations in the
calculation of three kinds of intermolecular interactions. Ne dimer,
because of its closed-shell electron configuration, serves as the
prototypical example of binding coming exclusively from
dispersion. Hartree-Fock (HF), which contains no correlation at all
(dispersion or otherwise), predicts a purely repulsive curve, while
MP2 and CCSD(T) predict roughly the correct binding. The CCSD(T)
minimum (our accurate reference) is at 3.09\AA\ interatomic
distance. The binding energy at equilibrium is small (0.084 kcal/mol)
but enough to crystallize neon into a closed-packed face-centered
cubic structure at low temperature (24.56 K\cite{kittel}) and zero
pressure.

All density functionals fail to correctly describe the dispersion
binding behavior in Ne dimer. LDA is spuriously attractive while GGAs
show a range of behaviors from overly attractive (PBE) to more
repulsive than HF (BLYP). The accuracy does not improve by using more
sophisticated functionals: meta-GGA functionals, hybrids, and
range-separated hybrids all fail to describe the binding in neon
dimer. This result is hardly surprising:\cite{lacks1993,kristyan1994}
LDAs and GGAs are semilocal approximations---it is not in their design
to account for dispersion interactions, which are long-range
correlation effects. Incorporating HF exchange in one form or another
does not help either.

The results are equally disappointing for $\pi$-$\pi$
interactions. Figure~\ref{fig:1c:repulsive} shows the binding energy
curve for the stacked configuration of the benzene dimer. Most of the
stabilization comes from dispersion, and CCSD(T) predicts a binding
energy of 1.681 kcal/mol (this configuration of benzene dimer is not
the most stable, but $\pi$-$\pi$ stacks are a particularly important
motif in biological systems\cite{johnson2009}). Again, all functionals
except LDA are purely repulsive to varying degrees. LDA gives an
answer close to the correct result, which explains its popularity in
modelling graphene-based systems in the
past.\cite{ortmann2006} Interestingly, MP2 grossly overbinds the
$\pi$-$\pi$ interaction---an effect that is found whenever the
monomers have low-lying excited states, for known
reasons.\cite{cybulski2007,tkatchenko2009b} The attractive or repulsive
character of the functionals follows the same trend as in the Ne dimer,
with PBE giving the most binding and BLYP the most repulsive. 

The third interaction type are hydrogen bonds, represented by the hydrogen
fluoride dimer in Figure~\ref{fig:1c:repulsive}. Because of their
strength relative to the rest of the van der Waals interactions,
hydrogen bonds usually dominate molecular aggregation and, as a
consequence, they are prevalent in supramolecular (e.g. molecular
crystal packing, crystal engineering) and biological systems (protein
folding, DNA structure and function), making their accurate
representation extremely important. The performance of various density
functionals for the hydrogen fluoride dimer, with a binding energy of
4.57 kcal/mol at equilibrium, is not as bad as for
dispersion-dominated dimers. Except for LDA, all density functionals
perform relatively well in absence of dispersion. This is reasonable
because hydrogen bonding is dominated by electrostatic and orbital
interactions, and HF alone obtains almost 4 kcal/mol of the binding
energy. A closer look reveals that all functionals are underbinding
except for PBE by an amount of up to 0.7 kcal/mol, which points to the
missing dispersion attraction.

Despite the reasonable performance for hydrogen bonding, common density
functionals have serious difficulties in bonding
hydrogen-bonded systems accurately. These problems surface in the modeling of more
complex systems, as, for instance, the overstructuring and overly small
diffusion coefficient in molecular dynamics simulations of liquid
water,\cite{grossman2004} the incorrect relative energies of the
phases of ice,\cite{santra2011} and the incorrect energy ranking of
water hexamer structures.\cite{santra2007,oterodelaroza2013b} Up to a
point, dispersion has been proven able to correct, at least partially,
some of these problems,\cite{santra2011,oterodelaroza2013b} but the
accurate modelling of extended hydrogen-bonded networks is still a
challenge in DFT.

In summary, traditional density-functionals perform reasonably well
for thermochemistry or reaction barriers but, as we have just shown,
are unreliable when it comes to the calculation of noncovalent
interaction energies. These difficulties are partially caused by
noncovalent binding being significantly weaker than covalent
binding. Errors from approximate functionals are much more significant
relative to a noncovalent bond energy than to a covalent bond. The
simplest way to account for the missing dispersion energy is to assume
that the density functional (in the following called the \emph{base}
functional) is approximately accounting for the other noncovalent
interactions and to simply add the dispersion term separately:
\begin{equation}
 \label{eq:etotal}
 E = E_{\text{base}} + E_{\text{disp}}  
\end{equation}
The quantity $E_{\text{disp}}$ is the \emph{dispersion correction} to
the base functional, and has to account for the missing dispersion
energy as well correct the behavior and uncontrolled effects on the
binding energies coming from the base functional. 

The erratic behavior of various functionals for different types of
interactions is exemplified in Figure~\ref{fig:interaction_types}. The
performance of different functionals improves, and the spread in the
average errors decreases, with the amount of binding that is accounted
for by the interaction of the ground-state charge distributions. In
this way, hydrogen bonds, which are dominated by electrostatic
interactions, are relatively well modeled. On the other hand,
dispersion interactions, which arise from instantaneous dipoles on
both molecules, present the largest average errors and spread.

Many dispersion corrections have been proposed over the last 15
years,\cite{johnson2009,grimme2011} some of which we consider in the
following sections. The simplest of those corrections is to use the
the asymptotic form of the dispersion energy, with a leading $R^{-6}$
term, to capture the long-range interaction between atoms and
molecules. This approximation is the basis of the pairwise dispersion
corrections, and works surprisingly well considering the simplicity of
the premise. We review this approach next.

\section*{Non-Covalent Interactions in DFT}
\label{s:ncidft}

\subsection*{Pairwise Dispersion Corrections}
\label{s:pairwise}
The accurate and efficient calculation of the dispersion energy
($E_{\text{disp}}$ in Eq.~\ref{eq:etotal}) is a complex problem
in the context of density-functional theory. The physical origin of
dispersion is the interaction between instantaneous dipoles in two
different molecules or fragments of the same molecule. These
instantaneous dipoles are created by short-lived molecular
excitations. As a consequence, dispersion is strictly a long-range
electron correlation effect, meaning it arises from the correlated
movement of electrons in two different molecules (or distant parts of
the same molecule). Common correlation functionals are based on local
approximations (i.e. the energy density at a point in space depends on
local properties such as the density or the gradient at that point)
and are, consequently, unable to model dispersion by design.

For two sufficiently-separated interacting neutral atoms, the
dispersion energy is always attractive, and decays as the sixth power
of the intermolecular distance:
\begin{equation}
 \label{eq:eqsimple6}
 E_{\text{disp}} = - \frac{C_6^{AB}}{R_{AB}^{-6}}
\end{equation}
The asymptotic behavior can be proven using simple arguments like, for
instance, the interaction of two coupled harmonic
oscillators,\cite{kittel} or second-order perturbation
theory.\cite{atkins} In Eq.~\ref{eq:eqsimple6}, the $C_6^{AB}$
are the dispersion coefficients (or simply the interaction
coefficients), and can be estimated using London's
formula:\cite{atkins}
\begin{equation}
\label{eq:london}
C_6^{AB} = \frac{3}{2} \left(\frac{I_AI_B}{I_A+I_B}\right)\alpha_A\alpha_B
\end{equation}
where $I$ are the atomic ionization potentials and $\alpha_A$ are the
atomic polarizabilities. 

The values calculated using London's formula are approximate estimates
with little practical value. The formula contains qualitative
information, namely, the interaction coefficient depends directly on
the polarizabilities of both atoms, which in turn are proportional to
the atomic size and the number of valence electrons.\cite{atkins} A
more accurate value for the interaction coefficients can be obtained
from second-order perturbation theory:\cite{atkins}
\begin{equation}
C_6^{AB} = \frac{2}{3}\sideset{}{'}\sum_{n_An_B} 
\frac{(\bm{\mu}_{A,0n_A}\cdot\bm{\mu}_{A,n_A0})(\bm{\mu}_{B,0n_B}\cdot\bm{\mu}_{B,n_B0})}{\varepsilon_{n_A} + \varepsilon_{n_B}}
\end{equation}
where the sum runs over the excited states of atoms A and B,
$\bm{\mu}_{A,0n_A} = \langle 0|\bm{\mu}_A|n_A\rangle$ is the
transition dipole moment, and $\varepsilon_{n_A} = E_{n_A} - E_0$ is
the excitation energy for the $n_A$ state of molecule A. The
$C_6^{AB}$ coefficients can be calculated more rigorously, using the
Casimir-Polder formula:\cite{casimir1948}
\begin{equation}
\label{eq:c6cp}
C_6^{AB} = \frac{3}{\pi}\int_0^{\infty} \alpha_{1,A}(i\omega) \alpha_{1,B}(i\omega) 
\end{equation}
that involves the atomic frequency-dependent polarizabilities
$\alpha(i\omega)$. This formula is consistent with the familiar
picture of dispersion arising from induced-dipole interactions on A
and B. These quantities model the response of the atom under
frequency-dependent electric fields, and are not directly available in
time-independent DFT, although models exist for their calculation
(e.g. ref.~\citenum{tao2012}). The Casimir-Polder formula can be
generalized to the calculation of higher-order dispersion
contributions:
\begin{equation}
\label{eq:casipol}
C_{2n}^{AB} = \sum_{l=1}^{n-2}\frac{(2n-2)!}{2\pi (2l)! (2n-2l-2)!}
\int_0^{\infty}\alpha_l^A(i\omega)\alpha_{n-l-1}^B(i\omega) d\omega
\end{equation}
that involves the higher order ($2^l$-polar) dynamic
polarizabilities. 

Equation~\ref{eq:eqsimple6} can be generalized to the interaction
between molecules, or distant fragments of the same molecule, by
considering that all atoms in the system interact with one another in
a pairwise fashion: 
\begin{equation}
 \label{eq:epairwise6}
 E_{\text{disp}} = - \sum_{A>B}C_6^{AB} R_{AB}^{-6} f_6(R_{AB}) - ...
\end{equation}
The atom-based calculation of the dispersion energy has proven to be
an excellent approximation, even possibly accounting for the missing
anisotropy in the dispersion interaction
coefficients\cite{zgarbova2010,grimme2011} (i.e., the dependence of
the dispersion interaction coefficients on the relative orientation of
the interacting molecules). Equation~\ref{eq:epairwise6} contains not
only the $R^{-6}$ contribution but also less important terms
involving interactions of order higher than the dipole-dipole
(dipole-quadrupole, quadrupole-quadrupole,...). These terms are
important,\cite{johnson2006a} and involve the coefficients in
equation~\ref{eq:casipol} for $n > 3$, but for simplicity we will
momentarily consider only the leading term. The higher-order terms are
considered in a forthcoming section titled ``The Exchange-Hole Dipole Moment (XDM) Model". 

The $f_6$ factor is called the \emph{damping function} and is a
one-dimensional function of the interatomic distance that goes to zero
when $R_{AB}\to 0$ and to one when $R_{AB}\to \infty$. This function has two roles in
dispersion-corrected DFT: one is to
correct for the error introduced by the approximations leading up to
equation~\ref{eq:epairwise6}, an important one being that the
interacting atoms are not infinitely separated. The other is to
deactivate the dispersion contribution at very short range to avoid
the singularity at $R_{AB} = 0$. As will be shown, the damping
function also performs the role of fixing the problems the base
functional has in reproducing other terms in the intermolecular
interaction energy (e.g. electrostatics) through its adjustable
parameters. Figure~\ref{fig:1c:repulsive} illustrates how different
functionals treat the non-dispersion part of the binding
energy. Taking, for instance, the example of the Ne dimer, the behavior of
the functional can range from spuriously attractive (PBE) to extremely
repulsive (BLYP) compared to the HF repulsive wall. The same effects
are observed in the benzene dimer and, to a lesser extent, in the HF dimer.

The energy correction in equation~\ref{eq:epairwise6} coupled with
first-principles simulations at the HF level was used for the first
time by Scoles et al. in the 1970s.\cite{hepburn1975,ahlrichs1977} In
parallel, pairwise dispersion corrections were also applied as
additions to the Gordon-Kim model\cite{gordon1972}---an approximate
model applied to the sum of frozen molecular electron densities that
uses the uniform electron gas exchange, correlation and kinetic energy
expressions in order to calculate intermolecular potentials. The
dispersion-corrected Gordon-Kim model was first proposed by
Rae,\cite{rae1973} and subsequently refined by Cohen and
Pack.\cite{cohen1974} The objective of these early studies was
simply to model the repulsive wall in noble gas dimers (and the
triplet of H$_2$) using the HF or the Gordon-Kim energies, and the
attractive part using the dispersion correction. It is also important
to note that a van der Waals term in the form of a Lennard-Jones
potential, which uses equation~\ref{eq:epairwise6} for the attractive
part, is also used in practically all classical molecular force
fields,\cite{halgren1992} such as 
CHARMM\cite{brooks2009} and AMBER.\cite{pearlman1995}

Because HF is missing all electron correlation it is not suitable for
the treatment of general intermolecular interactions, regardless of
the presence or absence of a dispersion correction, and this has limited
the applicability of the Hartree-Fock-Dispersion (HFD) methods. As a
consequence, in later years, the modelling of dispersion was replaced
by post-HF wavefunction calculations (e.g. MP2,
coupled-cluster, etc.). When the inability of DFT to model noncovalent
interactions became apparent in the 1990s, pairwise energy corrections
became popular as a straightforward and reasonably accurate way of
correcting for the missing dispersion. Among the first studies to 
include these terms are those by Gianturco et
al.\cite{gianturco1999} on the potential energy surface of the Ar--CO
dimer, and Elstner et al.\cite{elstner2001} and Wu and
Yang\cite{wu2002} on small molecules (rare gas dimers, DNA base
pairs, etc.), and Wu et al.\cite{wu2001}

At the heart of all pairwise dispersion corrections is the calculation
of the interatomic interaction coefficients $C_6^{AB}$. The
interaction coefficients traditionally used in classical force fields
are not adequate because they are treated as fitted parameters, rather
than physical quantities, so they account for other effects in
addition to dispersion\cite{wu2002} and therefore are widely variable 
across different classical force fields.\cite{halgren1992}

Early pairwise-dispersion-corrected DFT studies adapted a method
proposed by Halgren\cite{halgren1992} employing the
Slater-Kirkwood formula:\cite{slater1931}
\begin{equation}
\label{eq:slaterkirkwood}
C_6^{AB} = \frac{3}{2}\frac{\alpha_A\alpha_B}{
\left(\alpha_A/N_A\right)^{1/2} + 
\left(\alpha_B/N_B\right)^{1/2}}
\end{equation}
in which $N_A$ is the effective number of valence electrons ($N_A$ is
smaller than the actual number of valence electrons and not directly
calculable). Halgren proposed using empirical formulas for $N_A$,
together with accurate atomic polarizability data and the combination
rule derived from equation~\ref{eq:slaterkirkwood} for the mixed
coefficients:
\begin{equation}
\label{eq:slatercombination}
C_6^{AB} = \frac{2\alpha_A\alpha_BC_6^{AA}C_6^{BB}}{
\alpha_AC_6^{BB} + \alpha_BC_6^{AA}}
\end{equation}

In their seminal work, Wu and Yang obtained the interaction
coefficients by fitting the atomic $C_6^{AA}$ to molecular interaction
coefficients (that can be obtained as sums of the atomic
$C_6$\cite{halgren1992}), which in turn had been calculated from
experimental dipole-oscillator strength distribution measurements by
Meath et al.\cite{zeiss1977,margoliash1978,jhanwar1980,%
jhanwar1982,kumar1985a,kumar1985b,kumar1992,kumar1996,kumar1997,kumar2002a,%
kumar2002b,kumar2003a,kumar2003b} By using least-square fitting, the
authors obtained molecular $C_6$ in excellent agreement with
experimental data (1\% mean absolute errors for hydrocarbons) and
binding energies with an accuracy comparable to MP2. This method is
similar to the procedure followed in classical force field
calculations and, although not generalizable, the early articles
proved that the idea of adding a simple pairwise dispersion correction
to an unrelated density functional is not only valid, but gives
intermolecular interaction energies with an accuracy that is at least
as good as MP2.

In the last ten years, a number of approximations have been proposed
for the $C_6$ and higher-order coefficients with varying degrees of
accuracy and empiricism. Some of these have been turned into
full-fledged dispersion corrections by parametrization of an
associated damping function to routinely-used functionals. We will
review some of the most popular in the following sections. The list
includes:
\begin{itemize}
\item The exchange-hole dipole moment (XDM) model of dispersion, which
calculates the $C_6$, as well as higher-order coefficients, without
any empirical parameters.
\item Grimme's DFT-D\cite{grimme2004} and
DFT-D2,\cite{grimme2006} with empirical fixed $C_6$ coefficients, and
DFT-D3\cite{grimme2010} with a model that introduces the dependence
of the coefficients on the molecular geometry. The approach by Ortmann,\cite{ortmann2006} which was
popular in condensed-matter, and similar in spirit to DFT-D2.
\item The Tkatchenko-Scheffler approach in its first
version\cite{tkatchenko2009} (2009): the $C_6$ coefficients are
obtained from reference data but they are made geometry-dependent by
using the direct relation between polarizability and atomic volume.
\item The method by Tao, Perdew and Ruzsinszky\cite{tao2012} that
calculates the frequency-dependent polarizabilities non-empirically
using a model consisting of a metallic sphere of uniform density. The
frequencies yield the coefficients through equation~\ref{eq:c6cp}.
\end{itemize}

The pairwise dispersion correction in equation~\ref{eq:epairwise6} can
be generalized by considering the complete multipolar expansion of the
intermolecular interaction. The generalized dispersion energy is
written as a sum of 2-body, 3-body, etc. terms:
\begin{equation}
 \label{eq:enbody}
 E_{\text{disp}} = E_{\text{disp}}^{(2)} + E_{\text{disp}}^{(3)} + ... 
\end{equation}

The leading term in this expansion is the pairwise
interaction, which contains terms of order higher than the
dipole-dipole:
\begin{align}
 \label{eq:epairwise}
 E_{\text{disp}}^{(2)} & = E_6^{(2)} + E_8^{(2)} + E_{10}^{(2)} + ...
 \nonumber \\
 & = - \sum_{n=6,8,10,...} \sum_{A>B}C_n^{AB} R_{AB}^{-n} f_n(R_{AB})
\end{align}
The simple damping is replaced by a family of functions $f_n$ that,
again, account for the approximate nature of the multipolar expansion
at short range. 

Dispersion corrections with fixed interaction coefficients benefit
from the simplicity in the implementation. Indeed, taking the
derivatives of the energy (up to any order) is trivial and the
programming is equally easy, which has undoubtedly contributed to the
popularity of these methods. In addition, the dispersion contribution
is relatively minor in ``thermochemical'' cases, where there is
breaking or formation of covalent bonds. In those cases, the good
performance of the base functional is retained, which allows for the
treatment a wide range of chemical problems on equal footing.  Despite
the simplicity of the approximation, the results are surprisingly
accurate and methods where the parameters in the base functional are
optimized together with dispersion such as B97D\cite{grimme2006} and
$\omega$B97XD\cite{chai2008} see a widespread use nowadays.

A further advantage of pairwise dispersion corrections is that the
asymptotic $R^{-6}$ tail of the interaction energy is captured by
design. This is in contrast with methods based on modifications of the
existing base functionals (like the Minnesota functionals). The correct $R^{-6}$ dependence is
important in large condensed systems, such as molecular crystals or
biological macromolecules, but has little or no consequence in small
systems. A related limitation is that $R^{-6}$ is, for particular
systems, not the correct asymptotic limit of the dispersion
interaction. This happens in polarizable extended systems,
particularly in metal surfaces\cite{dobson2012} because of the
collective motion of the extensively delocalized electrons. The
pairwise-correction results for the binding in graphite at
equilibrium, however, are rather
accurate.\cite{barone2009,oterodelaroza2012a}

Another possible advantage of pairwise approaches is that the relative
values of the dispersion contribution to binding might give
``insight''\cite{grimme2011} into the nature of noncovalent bonding,
although the extent to which this insight is significant is arguable
since at equilibrium distances the base functional also contributes to
binding and might contain spurious contributions that are absorbed by
the damping function. The dispersion contribution to the binding
energy is always attractive. In particular cases involving
hydrogen-bonded systems, the base functional may already overestimate
the binding energy, in which case the dispersion correction will only
lead to poorer agreement with the reference binding energies,
regardless of the shape of the damping function (see
Figure~\ref{fig:1c:repulsive} and the last section of this chapter. 

Some aspects to consider regarding pairwise dispersion corrections
are: i) the interaction coefficients are known to depend upon the
chemical environment, as already noted by Wu and Yang\cite{wu2002} and
others,\cite{johnson2011} ii) higher-order two-body interactions
involving the $C_8$, $C_{10}$, etc. coefficients are known to give a
non-negligible contribution to the energy as
well,\cite{johnson2006a,grimme2011} and iii) depending on how the
interaction coefficients are calculated, there may be no simple way to
include the dispersion effects back into the density in the
self-consistent procedure. For a self-consistent implementation of the
dispersion functional in equation~\ref{eq:etotal}, it is necessary to
add a dispersion potential to the one-electron Hamiltonian
(Eq.~\ref{eq:hks1}), corresponding to the functional derivative of
$E_{\text{disp}}$ with respect to the density:
\begin{equation}
V_{\text{disp}} = \frac{\delta E_{\text{disp}}}{\delta\rho(\bm{r})}
\end{equation}
Because the dispersion forces are relatively small, the effect of the
dispersion potential on the self-consistent electron density is
relatively
minor,\cite{thonhauser2007,kong2009,ikabata2013,bremond2014}
justifying the calculation of the dispersion energy after the
self-consistent field procedure (post-SCF), which is far simpler.

In the following sections, we review the most popular approaches to
calculate the dispersion energy using pairwise energy expressions
(equations~\ref{eq:enbody} and~\ref{eq:epairwise}). These methods
comprise two components: i) a way of calculating or estimating the
interaction coefficients $C_n$, and ii) an expression for the damping
function that depends upon a number of adjustable coefficients, which
are empirical and must be fitted to a training set of
reference data (usually small dimers calculated using accurate
wavefunction methods). Normally, the training sets for the damping
function parametrization are small, in accordance with the likewise
small number of parameters in the models. The empirical parameters
for the damping function transfer relatively well to other
noncovalently bound dimers not in the parametrization set, making the
pairwise approach fairly easy to generalize to all atoms in the
periodic table.

\subsubsection*{The Exchange-Hole Dipole Moment (XDM) Model}
\label{s:xdm}
The exchange-hole dipole-moment (XDM) model of
dispersion\cite{becke2005a,%
becke2005b,johnson2005,becke2006a,johnson2006a,johnson2006b,becke2007a,%
becke2007b,kannemann2009,kannemann2010,oterodelaroza2012a,%
oterodelaroza2013a} was proposed in 2005 by Becke and
Johnson\cite{becke2005a,becke2005b} and developed in subsequent papers
into a practical approach to correct density functionals for
dispersion effects. The XDM model in its current formulation is a
semilocal functional (a meta-GGA) that gives the interaction
coefficients $C_n$ strictly from first principles, without intervening
empirical parameters.

An essential component of the XDM model is the exchange or Fermi hole: 
\begin{equation}
\label{eq:holewmat}
 h_{X\sigma}(\bm{r}_1,\bm{r}_2) = -\frac{
|\rho_{1\sigma}(\bm{r}_1,\bm{r}_2)|^2
}{
\rho_{\sigma}(\bm{r}_1)
}
\end{equation}
where $\rho_{1\sigma}$ is the one-electron spin density matrix, and
$\rho_{\sigma}$ is the $\sigma$-spin electron density. In the usual
one-determinant representation used in Kohn-Sham DFT,
\begin{equation}
\label{eq:holeworbs}
 h_{X\sigma}(\bm{r}_1,\bm{r}_2) =
-\frac{1}{\rho_{\sigma}(\bm{r}_1)}
\sum_{ij}\psi_{i\sigma}(\bm{r}_1)\psi_{j\sigma}(\bm{r}_1)
\psi_{i\sigma}(\bm{r}_2)\psi_{j\sigma}(\bm{r}_2)
\end{equation}
which involves a double sum over the occupied spin-orbitals
($\psi_{i\sigma}$). 

Given an electron of spin $\sigma$ at the reference point $\bm{r}_1$,
the exchange-hole represents the probability depletion of finding a
same-spin electron at $\bm{r}_2$. The exchange-hole is always
negative, and has well-known properties:
\begin{enumerate}
\item The on-top depth condition:
\begin{equation}
\label{eq:ontop}
h_{X\sigma}(\bm{r},\bm{r}) = -\rho_{\sigma}(\bm{r})
\end{equation}
establishes that, at the reference point, the hole excludes exactly
the amount of electron density at that point. This is a local version 
of the Pauli exclusion principle. 
\item The hole depletes exactly one electron:
\begin{equation}
\label{eq:holenorm}
\int h_{X\sigma}(\bm{r}_1,\bm{r}_2) d\bm{r}_2 = -1
\text{ for all } \bm{r}_1
\end{equation}
\item The associated exchange energy is:
\begin{equation}
E_x = \frac{1}{2} \sum_{\sigma} \int \rho_{\sigma}(\bm{r}_1)
\frac{h_{X\sigma}(\bm{r}_1,\bm{r}_2)}{r_{12}} 
d\bm{r}_1 d\bm{r}_2
\end{equation}
with $r_{12}$ the interelectronic distance.
\end{enumerate}

Let us assume two neutral non-overlapping atoms A and B. The key idea
in XDM is that the dispersion energy originates from the interaction
of the real-space electrostatic distributions generated by the
electrons and their associated exchange holes. At any point $\bm{r}$,
there is a negative charge equal to $\rho_{\sigma}(\bm{r})$, and an associated
positive distribution represented by the exchange-hole at that
reference point equal to
$\rho_{\sigma}(\bm{r})h_{X\sigma}(\bm{r},\bm{r}^{\prime})$. The hole integrates
to -1, so the leading contribution to the electrostatic potential from
that point is the dipole formed by the hole and the electron:
\begin{equation}
\bm{d}_{X\sigma}(\bm{r}) = \int \bm{r}^{\prime} h_{X\sigma}(\bm{r},\bm{r}^{\prime})
d\bm{r}^{\prime} - \bm{r}
\end{equation}
Hence, the dispersion interaction in XDM originates from the asymmetry of the
exchange-hole.\cite{becke2005a}

With the definitions above, and under the assumption that the
exchange-hole dipole is directed towards the closest nucleus, it is
relatively straightforward to apply classical electrostatic
arguments\cite{becke2007b} to calculate the square of the $l$-pole
operator: 
\begin{equation}
\langle M_l^2 \rangle_A = \sum_{\sigma} 
 \int \rho_{\sigma}(\bm{r}) [r_A^l - (r_A - d_{X\sigma})^l]^2 d\bm{r}
\label{eq:xdmmom}
\end{equation}
In this way, the multipoles can be calculated up to any order using
only the norm of the exchange-hole dipole. The squared moments are
then used to obtain the dispersion interaction 
coefficients:\cite{becke2007b}
\begin{align}
\label{eq:xdmc6}
 C_{6}^{\text{AB}} & = \frac{\alpha_A\alpha_B\langle M_1^2\rangle_A\langle
   M_1^2\rangle_B}{\langle M_1^2\rangle_A\alpha_B + \langle
   M_1^2\rangle_B \alpha_A} \\
\label{eq:xdmc8}
 C_{8}^{\text{AB}} & = \frac{3}{2}\frac{
\alpha_A\alpha_B\left(\langle M_1^2\rangle_A \langle M_2^2\rangle_B +
\langle M_2^2\rangle_A \langle M_1^2\rangle_B \right)
}{
\langle M_1^2\rangle_A \alpha_B + \langle M_1^2\rangle_B \alpha_A
} \\
\label{eq:xdmc10}
 C_{10}^{\text{AB}} & = 2 \frac{
\alpha_A\alpha_B \left(\langle M_1^2\rangle_A \langle M_3^2\rangle_B +
\langle M_3^2\rangle_A \langle M_1^2\rangle_B \right)
}{
\langle M_1^2\rangle_A \alpha_B + \langle M_1^2\rangle_B \alpha_A
} 
+ \frac{21}{5} \frac{
\alpha_A\alpha_B \langle M_2^2\rangle_A \langle M_2^2\rangle_B
}{
\langle M_1^2\rangle_A \alpha_B + \langle M_1^2\rangle_B \alpha_A
}
\end{align}
where $\alpha$ are the atomic polarizabilities (see below). By using a
model of dispersion based on the electrostatic interaction of
electrons and holes, the dispersion interaction in XDM can be
calculated without recourse to time-dependent or excited state
calculations.

The scheme above depends upon the definition of fragments A and B, but
it is far more practical to assign interaction coefficients to atoms
instead of molecules. To do this, Johnson and Becke\cite{johnson2005}
proposed to make use of the Hirshfeld partitioning
scheme:\cite{hirshfeld1977}
\begin{equation}
 \omega_A(\bm{r}) = \frac{\rho_A^{\text{at}}(\bm{r})}{\sum_B\rho_B^{\text{at}}(\bm{r})}
\end{equation}
where $A$ is an atom in a molecule, $\omega_A$ is the Hirshfeld
weight, $\rho_A^{\text{at}}$ is the \emph{in vacuo} atomic density of
A and the denominator is the promolecular density (the sum of the
\emph{in vacuo} atomic densities at the molecular geometry). The
Hirshfeld weights enter the moment equations:
\begin{equation}
\langle M_l^2 \rangle_A = \sum_{\sigma} 
 \int \omega_A(\bm{r}) \rho_{\sigma}(\bm{r}) [r_A^l - (r_A - d_{X\sigma}(\bm{r}))^l]^2 d\bm{r}
\end{equation}
and these are subsequently used to calculate atomic dispersion
coefficients using equations~\ref{eq:xdmc6} through~\ref{eq:xdmc10}. 

The atom-in-molecule polarizabilities are obtained in a similar way
by using the known direct proportionality between polarizability and
volume\cite{johnson2005} (see ref.~\citenum{kannemann2012} and
references therein for details): 
\begin{align}
\label{eq:polarizv}
 \alpha_A & = \frac{V_A}{V_A^{\text{at}}} \alpha_{A}^{\text{at}}\\
\label{eq:vxdm} V_A & = \int r^3\omega_A({\bf r})\rho({\bf r})d{\bf r} \\
 V_A^{\text{at}} & = \int r^3\rho_A^{\text{at}}({\bf r}) d{\bf r}
\end{align}
where $\alpha_{A}^{\text{at}}$ is the free-atom polarizability and the
fraction measures the volume occupied by atom A in the molecular
environment ($V_A$) in relation to the same atom in the vacuum
($V_A^{\text{at}}$).

The computation of the $C_n$ coefficients using the exchange hole in
equation~\ref{eq:holeworbs} involves a double sum over occupied
orbitals, which is computationally expensive, particularly in periodic
plane wave approaches that are used in condensed-matter
calculations. As a consequence, recent implementations of XDM do not
use the exact exchange hole but an approximation to it: the
Becke-Roussel (BR) model.\cite{becke1989} BR is a model of the
spherically-averaged exchange hole, $h_{X\sigma}(\bm{r},s)$. In it,
$h_{X\sigma}(\bm{r},s)$ is represented as an exponential $Ae^{-ar}$
located at a distance $b$ from the reference point $\bm{r}$. The three
parameters $A$, $a$, and $b$ are determined by imposing the on-top
depth condition (Eq.~\ref{eq:ontop}), the hole normalization
(Eq.~\ref{eq:holenorm}) and the exact curvature at the reference
point, which is:
\begin{equation}
 Q_{\sigma} = \frac{1}{6} \left(\nabla^2\rho_{\sigma} - 2D_{\sigma}\right) 
\end{equation}
where
\begin{align}
 D_{\sigma} & = \tau_{\sigma} -
 \frac{1}{4}\frac{(\nabla\rho_{\sigma})^2}{\rho_{\sigma}} \\
 \tau_{\sigma} & = \sum_i (\nabla\psi_{i\sigma})^2
\end{align}
with $\nabla^2\rho_{\sigma}$ the Laplacian of the electron density and
$\tau_{\sigma}$ the Kohn-Sham kinetic energy density. 

By using the BR model, and under the constraints above, the
exchange-hole dipole ($d_{X\sigma}$) reduces to the value of the
parameter $b$, which is calculated by solving for $x$ in:
\begin{equation}
 \frac{xe^{-2x/3}}{x-2} = \frac{2}{3}\pi^{2/3}\frac{\rho_{\sigma}^{5/3}}{Q_{\sigma}}
\end{equation}
and then substituting in:
\begin{equation}
 b^3 = \frac{x^3 e^{-x}}{8\pi\rho_{\sigma}}
\end{equation}
See ref.~\citenum{becke2005b} for details on the derivation.

The BR model has the computational advantage with respect to the exact
exchange-hole that determining the dipole depends only on local
quantities: the density and its derivatives and the kinetic energy
density. Hence, by using BR, XDM is formally a meta-GGA model of
dispersion, and the computational cost becomes negligible compared to
the base DFT calculation. In addition, the interaction coefficients
using the BR model give significantly better results in the
calculation of binding energies of small noncovalently bonded
dimers.\cite{kannemann2010} 

In the canonical implementation of XDM, the pairwise terms involving
$C_6$, $C_8$, and $C_{10}$ are used in the energy expression:
\begin{equation}
\label{eq:pwxdm}
 E_{\text{disp}} = -\sum_{A>B} \sum_{n=6,8,10} \frac{C_{n}^{\text{AB}}f_n(R_{AB})}{R_{AB}^n} 
\end{equation}
Generalized expressions for the pairwise coefficients up to any order
and for the coefficients involving more than two atoms have been
formulated.\cite{oterodelaroza2013a} However, using pairwise terms of
order higher than $n = 10$ gives, at first, a negligible contribution
to the energy and, ultimately, makes the dispersion series
diverge. The leading three-body term has the well-known
Axilrod-Teller-Muto expression\cite{axilrod1943,muto1943,bell1970}
that decays globally as $R^{-9}$ and involves a $C_9$ coefficient:
\begin{equation}
E^{(3)}_{\text{disp}} = C_9 \frac{3 \cos \theta_A \cos \theta_B \cos \theta_C +1}{R_{AB}^3 R_{AC}^3 R_{BC}^3}
\label{eq:atm}
\end{equation}
In XDM, the three-body dispersion coefficient is:\cite{oterodelaroza2013a}
\begin{equation}
C_9 = 
\langle M_1^2 \rangle_A \langle M_1^2 \rangle_B \langle M_1^2 \rangle_C 
\times \frac{Q_AQ_BQ_C}{(Q_A+Q_B)(Q_A+Q_C)(Q_B+Q_C)} 
\label{eq:c9xdm}
\end{equation}
with $Q_X = \langle M_1^2 \rangle_X/\alpha_X$. Despite the $C_9$
calculated with an accuracy similar to the $C_6$, no simple way of
conciliating this term with the pairwise correction
(Eq.~\ref{eq:pwxdm}) has been found,\cite{oterodelaroza2013a} mainly
because of uncertainties as to the shape of the damping function
$f_9$ (see below). 

To turn the dispersion coefficients into a practical energy
correction, an expression for the damping function in
equation~\ref{eq:pwxdm} is needed. The damping function traditionally
used in XDM is the Becke-Johnson damping function,\cite{johnson2006a}
that is defined as:
\begin{equation}
 f_n(R) = \frac{R^n}{R^n + R_{\text{vdw}}^n}
\end{equation}
This damping function depends naturally on the order of the
interaction, and the whole $f_n$ family has only two adjustable
parameters ($a_1$ and $a_2$) inside the van der Waals radii:
\begin{equation}
 R_{\text{vdw}} = a_1 R_c + a_2
\end{equation}
$R_{\text{vdw}}$ is related to the size of the associated atom. $R_c$
is the critical radus, which is defined as the arithmetic average of
the distances where the $C_6$, $C_8$ and $C_{10}$ terms acquire the
same magnitude:
\begin{equation}
R_{c} = \frac{1}{3} \left[
  \left(\frac{C_{8}}{C_{6}}\right)^{1/2} +
\left(\frac{C_{10}}{C_{6}}\right)^{1/4} +
\left(\frac{C_{10}}{C_{8}}\right)^{1/2} \right]
\end{equation}

The damping function parameters $a_1$ and $a_2$ are the only two
adjustable parameters in the XDM model. As mentioned before, these are
determined by fitting to a set of high-quality reference data (usually
at the CCSD(T) level extrapolated to the complete basis set limit). In
XDM it is customary to use the Kannemann-Becke (KB)
set.\cite{kannemann2009,kannemann2010} The dimers in the KB set are
made of small molecules with a mixture of interaction types (hydrogen
bonding, dipole-dipole, dispersion) and include the noble-gas
dimers. The latter have very small binding energies and many
functionals overbind, even in absence of a dispersion
correction. Because the parametrization is performed by minimizing the
mean absolute percent error (MAPE), for certain functionals the
determination of $a_1$ and $a_2$ is done on a smaller subset of KB
that does not contain noble-gas dimers (with 49 dimers instead of the
original 65). The dimers and the corresponding binding energies in the
KB set have been adapted from previous works, and subsequently
reviewed in later articles.\cite{johnson2013} The reader is pointed to
ref.~\citenum{gatsby} for the most recent energies, molecular
geometries, and the original literature references.

XDM has been implemented for use in molecular quantum chemistry
programs\cite{kannemann2010,oterodelaroza2013b} as well as in
condensed-matter plane-wave-based codes.\cite{oterodelaroza2012a}  It
has been extensively parametrized for common density-functionals in
both scenarios\cite{oterodelaroza2013b} (the $a_1$ and $a_2$
parameters are sensitive to the implementation, see
ref.~\citenum{gatsby} for the latest values), presents excellent
performance in molecular\cite{oterodelaroza2013b} and
solid-state\cite{oterodelaroza2012b} applications (see
the section titled ``Performance of Dispersion-Corrected Methods'' for detailed statistics), and has been used in
a number of real-life
applications,\cite{johnson2012,dilabio2013,ye2013,oterodelaroza2013c,oterodelaroza2014}
although to a lesser extent than other functionals like those in the
DFT-D family.

\begin{table}
\caption[$C_6$ calculated using different dispersion corrections]{$C_6$ Dispersion
Coefficients for the Carbon Atom in Different Bonding Situations
Calculated with Different Dispersion Corrections. The XDM Values Were
Computed using the LC-$\omega$PBE Functional at the aug-cc-pVTZ
level. The remaining values are from Johnson.\cite{johnson2011}
\label{tab:2a1:c6coef}}
\centering
\begin{minipage}{0.50\textwidth}
\begin{tabular}{cccccc}
\hline \hline
Molecule &    WY$^a$  &   TS$^b$ & B97D$^c$ &  D3$^d$ & XDM$^e$ \\
\hline
C free   &   ---  & 46.6 & 24.6 & 49.10 & 48.84 \\
C sp     &  29.71 & 30.6 & 24.6 & 29.36 & 31.31 \\
C sp$^2$ &  27.32 & 30.3 & 24.6 & 25.78 & 25.68 \\
C sp$^3$ &  22.05 & 24.1 & 24.6 & 18.21 & 23.89 \\
\hline \hline
\end{tabular}
\newline
\raggedright
${}^a$ Wu and Yang.\cite{wu2002}\\
${}^b$ Tkatchenko-Scheffler.\cite{tkatchenko2009}\\
${}^c$ DFT-D2-adapted functional by Grimme (B97D).\cite{grimme2006}\\
${}^d$ Grimme's DFT-D3.\cite{grimme2010}\\
${}^e$ XDM.
\end{minipage}
\end{table}

The advantages and disadvantages of the XDM model follow those
mentioned in earlier for pairwise dispersion
corrections. In addition, the interaction coefficients depend
naturally on the chemical environment, the importance of which was
already recognized in the early days of the DFT dispersion
corrections\cite{wu2002} and in the classical-force-field
community.\cite{halgren1992} A study of the variation of the
coefficients on the chemical environment for selected examples has
been presented by Johnson.\cite{johnson2011} As an illustration, the
$C_6$ values for carbon in different hybridization states are shown
in Table~\ref{tab:2a1:c6coef}. All variable-coefficient methods
predict the same trend, and roughly agree in the values. The
coefficients become smaller because they are proportional to the
square of the polarizability (Eq.~\ref{eq:london}), which in turn is
proportional to the atomic volume (Eq.~\ref{eq:polarizv}), which becomes 
smaller as more hydrogens sit around the carbon. B97D, which is based
on Grimme's DFT-D2 (see below), uses a fixed $C_6$ with an
average value.

The use of variable coefficients introduces the question of whether the
proper calculation of the nuclear forces is being done in the course of a geometry
optimization calculation. Since the $C_n$ depend on the geometry, the
differentiation of Eq.~\ref{eq:pwxdm} may no longer be an easy task,
particularly for XDM. Because the term coming from the $C_n$ nuclear
derivatives is relatively small in the canonical implementation of
XDM,\cite{kannemann2010,oterodelaroza2012a,oterodelaroza2013b} the
pragmatic assumption is made that the $C_n$ are fixed. This
approximation works if the optimization algorithm is robust enough to
handle small mismatches between energies and forces (as is the case,
for instance, for the Gaussian\cite{gaussian09} program), whereas in
other cases (e.g., Quantum ESPRESSO\cite{espresso}) the
geometry optimization is carried out with fixed $C_n$, and then needs to be
repeated after completion.

The atomic $C_n^{\text{AB}}$ in XDM are completely
non-empirical, and the model to obtain them is physically
motivated. Although it may look like a philosophical---rather than
practical---advantage, the plus in practice is that dispersion from
atoms in the whole periodic table can be treated on the same footing,
without concerns about the reliability of the empirical interaction
coefficients for ``exotic'' atoms.

The Hellmann-Feynman theorem establishes that atomic forces are
calculated using the nuclear positions and the electron density in the
classical electrostatics fashion. As a consequence, dispersion forces
must have an impact (albeit small) on the electron density
distribution. Dispersion functionals like XDM, where the dispersion
correction depends on the electron density, can be incorporated back
into the density by solving the self-consistent problem in the
presence of the dispersion potential contribution. This has been done
in the past by Kong et al.\cite{kong2009} though the implementation in
other software packages is still work in progress.

The XDM dispersion correction is available in the latest version of
Quantum ESPRESSO\cite{espresso} (post-SCF) and in Q-chem\cite{qchem}
(self-consistent). It is also provided as an external
program\cite{gatsby} (postg) that calculates the dispersion energy and
its derivatives and can be used with quantum chemistry codes,
particularly Gaussian.\cite{gaussian09} The code can be used to drive a
geometry optimization coupled with Gaussian's ``external'' keyword. A
collaboration aimed to achieve a robust self-consistent implementation
of the XDM dispersion energy functional and its derivatives in the
Gaussian program is underway.

\subsubsection*{The DFT-D Functionals}
\label{s:dftd}
The functionals in the DFT-D family, designed by Grimme and
collaborators, are the most widely-used dispersion corrections today
thanks to their relative accuracy, simplicity, and, particularly,
to its widespread implementation in popular software
packages. The DFT-D family consists of three generations: DFT-D
itself\cite{grimme2004} (proposed in 2004), which is seldom used
today, the very popular DFT-D2\cite{grimme2006} (2006), and the last
and ``final'' development, DFT-D3\cite{grimme2010} (2010), which is
replacing DFT-D2 in modern usage. The DFT-Dx functionals are all based
on the pairwise dispersion energy correction
(equation~\ref{eq:epairwise}) with increasing levels of complexity and
accuracy in later generations of the family. The design philosophy in
DFT-Dx sacrifices strong adherence to theoretical principles (many
design decisions in the DFT-D3 functional are \emph{ad hoc}) in
exchange for improved accuracy, flexibility, and simplicity in the
implementation. 

The DFT-Dx functionals are extensively parametrized---they have been
combined with tens of different functionals from all
levels\cite{grimme2010}---and benchmarked.\cite{goerigk2011} DFT-D2 has
been implemented in most software packages for quantum chemistry and
solid-state, including Gaussian,\cite{gaussian09}
GAMESS,\cite{gamess1,gamess2} NWChem,\cite{nwchem} Quantum
ESPRESSO,\cite{espresso} VASP,\cite{vasp1,vasp2} and
abinit.\cite{abinit1,abinit2} DFT-D3 has been implemented in most
quantum chemistry packages as well, although it is not available in
the mainstream solid-state codes yet.

The original DFT-D method\cite{grimme2004} uses a pairwise dispersion
energy correction involving only the $C_6$ coefficient and a global
scaling factor ($s_6$):
\begin{equation}
  \label{eq:dftd}
  E_{\text{DFT-D}} = -s_6 \sum_{i>j}\frac{C_6^{AB}}{R_{AB}^6} f_{\text{damp}}(R_{AB})
\end{equation}
where the damping function is:
\begin{equation}
 f_{\text{damp}}(R) = \frac{1}{1+\exp(-\alpha(R/R_0-1))}
\end{equation}
In this equation, $R_0$ is a quantity representing the atomic size
(akin to the sum of the van der Waals radii). For a particular pair of
atoms, $R_0$ is determined by assigning atomic radii to those atomic
species. These radii are calculated using the distance to the 0.01
a.u. iso-density envelopes of the \emph{in vacuo} atoms, scaled by an
\emph{ad hoc} factor of $1.22$. The value of the $\alpha$ parameter is
set to 23, as in the previous article by Wu and Yang.\cite{wu2002}

The homoatomic interaction coefficients were replicated from the
previous work of Wu and Yang,\cite{wu2002} but averaged over different
hybridization states in order to avoid the need to define atomic
types, which would be impractical. The heteroatomic coefficients are
calculated using the combination rule:
\begin{equation}
 C_6^{AB} = 2\frac{C_6^{AA}C_6^{BB}}{C_6^{AA}+C_6^{BB}}
\end{equation}

The global scaling coefficient $s_6$ is a fitted functional-dependent
parameter in the DFT-D method, with several values for different
functionals. The method is parametrized\cite{grimme2004} using a
collection of 18 gas-phase dimers for BLYP ($s_6 = 1.4$), B986 ($s_6 =
1.4$), and PBE ($s_6 = 0.7$). These values of the global scaling
factor conform to the overrepulsive behavior of the B88 exchange
functional and the overattractive behavior of PBE (see
Figure~\ref{fig:1c:repulsive}). The results presented in the original
paper for a collection of molecules improved upon MP2 for $\pi$-$\pi$
stacked interactions, although the results for hydrogen bonds were
somewhat unsatisfactory.

DFT-D was an early attempt at turning the work of Wu and
Yang\cite{wu2002} into a practical and general dispersion
correction. Although the success was limited, it showed that a
practical correction based on a pairwise expression of the energy
coupled with a common functional and parameterized appropriately gives
reasonably accurate results, fit for real-life applications. However,
DFT-D was limited by lack of $C_6$ data for general atomic pairs, by
systematic errors in molecules involving heavy elements (third row or
below), and by errors in the treatment of normal
thermochemistry.\cite{grimme2006}

The next development in the series, DFT-D2,\cite{grimme2006} was a
vast improvement and greatly popularized the whole approach. Two
things were proposed in the DFT-D2 paper: the standalone dispersion
correction itself, reviewed below, and a modification of the B97
semilocal functional proposed by Becke\cite{b97} but refitted at the
same time as the dispersion correction to a molecular set containing
noncovalent interaction energies as well as thermochemistry reference
data. The resulting functional, B97-D, proved to yield accurate
noncovalent interactions while, at the same time, improving the
thermochemistry of plain B97. Being a semilocal functional, B97-D was
also proposed as an efficient functional, coupled with the resolution
of the identity (RI) technique\cite{eichkorn1995,eichkorn1997} to the
computation of the Coulomb energy (already in the
TURBOMOLE\cite{turbomole} program at the time).

The DFT-D2 dispersion correction uses the same expression as DFT-D
(equation~\ref{eq:dftd}), with a number of minor differences: i) the
scaling factor in the van der Waals radius is reduced from $1.22$ to
$1.10$, and ii) the value of the $\alpha$ parameter is reduced from 23
to 20. The combination rule for obtaining the heteroatomic coefficients
is also replaced by:
\begin{equation}
C_6^{AB} = \sqrt{C_6^{AA}C_6^{BB}}
\end{equation}
and an \emph{ad hoc} formula based on London's (Eq.~\ref{eq:london})
is used to calculate the homoatomic coefficients:
\begin{equation}
\label{eq:magicc6}
C_6^{AA} = 0.05 N I_p^{A} \alpha^{A}
\end{equation}
where the $0.05$ coefficient is selected to adjust previous $C_6$
values, $N$ is the atomic number of the noble gas on the same period
as A, $I_p^{A}$ are the atomic ionization potentials, and $\alpha^{A}$
are the \emph{in vacuo} atomic static polarizabilities.

Grimme argues that DFT-D2 is less empirical than DFT-D, and that
there are fewer parameters than in other contemporaneous
methods.\cite{grimme2006} It is also claimed that DFT-D2 has CCSD(T)
accuracy on average which, following years of testing, seems not to be
the case. However, DFT-D2 does improve greatly upon the
accuracy of DFT-D and provides a functional that is in principle valid
for the whole periodic table. 

The refitted functional B97-D performs better in thermochemical tests
than both uncorrected B97 and the DFT-D2 dispersion-corrected version
of B97, but without re-fitting the base functional. For instance,
B97-D gives 3.8 kcal/mol average error on the G97/2 set of atomization
energies\cite{curtiss1997} as compared to the 3.6 kcal/mol for
B3LYP. This result is not surprising (and is observed for other
dispersion corrections\cite{oterodelaroza2013b}) since GGAs, in
general, slightly underbind molecules and solids. Addition of a
dispersion correction, which stabilizes molecules with respect to
atoms, tends to correct for those systematic deviations. In other
cases, such as reaction barriers affected by self-interaction error,
the incorporation of a dispersion correction can be
detrimental.\cite{grimme2006,oterodelaroza2013b} The improvement upon
B97 is argued to be associated with the ``avoidance of double counting
effect'' and the balance in the description of ``long-range'' and
``medium-range'' correlation effects in the original
article.\cite{grimme2006}

Subsequent works inspired by the performance of B97-D explored the
idea of refitting different functionals in combination with the dispersion
correction for noncovalent interactions as well as
for thermochemistry. For instance, the DFT-D2 scheme was used with minor
changes by Chai and Head-Gordon\cite{chai2008} in combination with a
refitted version of the long-range corrected B97
functional.\cite{wb97} The resulting functional, called
$\omega$B97X-D, goes to 100\%\ exact-exchange in the long-range
electron-electron interaction limit, while the amount of short-range
exchange is treated as an adjustable parameter. The parameters in the
dispersion correction as well as in the functional are fit to a set
that contains both thermochemical and noncovalent interaction
reference energies. The fitted parameters include the range-separation
parameter, as well as the coefficients in the enhancement factor and
the damping function. Unlike in plain DFT-D2, there is no global
scaling parameter in the $\omega$B97X-D approach.

Despite the good performance of DFT-D2, there are several notorious
disadvantages. The $C_6$ dispersion coefficients are fixed and
independent of the environment, which limits the accuracy of the
method (see Table~\ref{tab:2a1:c6coef}), although, as shown in
ref.~\citenum{grimme2006}, the damping function is flexible enough to
account for this shortcoming to some degree. DFT-D2 is also lacking
higher-order dispersion coefficients, which are known to give a
non-negligible contribution to the dispersion
energy.\cite{johnson2006a,grimme2010,grimme2011} Also, DFT-D2 is not
properly defined for metals because of the diversity in their bonding
environments,\cite{grimme2010} which precludes the use of a single
$C_6$ in all bonding situations.

In parallel to DFT-D2, a similar approach was presented by
Ortmann,\cite{ortmann2006} that became relatively popular in the
condensed-matter community. In this approach, the PW91 GGA was
combined with London's formula (Eq.~\ref{eq:london}), and experimental
polarizabilities and ionization potentials. A simple Fermi function
was used for damping:
\begin{align}
 f_{AB}(R) & = 1 - \exp(-\lambda x_{AB}) \\
 x_{AB} & = \frac{R}{R_{\text{cov}}^A + R_{\text{cov}}^B}
\end{align}
where $\lambda = 7.5\times 10^{-4}$, set to match the $c$
cell parameter of graphite. The correction gives excellent results for
graphite, and corrects systematic deviations of GGAs in the structures
and elastic properties (bulk moduli) of hard solids such as diamond
and NaCl, for a reason similar to why DFT-D2 improves results for
thermochemistry, that is, because GGAs are underbinding. The lattice
parameters of noble gas crystals are, however, overestimated because
PW91 is much more attractive than PBE.

The most recent development in the DFT-D family is DFT-D3, proposed by
Grimme and collaborators in 2010.\cite{grimme2010} DFT-D3 is more
complex than DFT-D2, and the interaction coefficients are dependent upon
the geometry, though not on the electron density. The formulation of
DFT-D3 is extensively based on pre-computed quantities using TDDFT
and \emph{ad hoc} recipes in order to determine the basic
components entering the model. In DFT-D3, the dispersion energy is
written as a sum of the $C_6$ term and the $C_8$ term:
\begin{equation}
E_{\text{disp}} = - \sum_{A>B} \frac{C_6^{AB}}{R_{AB}^6} f_6(R_{AB}) + 
s_8 \frac{C_8^{AB}}{R_{AB}^8} f_8(R_{AB}) 
\end{equation}
In this case, and contrary to DFT-D2, there is no $s_6$ scaling
parameter, and $s_8$ is an adjustable parameter. The higher-order
contributions ($C_{10}$, etc.) are omitted because the correction
becomes unstable.

The damping functions are the same as proposed in the previous work by
Chai and Head-Gordon:\cite{chai2008}
\begin{equation}
\label{eq:dampdftd3}
f_n(R_{AB}) = \frac{1}{1+6(R_{AB}/(s_{r,n}R_0^{AB}))^{-\alpha_n}}
\end{equation}
In this equation, the $s_{r,8}$ is set to $1$ and the $s_{r,6}$ is
treated as an adjustable parameter. The other parameters are set to
$\alpha_6 = 14$ and $\alpha_8 = 16$. This choice is made so that the
dispersion energy contribution is less than 1\%\ of the maximum total
dispersion energy for interatomic interactions at covalent distances.

The dispersion coefficients do not use the empirical formula of the
previous generation in equation~\ref{eq:magicc6}. Instead, they are
obtained by considering the hydrides of all the elements in the
periodic table, and by calculating their frequency-dependent
polarizabilities using TDDFT with PBE38 (the same as PBE0 but with
37.5\%\ of exact exchange instead of the physically-motivated 25\%;
this has been shown to give improved excitation energies). The
calculated frequency-dependent polarizabilities enter a
Casimir-Polder-like equation in the calculation of the atomic
interaction coefficients:
\begin{equation}
 C_6^{AB} = \frac{3}{\pi}\int_0^{\infty} \frac{1}{m}
\left[\alpha^{A_mH_n}(i\omega) - \frac{n}{2}\alpha^{H_2}(i\omega)\right]
\times \frac{1}{k}
\left[\alpha^{B_kH_l}(i\omega) - \frac{l}{2}\alpha^{H_2}(i\omega)\right]
d\omega
\end{equation}
where $m$, $n$, $k$, and $l$ are the stoichiometric numbers of
the corresponding hydrides. The formula involves the
frequency-dependent polarizability of the hydrogen molecule. 

Calculation of the higher-order coefficients makes use of
recurrence formulas. In particular,
\begin{align}
C_8^{AB} & = 3C_6^{AB}\sqrt{Q^AQ^B} \\
Q^A & = s_{42} \sqrt{Z^A} \frac{\langle r^4 \rangle^A}{\langle r^2\rangle^A}
\end{align}
where $\langle r^2\rangle$ and $\langle r^4\rangle$ are moments of the
electron density, $s_{42}$ is chosen so that the $C_8^{AA}$ for He,
Ne, and Ar are reproduced and the $\sqrt{Z^A}$ is an \emph{ad hoc}
term introduced to get consistent interaction energies for heavier
elements.  Coefficients of order higher than $C_8$ can be
calculated as well, using other recurrence relations but they are not
used in the energy expression. The three-body interaction coefficient
$C_9$, that enters the Axilrod-Teller-Muto term (Eq.~\ref{eq:atm}) is
calculated in DFT-D3 using another approximate formula:
\begin{equation}
C_9^{ABC} = -\sqrt{C_6^{AB}C_6^{AC}C_6^{BC}}
\end{equation}
However, the DFT-D3 authors recommend that the three-body term should
not be used.

The cutoff radii ($R_0^{AB}$) in the damping function
(Eq.~\ref{eq:dampdftd3}) are pre-computed for all pairs of atoms
independently rather than as the sum of radii for single atoms. As in
previous generations, the approach for obtaining $R_0^{AB}$ involves
obtaining the interatomic distance for which the first-order DFT
interaction energy (that is, the DFT energy obtained using the frozen
electron density distribution resulting from the sum of the two \emph{in vacuo}
atoms) is less than a certain cutoff value. The value of this cutoff
energy is chosen so that the $R_0$ of the carbon-carbon interaction is
the same as in DFT-D2.

The dependence of the interaction coefficients on the chemical
environment is obtained by an \emph{ad hoc} geometry dependence term that
is independent of the electron density distribution, and is 
based on a recipe for calculating the coordination number (CN) of an
atom:
\begin{equation}
\text{CN}^A = \sum_{A\neq B}\frac{1}{1+\exp{-k_1(k_2
(R_{\text{cov}}^A+R_{\text{cov}}^B)/R_{AB} - 1)}}
\end{equation} 
The $k_2$ parameter is set to $4/3$, but the covalent radii of all
metals are decreased by $10\%$, the $k_1$ parameters is set to $16$,
and the covalent radii are taken from a previous paper by Pyykk\"{o}
and Atsumi.\cite{pyykko2009} The CN recovers the ``chemically
intuitive'' coordination numbers for normal molecules.

The CN formula is used in the calculation of dispersion coefficients
by formulating a 2-dimensional space
$C_6^{AB}(\text{CN}^A,\text{CN}^B)$ where the $C_6$ coefficients are
calculated for a certain number of reference molecules (and
incorporated as fixed quantities within the model), and the $C_6$ for
unknown coordination numbers are 
interpolated. The parameters in the interpolation scheme are also
given. There is one more parameter in the interpolation scheme ($k_3$),
which is chosen to get smooth interpolation and plateaus for the
integer CN values.

DFT-D3 is widely implemented in popular software packages, and is
replacing DFT-D2. DFT-D3 provides a parametrization (on the same
footing) for all elements up to Pu, in principle solving the
shortcomings in DFT-D2 for heavy elements. The dependence on the
geometry and not on the electron density provides an energy expression
that is easier to differentiate, but the interatomic coefficients also
depend on the oxidation state, which is not directly addressed by
DFT-D3. It has also been extensively benchmarked (for instance,
see ref.~\citenum{goerigk2011}) and used in many applications to good
effect.

\subsubsection*{Other Approaches}
\label{s:otherpair}
\paragraph*{Tkatchenko-Scheffler model.}

The method proposed by Tkatchenko and Scheffler\cite{tkatchenko2009}
(TS) in 2009 is based on a pairwise dispersion energy correction. In
TS, the London formula (Eq.~\ref{eq:london}) is rewritten in order to
calculate the heteroatomic interaction coefficients from the
homoatomic $C_6^{AA}$:
\begin{equation}
C_6^{AB} = \frac{2C_6^{AA}C_6^{BB}}{
\frac{\alpha_B}{\alpha_A}C_6^{AA} + 
\frac{\alpha_A}{\alpha_B}C_6^{BB}}
\end{equation}
where $\alpha_A$ are the atomic static polarizabilities in the
molecular environment (atom-in-molecule polarizabilities). The
static polarizabilities are calculated using the same approach as XDM,
proposed by Johnson and Becke.\cite{johnson2005} The atom-in-molecule
polarizabilities are scaled according to equation~\ref{eq:polarizv},
with the \emph{in vacuo} atomic polarizabilities taken from the
accurate TDDFT results of Chu and Dalgarno.\cite{chu2004}  Contrary to
XDM, the homoatomic interaction coefficients are scaled from reference
atomic data. The scaling is defined using the same atomic partitioning
scheme (Hirshfeld) as the polarizabilities:
\begin{equation}
C_6^{AA} = \left(
\frac{\int r^3\omega_A({\bf r})\rho({\bf r})d{\bf r}}
{\int r^3\rho_A^{\text{at}}({\bf r}) d{\bf r}}
\right)^2 C_6^{AA,\text{at}}
\end{equation}
The atomic reference values ($C_6^{AA,\text{at}}$) are taken from the
same database as the polarizabilities.\cite{chu2004} The
intermolecular interaction coefficients computed in this way have an
average error of 5.5\% for the intermolecular $C_6$ coefficients
tested on the dipole oscillator strength distribution (DOSD)
experimental data of Meath et
al.\cite{zeiss1977,margoliash1978,jhanwar1980,%
  jhanwar1982,kumar1985a,kumar1985b,kumar1992,kumar1996,kumar1997,kumar2002a,%
  kumar2002b,kumar2003a,kumar2003b} The coefficients are sensitive to
the chemical environment (see Table~\ref{tab:2a1:c6coef}).

The damping function in TS is a Fermi function, similar to the
original used in Wu and Yang\cite{wu2002} and in
DFT-D.\cite{grimme2004} It is defined as:
\begin{equation}
f_{\text{damp}} = \frac{1}{1 + \exp{-d(R_{AB}/(s_RR_{AB}^0 - 1))}}
\end{equation}
where $R_{AB}^0$ is the sum of atomic van der Waals radii. The atomic
radius is defined as:
\begin{equation}
R^0_A = \left(\frac{V_A}{V_A^{\text{at}}}\right)^{1/3} R^{0,\text{at}}_A
\end{equation}
where $R^{0,\text{at}}_A$ is defined as the iso-density contour radius
corresponding to the density where the noble gas on the same period
equals the values by Bondi.\cite{bondi1964} The value of the parameter
$d$ is set to $20$, and the $s_R$ is fitted to the S22 database of
Jurecka et al.\cite{jurecka2006} The mean absolute error of the fit is
0.30 kcal/mol when the dispersion correction is coupled with the PBE
functional. 

The TS model of dispersion has been further revised to include
screening and anisotropy effects on the atomic polarizabilities as
well as many-body dipole-dipole dispersion
effects.\cite{tkatchenko2012,distasio2012,distasio2012b,tkatchenko2013} Screening
effects are important in systems with extensive electron
delocalization,\cite{rehr1975} for instance, on metal surfaces. The
revisions are based on a random-phase approximation (RPA) approach to
a model of interacting quantum harmonic oscillators located at the
atomic positions. The harmonic oscillators vibrate with a
characteristic frequency related to the effective atomic excitation
energy in London's formula and interact via a screened
(range-separated) Coulomb potential that is attenuated at short
distances using an adjustable parameter.

The atomic polarizabilities calculated using the volume scaling in
equation~\ref{eq:polarizv} enter a self-consistent equation derived
from the RPA treatment of the model system of coupled harmonic
oscillators. In this simplified system, the adiabatic connection
formula can be integrated analytically to yield a coupled set of
self-consistent equations that can be solved for the anisotropic
polarizabilities using matrix operations. The anisotropic static
polarizabilities present an improved agreement with experimental
reference data.\cite{distasio2012}

A drawback of the TS method, and subsequent revisions, is that it is
limited to dipole-dipole interactions and, therefore, does not take
into account the higher-order $C_8$ and $C_{10}$ pairwise
terms. Nevertheless, the energetics obtained by fitting the damping
and the range-separation parameters to standard datasets are
promising,\cite{tkatchenko2009,distasio2012,distasio2012b}
particularly in the formalism that includes many-body
interactions. The TS method with has been implemented in the
FHI-AIMS\cite{fhi-aims} program and in the latest version of Quantum
ESPRESSO.\cite{espresso}

\paragraph*{Density-dependent energy correction.}

The density-dependent energy correction (dDsC) proposed by Steinmann
and
Corminboeuf\cite{steinmann2009,steinmann2010,steinmann2011a,steinmann2011b}
is a dispersion correction based on XDM, but with modifications
pertaining to the calculation of the exchange-hole as well as a
density-dependent damping function that achieves excellent performance
in standard thermochemical and noncovalent interactions tests at a
low computational cost. 

The dispersion energy and the calculation of the interaction
coefficients in dDsC is the same as in XDM (equations~\ref{eq:pwxdm}
and~\ref{eq:xdmc6} to~\ref{eq:xdmc10}). The exchange-hole,
unlike the Becke-Roussel model, is based on a GGA approximation and
contains adjustable parameters:
\begin{equation}
b = Asr_se^{-Bs}
\end{equation}
where $s$ is the reduced density gradient
($s=\nabla\rho/[2(3\pi^2)^{1/3}\rho^{4/3}]$), and $r_s =
[3/(4\pi\rho)]^{1/3}$ is the Wigner-Seitz radius. The adjustable
parameters $A$ and $B$ are obtained by fitting to reference data for
the noble gas dimers. 

The second major difference of dDsC with respect to XDM is the density
dependence in the damping function. dDsC uses the Tang-Toennies
damping function,\cite{tang1984} which gives excellent results in
describing the potential energy curve of the noble gases. Its
expression is:
\begin{equation}
f_n(x) = 1 - \exp(-x) \sum_{k=0}^n\frac{x^k}{k!}
\end{equation}
The damping function enters the energy dispersion expression with a
scaling parameter: $f_n(bR_{AB})$ with $R_{AB}$ the interatomic
distance. The $b$ parameter depends on the system electron density and
contains two adjustable parameters $a_0$ and $b_0$. A further minor
change from XDM is that dDsC uses the Hirshfeld-dominant scheme: the
atomic weights $\omega_A(\bm{r})$ for atom $A$ assigned to a point
$\bm{r}$ are either $1$ if in the normal Hirshfeld partition the
weight is greater for atom $A$ than for any other atom or $0$
otherwise.

The dDsC model has been parametrized for a number of popular base
functionals\cite{steinmann2011b} using noncovalent interactions as
well as thermochemical reference data. The method presents good
accuracy in the calculation of intermolecular
coefficients\cite{steinmann2011a} (errors of slightly less than 10\%)
as well as in the energetics of noncovalent
dimers,\cite{steinmann2011b} slightly improving upon B2PLYP-D3
(described earlier) and M06-2X (described later) for the
tests presented in ref.~\citenum{steinmann2011b}. The dDsC method is
implemented in recent versions of ADF,\cite{adf1,adf2,adf3}
Q-Chem\cite{qchem} and there is a patch for
GAMESS\cite{gamess1,gamess2} on the authors'
webpage.\cite{clemenceweb}

\paragraph*{Local-response dispersion model.}

The local-response dispersion (LRD) model proposed by Sato and
Nakai\cite{sato2009,sato2010} is based on the second-order
perturbation-theory intermolecular interaction. In the latest version,
the dispersion energy is written as a generalized multicenter
approach, obtained from the Casimir-Polder equation
(Eq.~\ref{eq:casipol}) by expanding the Coulomb interaction operator
in a multicenter atomic partition:\cite{stone1997}
\begin{equation}
E_{\text{disp}}^{AB} = -\sum_{aa^{'}bb^{'}}
\sum_{tt^{'}uu^{'}}
T_{tu}^{ab}T_{t^{'}u^{'}}^{a^{'}b^{'}}
\int_0^{\infty}
\alpha^{aa^{'}}_{tt^{'}}(i\omega)
\alpha^{bb^{'}}_{uu^{'}}(i\omega)
d\omega
\end{equation}
where $t$ and $u$ are indices corresponding to different angular
momentum contributions ($t=lm$), $T$ is a damped interaction tensor
that depends only upon the relative position of the
atoms,\cite{stone1997} and $\alpha^{aa^{'}}_{tt^{'}}(i\omega)$ are the
generalized atom-pair dynamic polarizabilities. These are calculated
in the local-response approximation proposed by Dobson and
Dinte:\cite{dobson1996}
\begin{equation}
\alpha_{tt^{'}}^{aa^{'}}(iu) = \int
w_a(\bm{r})w_{a^{'}}(\bm{r})
\bar{\alpha}(\bm{r},i\omega)
\bm{\nabla}R_t(\bm{r}-\bm{R_a})
\bm{\nabla}R_{t^{'}}(\bm{r}-\bm{R_{a^{'}}})
\end{equation}
with $R_t$ a solid harmonic, $w_a$ an atomic partition function (in
this case, the Becke integration weights\cite{becke1988}) and
$\bar{\alpha}$ is the polarization density. For the latter, Sato and
Nakai use the approximation proposed by Vydrov and van
Voorhis:\cite{vydrov2009b}
\begin{align}
\bar{\alpha}(\bm{r},i\omega) & = 
\frac{\rho(\bm{r})}{\omega_0^2(\bm{r}) + \omega^2} \\
\omega_0(\bm{r}) & = \frac{q_0^2(\bm{r})}{3} \\
q_0(\bm{r}) & = k_F(1+\lambda s^2)
\end{align}
with $k_F = (3\pi^2\rho)^{1/3}$ being the Fermi wave-vector and $s
= \nabla\rho / (2k_F\rho)$ being the reduced density gradient. $\lambda$ is
an adjustable parameter that is fit to a training
set. The damping of the dispersion interaction occurs via the
interaction tensor $T_{t^{'}u^{'}}^{a^{'}b^{'}}$, which corresponds to
the usual geometric function\cite{stone1997} times a damping factor
that depends on the atoms and the angular momenta in $t$ and $u$
(the compound $t$ and $u$ indices contain $l_1$ and $l_2$ respectively):
\begin{equation}
f^{ab}_{l_1l_2} = \exp\left[-\frac{l_1+l_2-1}{2}\left(\frac{R_{ab}}{\bar{R}}\right)^{-6}\right]
\end{equation}
with $R_{ab}$ the interatomic distance and $\bar{R}$ is an atomic
radius that contains the adjustable parameters. 

The method described above is equivalent to calculating the
interatomic interaction coefficients as:
\begin{equation}
C_n^{ab} = \frac{1}{2\pi}
\sum_{tt^{'}uu^{'}}
S_{tu}^{ab}S_{t^{'}u^{'}}^{ab}
\int_0^{\infty}
\alpha^{aa}_{tt^{'}}(i\omega)
\alpha^{bb}_{uu^{'}}(i\omega)
d\omega
\end{equation}
with $S_{tu}^{ab}$ a geometric factor that depends on the positions of
$a$ and $b$ and the angular momenta of the interaction. The order $n$
is determined as the sum of the angular momenta in $t$, $t^{'}$, $u$,
and $u^{'}$ plus $2$. The atom-atom interaction coefficients and the
closely-related intermolecular dispersion coefficients are relatively
accurate with errors averaging 6.0\%\ in a test of more than a
thousand interaction coefficients.\cite{sato2010}

The LRD functional has been combined with long-range-corrected
B88\cite{b88} plus the one-parameter progressive (OP) correlation
functional\cite{op} (LC-BOP), and variants thereof.\cite{kar2013} The
self-consistent implementation has been developed,\cite{ikabata2013}
and the energies and general performance of the method are essentially
unaffected by the relatively minor changes in the electron density
distribution caused by the dispersion potential. The three parameters
in the LC-BOP-LRD functional were obtained first by fitting to
rare-gas atoms,\cite{sato2009,sato2010} then to the S66 database of
dimer binding energies.\cite{rezac2011} Tests on the S22 give an
accuracy of 0.22 kcal/mol (4.6\%), where the interactions
corresponding to higher-order coefficients are relatively
important. The functional has been tested in other benchmark
datasets\cite{kar2013} with relative success. To our knowledge, the
only software package implementing the LRD dispersion model is
GAMESS.\cite{gamess1,gamess2}

\paragraph*{Solid-sphere model.}

The solid-sphere model (SSM) by Tao, Perdew and
Ruzsinszky\cite{tao2010,tao2012,tao2013} relies on calculating 
the dynamic polarizabilities (including those of higher-order) using a
uniform-density metallic sphere model. The dynamic polarizabilities,
and the dispersion coefficients through the Casimir-Polder formula, are
non-empirical.

The SSM model was first proposed in the context of correcting the
overbinding behavior of GGAs for alkali metals.\cite{tao2010} The
authors argue that alkali metals are ``soft matter'', and that an
adequately-screened dispersion correction is necessary to correct for
the incorrect behavior of the GGAs, in agreement with previous work by
Rehr, Zaremba and Kohn.\cite{rehr1975} The SSM model was subsequently
extended to the calculation of the $C_8$ and $C_{10}$ coefficients as
well.\cite{tao2012,tao2013} Although it has not been transformed into
a general-purpose dispersion energy functional, the SSM model has been
used successfully in the calculation of fullerene interaction
coefficients.\cite{ruzsinszky2012}

The SSM model is based on the calculation of the $2^l$-pole dynamic
polarizabilities $\alpha_l(i\omega)$. For a metallic sphere of uniform
density and radius $R$, and assuming a uniform electron gas expression
for the dielectric function ($\varepsilon = 1 +
\frac{\omega_p^2}{\omega^2}$), this gives: 
\begin{equation}
\alpha_l(i\omega) = \left(\frac{\omega_l^2}{\omega_l^2+\omega^2}\right)R^{2l+1}
\end{equation}
where $\omega_l$ is the multipole resonance frequency of the sphere,
equal to:
\begin{equation}
\omega_l = \omega_p\sqrt{\frac{l}{2l+1}}
\end{equation}
with $\omega_p = \sqrt{4\pi\rho}$ being the plasmon frequency.

Three constraints are imposed on the model: having the correct static
polarizability ($\alpha_l(0)$), reproducing the correct high frequency
limit ($\alpha_l(iu) \to l\int_0^{\infty}4\pi r^2n(r)r^{2l-2}/u^2 dr$)
and the model must be exact for a metallic sphere of uniform
density and radius $R$. Under these constraints, the model dynamic
polarizability is:
\begin{equation}
\alpha_l(i\omega) = \frac{1}{4\pi a_l}\left(\frac{2l+1}{l}
\int \Theta(R_l-r)\frac{lr^{2l-2}a_l^4\omega_l^2}{a_l^4\omega_l^2 +
  \omega^2}
d\bm{r}\right)
\end{equation}
where $\Theta$ is a step function and the $a_l$ and $R_l$ are
obtained by self-consistently solving the equations:
\begin{align}
R_l & = \left(a_l\alpha_l(0)\right)^{1/(2l+1)} \\
a_l & = \left[\frac{\int_0^{\infty}4\pi r^2 r^{2l-2} \rho(r) dr}
{\int_0^{\infty}4\pi r^2 r^{2l-2} \rho(r) dr}\right]^{1/3}
\end{align}
with $\rho(r)$ being the electron density of the system. Interatomic
interaction coefficients are reproduced with an accuracy of 3\%,
similar to XDM for the $C_6$ coefficients, but significantly better for
the higher-order coefficients. The SSM model has been used to
calculate the interaction coefficients of nanoparticles and large
systems,\cite{ruzsinszky2012,tao2013} but it has not
been coupled with functionals in order to obtain a dispersion energy
correction. 

\paragraph*{Miscellaneous approaches.}

The dispersion correction proposed by Alves de
Lima\cite{alvesdelima2010} is based on a local approximation to the
density response function. The calculation of the dispersion
coefficients is based on the generalized Casimir-Polder formula
(Eq.~\ref{eq:casipol}), with the dynamic polarizabilities calculated
as: 
\begin{equation}
\alpha_l^A(i\omega) = \int_{V_A} \chi_l^A(i\omega,n) d\bm{r}_A
\end{equation}
where $V_A$ symbolizes an atomic partitioning (in this model,
Hirshfeld), and $\chi_l$ is the dynamic $2^l$-polar susceptibility,
which is calculated using a Pad\'e approximant. The model gives good
results for the interatomic coefficients of noble gases and alkali
metals. We are not aware of any software package that implements this
dispersion correction.

In the spherical atom model (SAM) correction, proposed by Austin et
al.,\cite{austin2012} the dispersion energy is calculated by assuming
that every atomic site has a shell attached that contributes to the
dispersion interaction. The SAM is reminiscent of the Drude shell model used
to capture polarization in force-field calculations. In the SAM model,
the energy is calculated using a modified pairwise contribution:
\begin{equation}
 E_{\text{disp}} = - \sum_{A>B} \frac{C_6^{AB}f(R_{AB})g(R_{AB})}
{ \left(R_{AB}^2 - R_{s,AB}^2\right)^3 }
\end{equation}
where $f$ and $g$ are two different damping functions and $C_6^{AB}$
are obtained using London's formula. The dispersion energy is
nullified if the interatomic distance is less than $R_{s,AB}$.  The
functional describes accurately noble gas dimers and small
dispersion-bound complexes, although the results are less satisfactory
for $\pi$-$\pi$ interactions. It is implemented in the Gaussian
package.\cite{gaussian09}

\subsection*{Potential-Based Methods}
\label{s:potential}
Another approach to tackling the dispersion problem in DFT that differs
from pairwise corrections is through the use of atom-centered
potentials. The  philosophy behind methods of this kind is
that the densities in the intermolecular regions associated with
noncovalently-interacting systems predicted by conventional DFT
methods are not correct, and applying potentials to some or all
of the atoms in a system can adjust the density in such a way that
noncovalent interactions are better reproduced. The potentials are
completely empirical in the sense that they are generated through a
fitting procedure for which the goal is to minimize the error in
DFT-calculated noncovalent properties relative to a set of reference
data. In the next two subsections, we review two
atom-centered potential approaches that have been described in the
literature. We first discuss the dispersion-correcting potential (DCP)
approach, which has been developed for use with computational
chemistry programs that employ atom-centered basis sets. The second
 is the dispersion-corrected atom-centered
potential (DCACP) approach, which is a plane-wave-based method.

\subsubsection*{Dispersion-Correcting Potentials (DCP)}
\label{s:dcp}
The notion that the success in modeling noncovalent interactions with
DFT methods significantly improves when the leading contribution to
the binding energy goes from dispersion to
electrostatics~\cite{johnson2006} drove the initial development of
dispersion-correcting potentials (DCPs). These were first developed
for the carbon atom with the motivation that most of the problems with
dispersion in DFT methods are made obvious in the interactions between
hydrocarbon molecules, which interact mainly via
dispersion.~\cite{dilabiodcp2008a,dilabiodcp2008b} Later efforts
extended the library of DCPs to include the H, N, and O
atoms,~\cite{dilabiodcp2012,dilabiodcp2013} and an on-line tool is
available to help users build input files containing
DCPs.~\cite{ginoweb} The DCP approach itself is based on a philosophy
associated with earlier efforts to develop new approaches to bridging
quantum and classical regions in quantum mechanics/molecular
mechanics (QM/MM) simulations.~\cite{dilabioqcp2002,dilabioqcp2005}

DCPs are composed of atom-centered Gaussian-type functions having
the same form as effective core potentials (equation~\ref{eq:ecp}
below). Effective core potentials are atom centered potentials that
are normally used to replace the core electrons during simulations of
heavy elements. An example is provided in
ref.~\citenum{christiansen1979}. To efficiently model systems
containing atoms with many electrons (such as, for instance, lead),
computational advantages are obtained by modeling such atoms using
only their valence electrons. Core electrons do not participate
directly in chemical bonding, but if the core electrons of an atom
were removed, the remaining electrons in the valence space would
collapse into the core owing to the strongly attractive Coulomb
attraction between the positive nucleus and the negatively-charged
electrons. By including a potential in place of the core electrons,
the collapse can be prevented and the atoms can be made to behave as
though the core electrons were present. This approach not only reduces
the computational expense associated with simulating these systems but
also has the added advantage of introducing the effects of relativity
in the simulation, which is important for obtaining reasonably
accurate atomic and molecular
properties.~\cite{wildman1997,dilabio1998} The effective core
potentials are thus developed to reproduce some of the valence
properties of the atoms for which they are designed and modify the
energy landscape in which valence electrons move through a direct
modification of the Hamiltonian of the system.

DCP functions are the same as those used for effective core potentials
but they do not replace core electrons. Instead, DCPs modify the
potential in which all of the electrons move so that noncovalent
interactions are reproduced, as manifested by binding energies and
structural properties. The functional form of an effective core
potential is:
\begin{equation}
\label{eq:ecp}
U^{\text{ECP}} = U_{l_{\text{max}+1}}(r) +  \sum_{l=0}^{l_{\text{max}}} \sum_{m=-l}^{l} | Y_{lm}\rangle U_l(r) \langle Y_{lm} |
\end{equation}
where $Y_{lm}$ are spherical harmonics that allow for applying
potentials to the electron density associated with different angular
momenta $l$ (viz., s-, p-, d-density). In the case of DCPs, there are
no limits or requirements associated with equation~\ref{eq:ecp} in
terms of the number of functions utilized to build a set of DCPs.

Each $U_l(r)$ in equation \ref{eq:ecp} is built from Gaussian-type
functions of the form:
\begin{equation}
\label{eq:dcp-ecp}
U_l(r) = r^{-2} \sum_{i}^{N_{l}} c_{li} r^{n_{li}} e^{-\zeta_{li} r^{2}} 
\end{equation}
where $N_{l}$ is the number of Gaussian functions, $n_{li}$ is an
integer power of $r$, $c_{li}$ is the coefficient of the Gaussian and
$\zeta_{li}$ is its exponent.  

DCPs are generally developed for specific atoms by optimizing the
exponents and coefficients in equation \ref{eq:dcp-ecp}. The values of
$n_{li}$ are usually set to 2. In principle, they can take on any
integer, although some computational chemistry programs limit these to
integer values. The optimizations are performed by minimizing the mean
absolute error (MAE) between the binding energies calculated by the
DFT method and basis set with the DCPs and those obtained from
high-level {\it ab initio} fitting data. The binding energies for fitting 
are most often obtained from CCSD(T) calculations with
complete-basis-set (CBS) extrapolation and there is a growing body of
these data available.~\cite{rezac2011,gmtkn302011} These fitting data
contain one dimensional potential energy surfaces for noncovalently
bonded dimers that span the range from just inside the minimum, out
to complete dissociation. This is done to ensure that the DCPs are
able to reproduce the correct dissociation behavior of
noncovalently-bonded systems, though not necessarily the correct
$\frac{1}{r^{-6}}$ behavior in the extremely long-range. Accumulated
experience indicates that for most problems of practical interest,
obtaining the correct $\frac{1}{r^{-6}}$ behavior is not required to
obtain good performance.

\begin{figure}
\includegraphics[width=0.70\textwidth]{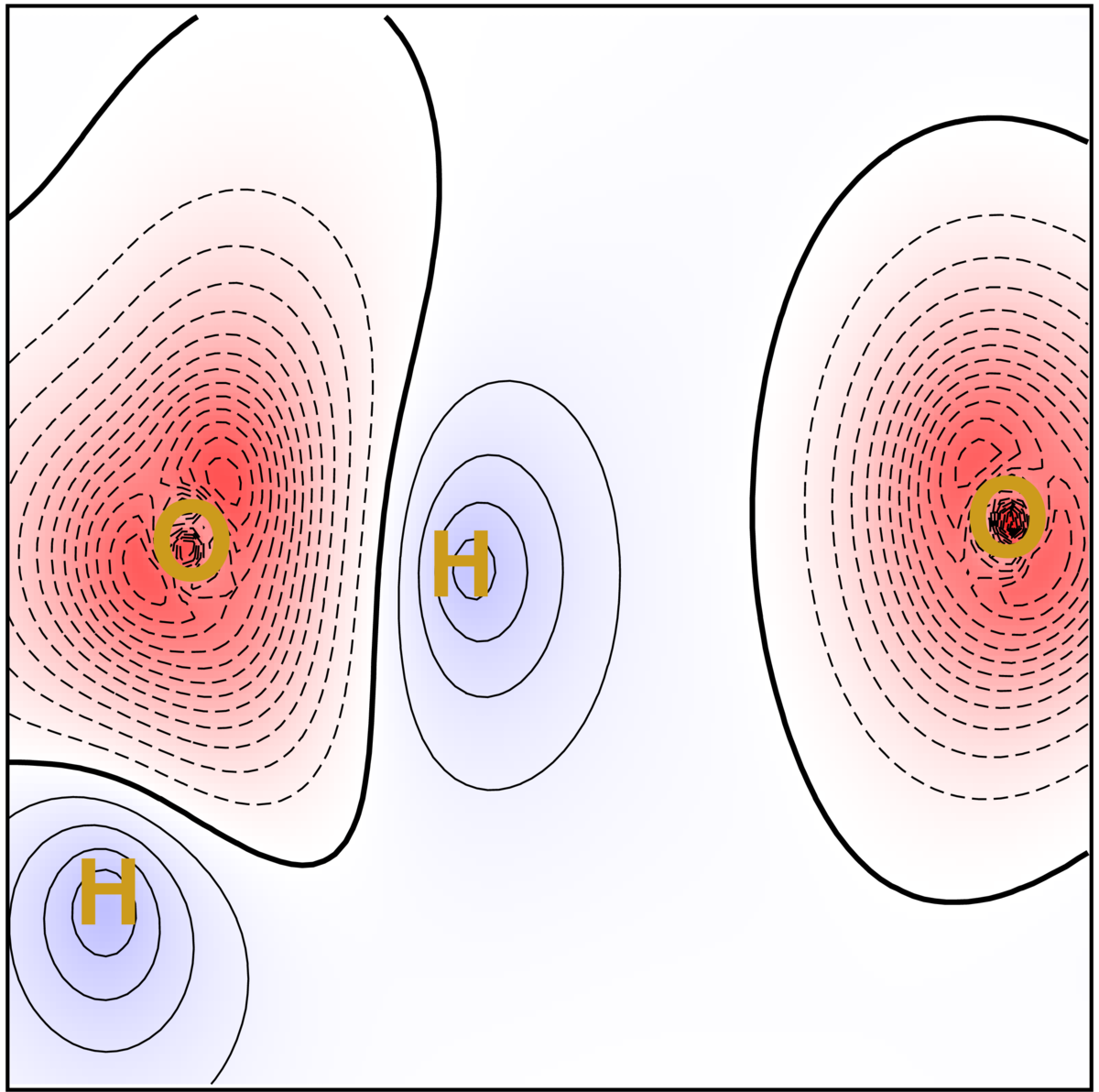}
\caption{Changes induced by a DCP on calculated electron density of
the water dimer (B3LYP-DCP/6-31+G(2d,2p)). The plot shows the contours
(in $5\times 10^{-5}$ steps, positive contour are full lines, negative
contours are stippled lines) as well as a color map (blue is positive,
red is negative) resulting from subtracting the non-DCP
density from the density obtained in the equivalent calculation using
DCPs. The main effect is that the electron density migrates from the
oxygen towards the hydrogens and the interstitial space. The thick
full line represents the zero-value contour. Colour is available in e-book form only.
\label{fig:rhodcp}}
\end{figure}

To develop DCPs that perform well for a broad set of systems
containing H, C, N, and O atoms, a set of about 16 different
noncovalently interacting dimers is required. DCP development follows
a bootstrapping approach where the DCPs for carbon and hydrogen are
developed together and then used in the generation of DCPs for other
atoms such as nitrogen and oxygen. For non-hydrogen atoms, a full set
of DCPs will have functions for each angular momentum channel, from s
to f. The f-functions operate on all of the electron density in the
system and this tends to introduce most of the changes in electron
density distribution needed to improve noncovalent properties. Fine
adjustments are achieved through the use of the lower angular momentum
functions, which affect the distribution associated with s, p, and d
electron density. Figure~\ref{fig:rhodcp} shows the change in
electron density upon the application of DCPs to the water dimer.

There are many positive attributes of DCPs. For example,
because DCPs have the same expression, they can be employed in the
same way as effective core potentials. Many computational chemistry
programs allow effective potentials to be specified by the user
through the modification of input files and DCPs may also be given in
this fashion. Therefore, DCPs can be employed in many computational
chemistry packages without the need for reprogramming. Furthermore,
since DCPs modify the Hamiltonian associated with the
system being modeled, all of the properties are determined
self-consistently, meaning that DCPs introduce changes to the energy
(or other properties) by altering the electron
density. Hence, all of the properties are affected by the presence of
the DCPs. Consequently, DCPs can be used with all of the
''machinery'' of computational chemistry packages, and included in the
calculation of properties like solvation energies, NMR chemical
shifts, and others. 

DCPs can also be used to mitigate, to some extent, the errors
associated with basis set incompleteness. The most recently-developed
set of DCPs are associated with the
B3LYP~\cite{dilabiodcp2012,dilabiodcp2013} and LC-$\omega$PBE
~\cite{dilabiodcp2014} functionals and 6-31+G(2d,2p) basis sets. DCPs
generated from the fitting procedure make up for not only the
shortcomings of the underlying functional with respect to noncovalent
interactions but also for the errors induced by the reduced size of
the basis set. Another positive aspect of DCPs is that they can be
designed to incorporate n-body effects simply through the inclusion of
the appropriate data in the fitting set. Finally, it is possible to
develop DCPs for any DFT method, including those that include other
modes of corrections for noncovalent interactions, such as pair-wise
schemes or non-local functionals.

\begin{figure}
\includegraphics[width=0.70\textwidth]{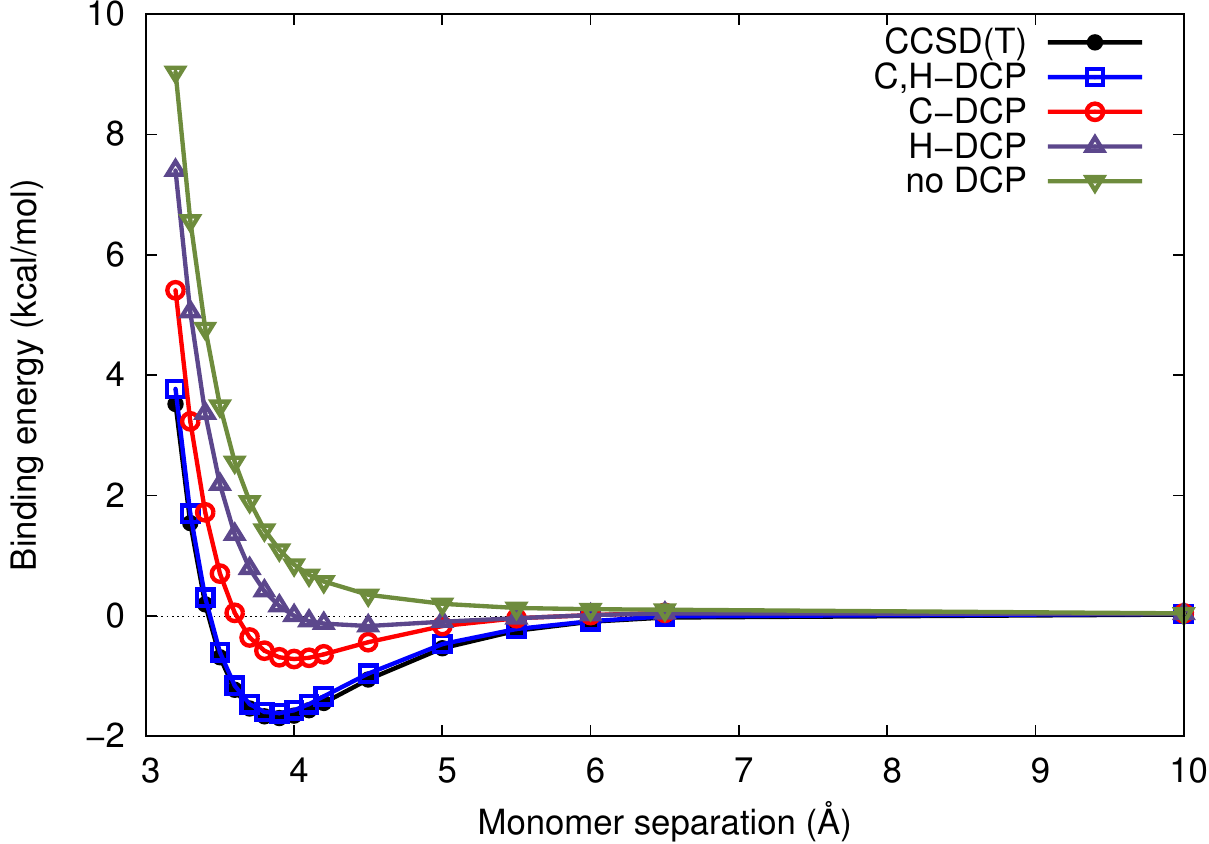}
\caption{Potential energy curve calculated using LC-$\omega$PBE/6-31+G(2d,2p)
compared to reference data for the stacked benzene dimer. The curves
shown include the DCPs for both C and H, only C, only H, and no
DCPs. When the DCPs for both atoms are used, the reference curve is
accurately reproduced including its long-range behavior.
\label{fig:FigureDCP}}
\end{figure}

One of the most important limitations of the DCP approach is that it 
is empirical and requires high level {\it ab initio} and/or
experimental data for DCP generation. This means that the development
of DCPs for particular atoms may not be possible without first
investing heavily in the generation of fitting data. However, unlike
pairwise dispersion correction approaches, improvements can be
obtained in systems in which only some of the atoms have DCPs applied. Figure~\ref{fig:FigureDCP} demonstrates this feature of DCPs
for the potential energy surface of the benzene dimer. 

Another drawback of DCPs is that the method is difficult to
generalize. Moreover, it is also difficult to understand its mode of action
because of the limited theoretical foundation. One possible outcome of
this limitation is that DCPs developed to fix one problem may cause
another.~\cite{dilabiodcp2013, goerigk2014} Furthermore, 
a new set of DCPs is in principle required for every DFT method and basis set
combination, just as this is the case for pair-wise dispersion
correction techniques. However, in practice, it has been found that
the DCPs associated with certain families of density functionals tend
to be similar,~\cite{dilabiodcp2008b} and the dependence of
performance on basis set size is muted for basis sets larger than
6-31+G(d,p). These observations suggest that some generalizations are
possible.

It is also not possible for DCPs to correct for certain underlying
deficiencies in particular functionals. For cases in which the
underlying DFT method erroneously predicts the energy level of the
highest occupied molecular orbital (HOMO) of an electron donor to be
too close to the energy level of the lowest unoccupied molecular
orbital (LUMO) of an electron acceptor, there will be too much charge
transfer and this will result in significant overbinding. This problem
stems from the ``fractional charge'' or
``self-interaction''\cite{vydrov2006,zhang1998,ruzsinszky2006,ruiz1996,cohen2011}
and can occur in molecule-radical
complexes~\cite{johnsonradicals2013,frison2011} in addition to
molecule-molecule complexes. At this point in time, it appears that
DCPs cannot easily correct for self-interaction errors in functionals
in which this shortcoming is present. The only reasonable approach is
to develop DCPs for DFT methods that are less susceptible to charge
transfer problems, i.e. range-separated corrected DFT
methods~\cite{dilabiodcp2014} or those with more (ca. 50\%)
Hartree-Fock exchange.

\subsubsection*{Dispersion-Corrected Atom-Centered Potentials (DCACP)}
\label{s:dcacp}
The dispersion-corrected atom-centered potential (DCACP) approach is
very similar to that for DCPs but was first described several years
earlier by von Lilienfeld et al.~\cite{vonlilienfeld2004} The
corrections are implemented through analytic pseudopotentials of a
type similar to those of Goedecker et al.~\cite{goedecker1996,
hartwigsen1998} DCACPs have the same form as the non-local component
of Goedecker pseudopotentials but, like DCPs, do not replace core
electrons. Instead, a single high angular momentum function is
introduced to existing Goedecker pseudopotentials, and the parameters
of this function are optimized to reproduce noncovalent interactions.

The non-local part of the Goedecker pseudopotentials have the
form:\cite{vonlilienfeld2004}
\begin{equation}
\label{eq:Goedecker-NL}
V_l^{\text{nl}}(\bm{r},\bm{r}^{'}) = 
\sum_{m=-l}^{+l} Y_{lm}(\hat{\bm{r}})
\sum_{j,h} p_{lh}(r) h_{lhj} p_{lj}(r^{'})
Y_{lm}^{\ast}(\hat{\bm{r}}^{'})
\end{equation}
where $Y_{lm}$ are spherical harmonics, $\hat{\bm{r}}$ are unit
vectors in the direction of $\bm{r}$, $h_{lhj}$ are
adjustable parameters, and
\begin{equation}
p_{lh}(r) \propto r^{l+2(h-1)} \exp^{-r^{2}/(2r_{l}^{2})}
\end{equation}
These ECPs have the same form as the effective core potentials
described in equation~\ref{eq:ecp}. The DCACPs developed in
reference~\citenum{vonlilienfeld2004} and in subsequent
work~\cite{lin2007} use a single high angular momentum function of
this type. This simplifies Eq.~\ref{eq:Goedecker-NL}
to a potential with a proportionality:
\begin{equation}
\label{eq:dcacp}
V_l \propto r^{-2} \sum \sigma_1 \exp{-\frac{r^2}{2\sigma_{2}^{2}}}
\end{equation}

Again, the form of Eq.~\ref{eq:dcacp} is similar to that of
equation \ref{eq:dcp-ecp}. For the DCACPs, values of the $\sigma_1$
and $\sigma_2$ parameters are optimized by minimizing the errors in
calculated binding energies and forces in a very small set of
reference systems. For example, for hydrogen and carbon, only the
parallel (H$_2$)$_2$ and the stacked benzene dimers, respectively,
were used as reference systems.~\cite{lin2007} Recently, Jordan's
group found that DCACPs containing two functions offer much better
performance than the single function approach.~\cite{karalti2014}

The advantages of DCAPCs are analogous to those of DCPs. For example,
DCACPs can be used with plane wave computational packages without the
need for reprogramming and are able to take advantage of the full
machinery of such packages, including {\it ab initio} molecular
dynamics. They can also be developed for any functional, including
those that incorporate other modes of dispersion corrections. However,
not all functionals, in particular those that contain Hartree-Fock
exchange, are efficiently implemented in all plane wave
programs. DCACPs can also be designed to include n-body effects. A
unique advantage associated with the use of DCACPs is that is that plane wave
basis sets are effectively complete, obviating the problem of basis set deficiency.

The drawbacks to the DCACP approach also mirror those of DCPs. The
DCACP approach is empirical and requires high level {\it ab initio}
and/or experimental data in order to generate the potentials. That
said, von Lilienfeld et al. used only a very small fitting set (one
per atom type) to generate a seemingly highly transferable set of
DCACPs.~\cite{vonlilienfeld2004,lin2007} This drawback is further
mitigated by the fact that it is highly likely that incremental
improvements in the treatment of noncovalent interactions can be
achieved by applying DCACPs to a subset of atoms in a system of
interest, although this has not, to our knowledge, been
demonstrated explicitly. In addition, and like DCPs, new DCACPs need to be
developed for each density functional.

\subsection*{Minnesota Functionals}
\label{s:minnesota}
The Minnesota family are a collection of functionals proposed by
Truhlar and collaborators. These functionals employ a
heavily-parametrized functional form in order to
model, on the same footing, diverse types of problems, including
barrier heights (chemical kinetics), metal-ligand and metal-metal bond
dissociation, main-group thermochemistry and noncovalent
interactions. The design procedure of all Minnesota functionals is
similar. First, a functional form with a large number of parameters
is adopted, possibly fixing some of the parameters by applying some
physical constraints (like the correct uniform electron gas
behavior). A training database is then constructed using high-level
reference data and, associated to it, a cost function is proposed by
mixing the root mean square errors from each of the component subsets
of the database. Finally, the parameters in the functional are
determined by minimization of the cost function. The functionals are
routinely tested on a dataset larger than the training database for
consistency. 

The Minnesota functional family contains twelve members, that are
labeled by the year of publication. In chronological order, they are:
M05,\cite{m05} M05-2X,\cite{m052x} M06-L,\cite{m06l}
M06-HF,\cite{m06hf} M06,\cite{m06} M08,\cite{m08}
SOGGA11,\cite{sogga11} M11,\cite{m11} M11-L,\cite{m11l}
MN12-L,\cite{mn12l} N12,\cite{n12} N12-SX,\cite{n12sx} and
MN12-SX.\cite{n12sx} All of these (except M05) contain noncovalent
interactions in the training set. Although the performance of some of
the functionals for noncovalent interactions is better than
that of other base functionals, only M06-2X performs with reasonable
accuracy in standard noncovalent interactions tests (see
the section below titled ``Performance of Dispersion-Corrected Methods'').

The first functional in the Minnesota family is M05,\cite{m05} a hybrid with 28\%\ exact exchange. The semilocal exchange functional
is a meta-GGA based on PBE exchange:
\begin{equation}
\label{eq:m05x}
E_x^{\text{semilocal}} = \sum_{\sigma}\int
\varepsilon_{x\sigma}^{\text{PBE}}
\left(\sum_{i=0}^m a_i w_{\sigma}^i\right)
d\bm{r}
\end{equation}
where $\varepsilon_{x\sigma}^{\text{PBE}}$ is the PBE exchange energy
density for spin $\sigma$, and $w_{\sigma}$ is Becke's measure of the
exchange-hole non-locality:\cite{becke2000}
\begin{equation}
\label{eq:truhlarw}
w_{\sigma} = \frac{t_{\sigma}-1}{t_{\sigma}+1}
\end{equation}
with $t_{\sigma} = \tau^{\text{LDA}}/\tau^{\text{exact}}$ being the
ratio between the LDA and Kohn-Sham kinetic energy densities. The
$a_i$ are adjustable parameters with $a_0 = 1$ to recover the correct
uniform electron gas limit and $m = 11$. The correlation functional is
based on the $\tau$HCTH\cite{tauhcth} and the BMK\cite{bmk}
functionals but using the self-interaction correlation correction
proposed by Becke in the B95 functional.\cite{b96c1,b97} The
correlation functional contains another $10$ adjustable parameters,
two of which are independently fitted to atomic reference data for the
noble gases. The functional form of M05 (and some of the subsequent
functionals) is designed so as to preserve the correct behavior in the
uniform electron gas limit.

The key in the performance of the Minnesota functionals is the
parameter fitting to an extensive set of reference data. In the case
of M05, the data set employed comprises 35 data points including
atomization energies, ionization potentials, electron affinities,
barrier heights, total energies of atoms, bond dissociation energies
and noncovalent binding energies. The $20$ parameters (including the
fraction of exact exchange) are optimized by minimizing a target
function that mixes the root mean square deviations of the different
sets using a genetic algorithm. The functional was subsequently
tested, with relatively satisfactory results, in a larger set
containing $231$ data points.

The M05-2X functional,\cite{m052x} proposed shortly after M05,
approximately doubles the amount of exact exchange in M05 (56\%),
hence the name. The functional form is the same as M05, but the
training set used in the determination of the parameters is different,
and does not include metallic systems. As a result, M05-2X improves
the treatment of thermochemistry, kinetics and noncovalent
interactions by sacrificing the good performance in metallic systems.
The M05 and M05-2X functionals have also been tested for interactions
in noble gas dimers, group 2 dimers, the Zn dimer and Zn-rare gas
dimers.\cite{zhao2006} In general, they outperform other
non-dispersion-corrected functionals for those systems.

The M06-L functional\cite{m06l} is a pure meta-GGA functional (no
exact exchange). It is the basis of the M06 suite of functionals. The
exchange functional is more involved than in the M05 family, and
includes a contribution based on the van Voorhis and Scuseria
functional.\cite{vanvoorhis1998}
\begin{equation}
E_x^{\text{semilocal}} = \sum_{\sigma}\int
\left[
\varepsilon_{x\sigma}^{\text{PBE}}
\left(\sum_{i=0}^m a_i w_{\sigma}^i\right)
+ \varepsilon_{x\sigma}^{\text{LDA}} h_x(x_{\sigma},z_{\sigma})
\right]
d\bm{r}
\end{equation}
where the additional term not present in M05 (Eq.~\ref{eq:m05x})
contains the LDA exchange energy density and a function that depends
on the dimensionless density gradient ($x_{\sigma} =
\nabla\rho_{\sigma}/\rho_{\sigma}^{4/3}$) and on the variable
$z_{\sigma} = \tau_{\sigma}/\rho_{\sigma}^{5/3} -
3/5(6\pi^2)^{2/3}$. The expression for $h_x$ is:
\begin{equation}
h_x(x_{\sigma},z_{\sigma}) = \left(
\frac{d_0}{\gamma(x_{\sigma},z_{\sigma})} + 
\frac{d_1 x_{\sigma}^2 + d_2 z_{\sigma}}{\gamma^2(x_{\sigma},z_{\sigma})} + 
\frac{d_3 x_{\sigma}^4 + d_4 x_{\sigma}^2 z_{\sigma}}{\gamma^3(x_{\sigma},z_{\sigma})}
\right)
\end{equation}
with $\gamma(x,z) = 1 + \alpha (x^2+z)$, and $\alpha$ is a parameter
that is different depending on whether $\gamma$ appears in the
exchange or in the correlation functionals. The values for the
different $\alpha$ parameters are taken from previous
works.\cite{vanvoorhis1998} The correlation functional in M06-L is 
similarly based on M05 but augmented with terms coming from the work
of van Voorhis and Scuseria.\cite{vanvoorhis1998}

In its definition, M06-L contains $32$ parameters distributed within
different parts of the exchange and correlation functionals. As in
M05, the parameters are fitted to a training set by defining a cost
function in terms of the weighed root mean square deviations on the
different sets. The data sets include atomization energies, ionization
potentials, electron affinities, proton affinities, barrier heights,
noncovalent interactions, transition metal ligand removal energies,
alkyl bond dissociations, isomeric reactions and proton affinities
between unsaturated hydrocarbons, excitation energies, bond lengths
and bond frequencies. The training set contains $314$ data points and
the optimization is carried out under constraints to preserve the
physical soundness of some parameters and the correct behavior at the
uniform gas limit. M06-L performs better than other semilocal
functionals or even hybrids in standard thermochemical and kinetics
tests (with the possible exception of $\pi$-system proton affinities
and electron affinities). Being a pure meta-GGA it also presents the
advantage of being more computationally efficient than any hybrid
functional, and the possibility of implementation in plane wave codes
for condensed matter applications.

The closely-related M06-HF functional\cite{m06hf} contains a full
(100\%) exact exchange contribution plus a term equal to the M06
exchange-correlation energy but with different fitted parameters. The
functional is designed for the calculation of excitation energies
using TDDFT. The parametrization includes data sets of
noncovalently-bound systems, but the results for ground-state
properties, including noncovalent binding energies, are, in general,
worse than M05-2X.

Zhao and Truhlar subsequently used the same functional form as in
M06-L and a greatly enlarged training database to define a hybrid
with 27\%\ exact exchange (M06) and a hybrid with double the amount of
exact exchange (54\%, M06-2X).\cite{m06} As in the case of M05 and
M05-2X, the former is recommended by the authors for all applications
including main-group thermochemistry, kinetics, transition metal
chemistry, and noncovalent interactions, whereas the 2X version is
not appropriate for organometallic applications.

Because this chapter is about noncovalent interactions, a digression is
necessary at this point. All the functionals above have been fitted to
noncovalent interactions, but no measure of the applicability of
these has been given. For comparison, in ref.~\citenum{m06}, the
(balanced) mean average errors in the S22 set (see
the section titled ``Description of Noncovalent Interaction Benchmarks'') using the average of the counterpoise and
non-counterpoise-corrected results are 0.47 (M06-2X), 0.71 (M06-HF),
0.75 (M05-2X), 0.77 (M06-2L), 0.85 (M06), and 1.83 (M05)
kcal/mol. These results are much better than those obtained using
other non-dispersion corrected, less-parametrized density-functionals
(see the end of the benchmarks section below), but the performance is not as
good as that of other dispersion-corrected functionals using D3, XDM,
DCPs or most of the other dispersion corrections
(described in the section titled ``Performance of Dispersion-Corrected Methods''). Of the Minnesota family, M06-2X, with
0.47 kcal/mol error on average for the S22 set, displays the best
performance for noncovalent interactions.

For completeness, the remaining functionals in the Minnesota family
are: 

\begin{itemize}
\item The somewhat less-popular M08-HX and M08-SO
functionals.\cite{m08} The authors' motivation was to
``improve'' the functional expression in the M06 family by
increasing its flexibility, resulting in two functionals that depend
on 44 parameters each. The M08 functionals are meta-GGA hybrids, where
the semilocal exchange is written as:
\begin{equation}
\label{eq:m08x}
E_x^{\text{semilocal}} = \int \varepsilon_x^{\text{LDA}} \left(
f_1(w) F_x^{\text{PBE}} + 
f_2(w) F_x^{\text{RPBE}}\right)
d\bm{r}
\end{equation}
with $\varepsilon_x^{\text{LDA}}$ the LDA exchange energy density,
$F_x^{\text{PBE}}$ the PBE enhancement factor\cite{pbe} and
$F_x^{\text{RPBE}}$ the RPBE enhancement factor.\cite{rpbe} For the
correlation functional,
\begin{equation}
E_c = \int \left( \varepsilon_c^{\text{LDA}} f_3(w) + 
H_c^{\text{PBE}} f_4(w) \right) d\bm{r}
\end{equation}
where $\varepsilon_c^{\text{LDA}}$ is the LDA correlation energy and
$H_c^{\text{PBE}}$ the gradient correction to correlation from the PBE
functional.\cite{pbe} The $f_n$ factors are 11-degree polynomials in
$w$, which depends on the kinetic energy density
(Eq.~\ref{eq:truhlarw}). 

The difference between M08-HX and M08-SO is in the constraints imposed
on the parameters. For the former, the correct behavior in the uniform
electron gas limit is imposed, whereas the latter presents the correct
behavior to second-order in the reduced-density gradient expansion
(proportional to $s^2$) when $s\to 0$. The functionals are fitted
using the same procedure as for the other Minnesota functionals, that
is, designating a cost function and minimizing it against a very
extensive training set. The resulting functionals are hybrids with
about 50\%\ exchange (52.23\%\ for M08-HX and 56.79\%\ for M08-SO)
that improve slightly on the previous M06 functionals, except for
transition metals and systems with multireference character.

\item The SOGGA11\cite{sogga11} functional is a semilocal GGA, with
parametrized enhancement factors for exchange and correlation (10
parameters each, 20 parameters in total). The SOGGA11 functional is
fitted to a training data set and it preserves the correct uniform
electron gas limit and the exchange enhancement factor has the correct
$s^2$ behavior in the $s \to 0$ limit. However, the enhancement
factors for non-zero $s$ are fluctuating, which is a result of
oscillatory parameters coming from the fit (the same effect is
observed in the other Minnesota functionals of the pure GGA variety).

\item The M11\cite{m11} functional is the first range-separated hybrid
functional in the Minnesota set. M11 has 42.8\%\ exact-exchange hybrid
at short range and 100\%\ at long range, with a range-separation
parameter of 0.25 (both the amount of exact exchange and the
range-separation parameter are chosen to minimize the mean absolute
error in the dataset). The functional form of M11 is similar to
M08. The semilocal part of the short-range exchange functional being
the same as in M08 (Eq.~\ref{eq:m08x}), except in that the
range-separation enters the LDA exchange energy density according
using the expressions of Chai and Head-Gordon.\cite{wb97} The
correlation functional is exactly the same as in M08. After
application of some constraints on the parameter---uniform electron
gas limit, quadratic coefficient of the reduced-density gradient
expansion, and independence of the kinetic energy density in the bond
saddle points and density tails---the resulting functional has 38
coefficients that are obtained by the usual minimization procedure
employing a large training set.

\item The closely-related M11L functional uses the same
range-separated approach and the same functional form as M11, but it
replaces exact exchange by a different semilocal functional for the
long range exchange. M11L is, therefore, also semilocal. The
long-range exchange functional has the same expression as short-range
M11 exchange, but the short-range LDA exchange energy density is
replaced by its long-range counterpart. This functional is intended to
be used in solid-state calculations under plane waves and, as a
consequence, is parametrized using a collection of lattice constants
of simple solids in the training set.

\item The N12 functional\cite{n12} is the basis for the latest series
in the Minnesota family of functionals. N12 drops all previous
constraints on the exchange-correlation functional (including the
uniform electron gas limit and the spin-scaling relations for
exchange) and gives the exchange energy as:
\begin{equation}
\label{eq:n12}
E_{\text{nxc}} = \sum_{\sigma}\int 
\left[
\varepsilon_{x\sigma}^{\text{LDA}}
\sum_{i=0}^m \sum_{j=0}^{m^{'}} a_{ij} u_{x\sigma}^i v_{x\sigma}^j
\right]
\end{equation}
where $a_{ij}$ are adjustable parameters and the $u$ and $v$ variables
are defined as in the B97\cite{b97} and the Liu-Parr
functional\cite{liu1997} respectively:
\begin{align}
u_{x\sigma} & = \frac{\gamma_{x\sigma}x_{\sigma}^2}
{1+\gamma_{x\sigma}x_{\sigma}^2} \\
v_{x\sigma} & = \frac{\omega_{x\sigma}\rho_{\sigma}^{1/3}}
{1+\omega_{x\sigma}\rho_{\sigma}^{1/3}}
\end{align}
where $\gamma$ and $\omega$ are parameters and $x_{\sigma}
= \nabla\rho_{\sigma} / \rho_{\sigma}^{4/3}$ is the dimensionless
density gradient. The exchange functional is labeled ``nxc'' because,
since it does not obey the usual exchange spin-scaling relations, the
authors argue that it is also be accounting for correlation. The
correlation functional in N12 has a functional form similar to
exchange, with a Taylor expansion in $u$ as a factor of the uniform
electron gas correlation energy density. The resulting N12 functional
is a pure GGA
and the intent of the authors is to provide a GGA that gives good
structures (lattice constants, bond lengths) and energetics
(atomization energies, cohesive energies) in both molecules and
periodic solids.

\item The exact exchange expression in equation~\ref{eq:n12} was
extended by employing the variable $w$ that depends on the kinetic
energy density (Eq.~\ref{eq:truhlarw}). The resulting functional
(MN12-L) is a semilocal meta-GGA with the exchange energy being:
\begin{equation}
E_{\text{nxc}} = \sum_{\sigma}\int 
\left[
\varepsilon_{x\sigma}^{\text{LDA}}
\sum_{i=0}^3 \sum_{j=0}^{3-i} \sum_{k=0}^{5-i-j}
a_{ijk} u_{x\sigma}^i v_{x\sigma}^j w_{x\sigma}^k
\right]
\end{equation}
The correlation functional is the same as in N12. This functional
provides a small improvement over the older M11-L.

\item The two last functionals in the Minnesota family are N12-SX and
MN12-SX, proposed recently by Peverati and Truhlar.\cite{n12sx} These
functionals are based on N12 and MN12-L respectively, but use a
range-separated approach with short-range exact exchange (similar to
the HSE functionals\cite{hse03,hse06}). At long range, the two
functionals have 25\%\ exact exchange and the range separation
parameter is $\omega = 0.11$ in both cases.
\end{itemize}

Among the advantages of the Minnesota functionals is their widespread
implementation in most popular software packages, including
Gaussian,\cite{gaussian09} GAMESS,\cite{gamess1,gamess2} and
NWChem.\cite{nwchem} The M06-L semilocal functional is also
implemented in VASP\cite{vasp1,vasp2} and Quantum
ESPRESSO\cite{espresso} for calculations in periodic solids using
plane waves. Their popularity is justified by their good performance
(relative to other less-parametrized functionals) in dealing with
thermochemical and kinetics problems. Regarding noncovalent
interactions, however, one should be aware that the only functional in
the family that competes with the other methods presented in this
chapter is M06-2X and, even then, this functional has particular
disadvantages of its own. 

\begin{figure}
\includegraphics[width=0.80\textwidth]{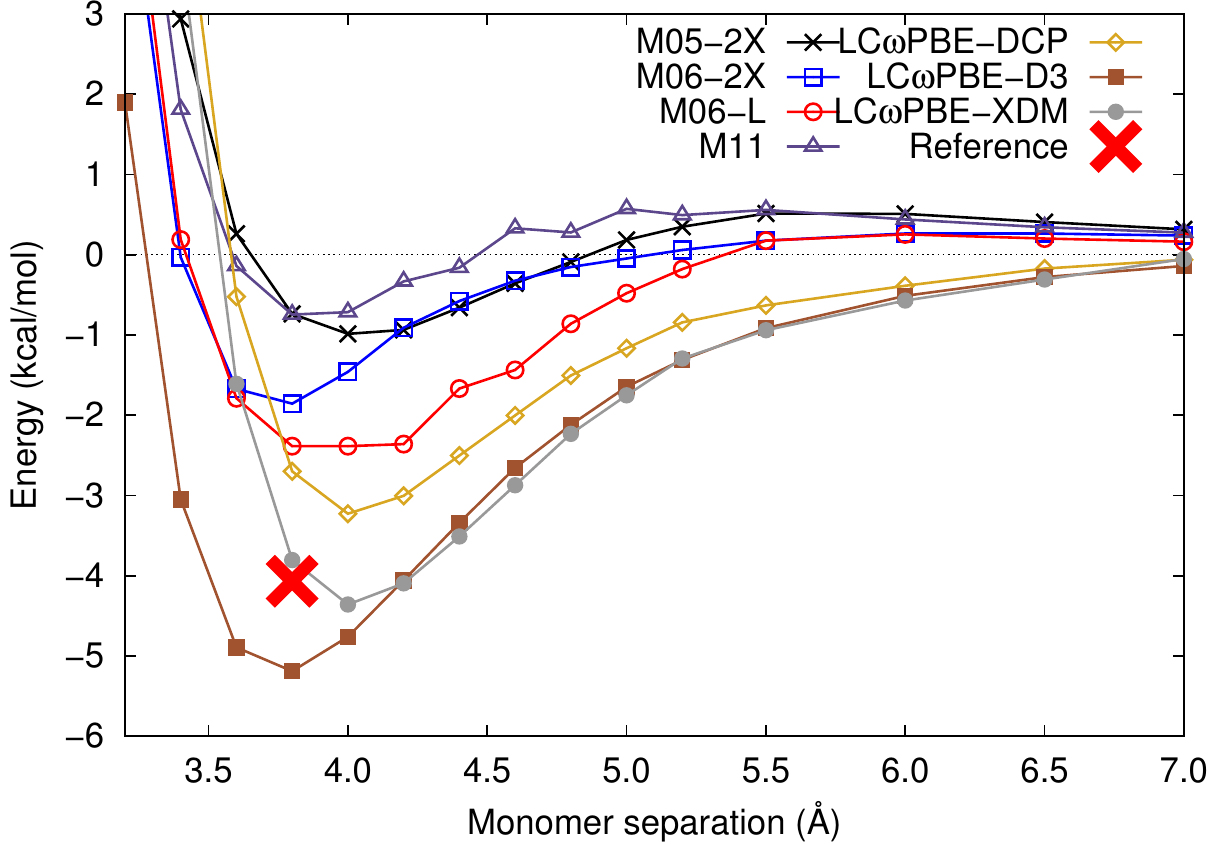}
\caption{Potential energy curves for the parallel naphthalene dimer,
calculated using different dispersion-corrected methods: XDM, one functional corrected with DCPs, DFT-D3 (using Becke-Johnson damping),
and four Minnesota functionals, including
M06-2X. The calculations were run using aug-cc-pVTZ and a pruned grid
with 99 radial and 590 angular (Lebedev) points (the ``ultrafine''
setting in Gaussian09\cite{gaussian09}). The reference value comes
from ref.~\citenum{johnson2013}.
\label{fig:m0sucks}}
\end{figure}

Figure~\ref{fig:m0sucks} illustrates clearly some of the problems with
the Minnesota functionals. The figure represents a potential energy
surface of the parallel naphthalene dimer. All Minnesota functionals
(including M06-2X) are strongly underestimating the binding energy. In
addition, the potential energy surface is uneven with multiple
spurious minima precluding the use of these functionals anywhere
except at the equilibrium geometry. The roughness of the potential is
smaller in M06-2X but the functional is underestimating the correct
binding energy by 2 kcal/mol (approximately 50\%\ error).  Another
spurious feature of the Minnesota potential energy curves for the
naphthalene dimer is that they display repulsive behaviour at large
monomer separations. Not only do the Minnesota functionals have the
incorrect long-range behavior, but they are repulsive except at
geometries close to equilibrium. The difficulties of the Minnesota
functionals away from equilibrium have been noted by other
authors.\cite{johnson2009,vydrov2012} Sensitivity to the integration
grid has also been reported.\cite{johnson2009}

\begin{table}
\caption{Mean Absolute Errors (MAE) with (CP) and without (no CP) Counterpoise
Corrections for Different Minnesota Functionals and Basis Sets (aNZ
$=$ aug-cc-pVNZ). 
\label{tab:m0sucks}}
\begin{tabular}{ccccc}
\hline \hline
Functional & Basis set & MAE (CP) & MAE (no CP) & Factor$^a$ \\
\hline
M06       & aQZ  &   0.79 &     0.44 &   1.79 \\
M06-2X    & aQZ  &   0.28 &     0.23 &   1.23 \\
M06-HF    & aQZ  &   0.83 &     0.49 &   1.70 \\
M06-L     & aTZ  &   0.73 &     0.28 &   2.62 \\
M06-L$^b$ & aQZ  &   0.90 &     0.51 &   1.77 \\
M06-L$^b$ & a5Z  &   0.84 &     0.49 &   1.73 \\
M11       & aQZ  &   0.79 &     0.44 &   1.79 \\
M11L      & aQZ  &   1.37 &     0.67 &   2.05 \\
MN12SX    & aQZ  &   1.18 &     0.67 &   1.75 \\
N12       & aQZ  &   3.87 &     3.72 &   1.04 \\
N12SX     & aQZ  &   2.10 &     1.97 &   1.07 \\
\hline \hline
\end{tabular}
\begin{minipage}{0.7\textwidth}
\raggedright
{$^a$ Quotient between both MAEs.}\\
{$^b$ SCF convergence problems were found for 2 dimers in
M06-L/aug-cc-pVQZ and for 21 dimers in M06-L/aug-cc-pV5Z. The MAEs for
those have been calculated using a reduced set.}
\end{minipage}
\end{table}

All Minnesota functionals except M06-2X are also extremely sensitive
to basis-set incompleteness errors in the calculation of noncovalent
interactions (see the section titled ``Performance of Dispersion-Corrected Methods'' and
Table~\ref{tab:m0sucks}). The table shows that the sensitivity is
lower in M06-2X, N12, and N12SX, for which it is similar to other
density functionals but, for the remaining members in the family,
the counterpoise correction has a non-negligible effect, even for a
basis set as large as aug-cc-pVQZ (or aug-cc-pV5Z in the case of
M06-L).

In addition, noncovalent binding in these functionals arises not from
a physically-justified model, but from parameter fitting to a database
that includes noncovalently bound dimers and also other
systems. Dispersion in the Minnesota functionals is partially
accounted for by adjusting a semilocal functional (a method that is
similar to using dispersion-correcting potentials,
described earlier), which is by design unable to describe long-range
intermolecular interactions such as dispersion. As a consequence, and
in contrast to other dispersion-including methods, dispersion-related
binding arises from a semilocal description of the electron density in
the intermolecular region. Molecules behave as if they are
``sticking'' to each other, but the interaction decays quickly
when separating the molecules as the electron density becomes the same
as in the non-interacting monomers. These observations may explain the
sensitivity of the Minnesota functionals to the grid and the basis
set, since the representation of the intermolecular regions is done by
atom-centered Gaussians.

\subsection*{Non-Local Functionals}
\label{s:nonlocal}
The use of non-local dispersion functionals is an approach that is
better rooted in traditional density functional theory development
than the functionals described earlier. Non-local functionals
model the dispersion energy contribution arising from electron density
fluctuations in distant parts of a system by explicitly taking into
account the effects those distant regions have on one another. The
non-local correlation contribution is:
\begin{equation}
\label{eq:nonlocalec}
E_{\text{c}}^{\text{nl}}
= \frac{1}{2} \int\int \phi(\bm{r},\bm{r}^{'}) \rho(\bm{r})\rho(\bm{r}^{'})
d\bm{r} d\bm{r}^{'}
\end{equation}
where $\phi(\bm{r},\bm{r}^{'})$ is the correlation
kernel. Conceptually, non-local functionals calculate the same
dispersion attraction as the simpler pairwise approaches, but they do
so without using an asymptotic expression. Non-local functionals are
``seamless'', meaning that the correlation energy defined in
Eq.~\ref{eq:nonlocalec} is added to the rest of the functional without
the need of specifying fragments or using an atomic partition, as is
the case in the pairwise approach. In addition, and unlike similar
approaches like RPA, all non-local functionals depend only on the
density and its derivatives and not on the orbitals (neither occupied
nor virtual). Non-local functionals are usually designed with little
or no empiricism.

Early studies that were fundamental in the development of non-local
functionals for dispersion are those of Zaremba and
Kohn\cite{zaremba1976} and of Rapcewicz and
Ashcroft.\cite{rapcewicz1991} In the limit of separated
(non-overlapping) fragments, the second-order perturbation theory
expression for the dispersion interaction is:\cite{zaremba1976}
\begin{equation}
\label{eq:response}
E_{\text{disp}} = -\frac{1}{2\pi}
\int
\frac{
\chi(\bm{r}_1,\bm{r}_1^{'},i\omega)\chi(\bm{r}_2,\bm{r}_2^{'},i\omega)
}{
r_{12} r_{12}^{'}
} d\bm{r}_1 d\bm{r}_2 d\bm{r}_1^{'} d\bm{r}_2^{'} d\omega
\end{equation}
where the integration over $\bm{r}_1$ is restricted to the first
fragment and $\bm{r}_2$ to the second. The $\chi$ functions are the
density-density response functions\cite{zangwill1980} (usually simply
response functions or electric susceptibilities) defined as the linear
response of the electron density with respect to a perturbation in the
external potential with frequency $\omega$:
\begin{equation}
\delta \rho(\bm{r},\omega) = \int \chi(\bm{r},\bm{r}^{'},i\omega) \delta
V^{\text{ext}}(\bm{r}^{'},\omega) d\bm{r}^{'}
\end{equation}
These quantities are central to the current section and are related to
the dynamic polarizabilities (see ``Pairwise Dispersion Corrections'' above)
by:\cite{zangwill1980,dobson2012}  
\begin{equation}
\alpha_{ij}(i\omega) = \int \bm{r}_i\bm{r}^{'}_j
\chi(\bm{r},\bm{r}^{'},i\omega) d\bm{r} d\bm{r}^{'}
\end{equation}
where $i$ and $j$ are the components of the polarizability
tensor. Likewise, they can be used to rewrite the adiabatic connection
formula (Eq.~\ref{eq:acf}):
\begin{equation}
\label{eq:acf_chi}
E_c = \frac{-1}{2\pi}
\int_0^1 d\lambda
\int d\bm{r} d\bm{r}^{'} \frac{1}{|\bm{r}-\bm{r}^{'}|}
\int_0^{\infty}
d\omega
\left[
\chi_{\lambda}(\bm{r},\bm{r}^{'},i\omega)-\chi_0(\bm{r},\bm{r}^{'},i\omega)
\right]
\end{equation}
where $\chi_0$ is the response function for the non-interacting system
that has an analytical expression in the Kohn-Sham scheme:
\begin{equation}
\chi_0(\bm{r},\bm{r}^{'},i\omega) = -4
\sum_i^{\text{occ}} \sum_a^{\text{unocc}}
\frac{\varepsilon_{ai}}{\varepsilon_{ai}^2 + \omega^2}
\psi_i(\bm{r})\psi_a(\bm{r})\psi_a(\bm{r}^{'})\psi_i(\bm{r}^{'})
\end{equation}
with $i$ running over occupied and $a$ over unoccupied orbitals, and
$\varepsilon$ represents the orbital energy differences.

The seminal works upon which non-local dispersion functionals rest
are the original functionals of Dobson and Dinte (DD)\cite{dobson1996}
and Andersson, Langreth and Lundqvist (ALL),\cite{andersson1996} which
have the same expression and were published independently in 1996. The
DD/ALL functional is a variant of the Rapcewicz-Ashcroft functional
that uses the density response function of the uniform electron gas:
\begin{equation}
\label{eq:chiueg}
\chi(\omega)
= \frac{1}{4\pi}\left[1-\frac{1}{\varepsilon(\omega)}\right]
= \frac{1}{4\pi}\frac{\omega_p^2}{\omega_p^2-\omega^2}
\end{equation}
where $\varepsilon(\omega)= 1 - \omega_p^2/\omega^2$ is the dielectric
function and $\omega_p = \sqrt{4\pi\rho}$ is the plasma frequency of
the uniform electron gas with density $\rho$. By using the local
approximation to the plasma frequency ($\omega_p(\bm{r}) =
\sqrt{4\pi\rho(\bm{r})}$), the resulting dispersion energy is:
\begin{equation}
\label{eq:all}
E_{\text{disp}} = -\frac{3}{32\pi^2}\int
\frac{1}{r_{12}^6}
\frac{\omega_{p}(\bm{r})\omega_{p}(\bm{r}^{'})}{\omega_{p}(\bm{r})+\omega_{p}(\bm{r}^{'})} 
d\bm{r} d\bm{r}^{'}
\end{equation}
where $\omega_p$ is the local plasma frequency. This functional
requires that the interacting systems are non-overlapping. The
dynamical polarizabilities and the dispersion interaction coefficients
can be calculated in the same fashion.\cite{andersson1996} Note the
resemblance between equation~\ref{eq:all} and London's formula
(Eq.~\ref{eq:london}).

Langreth's group subsequently proposed variations of this functional
for different systems, including the study of the interaction between
parallel infinite jellium surfaces,\cite{rydberg2000} a
non-overlapping formulation with a cutoff to account for
overlaps,\cite{hult1999} and a functional for layered
structures.\cite{rydberg2003} However, the first truly
geometry-independent functional is vdw-DF, proposed in 2004 by Dion et
al.\cite{dion2004,dion2004err}

In vdw-DF, the correlation energy is written as the sum of a semilocal
part, represented by LDA correlation, and a long-range non-local
energy according to equation~\ref{eq:nonlocalec}.
\begin{equation}
E_c = E_c^{\text{sr}} + E_c^{\text{nl}}
\end{equation}
The key to a ``seamless'' functional is that the non-local part of the
correlation energy vanishes for the uniform electron gas, therefore
allowing the treatment of long-range and short-range interactions on
the same footing and preventing double counting of correlation
effects. The vdw-DF functional makes approximations to the adiabatic
connection formula (equation~\ref{eq:acf_chi}) based on a second-order
expansion of the $S = 1-\varepsilon^{-1}$ variable ($\varepsilon$
being the dielectric function) and a plasmon pole approximation to the
plane wave representation of $S$. By virtue of those approximations,
the adiabatic connection formula can be integrated in the coupling
constant. After some algebra the kernel in
equation~\ref{eq:nonlocalec} is written as a function of
two variables $d$ and $d^{'}$ that depend only on the distance between
$\bm{r}$ and $\bm{r}^{'}$ and on the electron density and its gradient
at those points. Their expressions are:
\begin{align}
d & = |\bm{r}-\bm{r}^{'}|q_0(\bm{r}) \\
d^{'} & = |\bm{r}-\bm{r}^{'}|q_0(\bm{r}^{'}) 
\end{align}
with:
\begin{equation}
\label{eq:vdwdfq}
q_0(\bm{r}) = -\frac{4\pi}{3}
\left(
\varepsilon_c^{\text{LDA}}(\bm{r}) + 
\varepsilon_x^{\text{LDA}}(\bm{r}) [1+\lambda s(\bm{r})^2]
\right)
\end{equation}
involving the LDA exchange and correlation energy densities, the
reduced density gradient:
\begin{equation}
s = \frac{\nabla\rho}{2(3\pi^2)^{1/3}\rho^{4/3}}
\end{equation}
and the parameter $\lambda = 0.8491/9$ that controls the relative
importance of the gradient correction. The expression of the kernel
$\phi(d,d^{'})$ is complicated, involving a double integral, but the
existence of the intermediate $d$ variables allow a pre-computation of
a lookup table for $\phi$, which is used to interpolate its values and
derivatives in the actual SCF calculation.

The vdw-DF functional has been implemented self-consistently in the
plane wave approach by Thonhauser et al.\cite{thonhauser2007} and for
Gaussian basis sets by Vydrov et al.\cite{vydrov2008} Likewise, the
analytic energy gradients that are required for geometry optimizations
have been implemented in both cases. The computational cost of
evaluating the non-local correlation energy is the bottleneck if a
semilocal exchange functional is chosen, but it is not more expensive
than calculating the exact exchange energy contribution in a hybrid or
range-separated hybrid.\cite{vydrov2008,vydrov2012} Efficient
implementations in the particular case of plane wave basis sets have
been proposed as well.\cite{romanperez2009}

The primary target for the vdw-DF functional are systems with extensive
electron-electron delocalization such as metal surfaces,
physisorption, interactions with graphene, etc. A review of some
applications has been published by Langreth et
al.\cite{langreth2009} In the case of molecular interactions at
equilibrium, however, the functional is plagued by problems coming
from spurious binding caused by the semilocal exchange and correlation
components.\cite{vydrov2009b} In the original implementation, the
authors used the revPBE functional.\cite{revpbe} Several other options
for the exchange functional have been explored by other
authors.\cite{klimes2010} For comparison, the mean average error of
vdw-DF on the S22 is 1.44 kcal/mol with revPBE,\cite{vydrov2010c} 1.03
kcal/mol with revised PW86,\cite{vydrov2010c} and 0.23 kcal/mol with a
specifically-adapted version of the B88 functional called
opt-B88.\cite{klimes2010} In addition, vdw-DF has a tendency to
overestimate molecular separations and to underestimate the strength
of hydrogen bonds.\cite{vydrov2009b,lee2010}

To address some of the problems in vdw-DF, and with an eye on
molecular interactions in the overlapping regime, Lee et al. proposed
an improved functional, vdw-DF2.\cite{lee2010} The authors replaced the over-repulsive revPBE functional with a
revised version of the PW86 functional.\cite{murray2009,lee2010} In
addition, the coefficient in the internal coefficient controlling the
gradient correction to LDA in equation~\ref{eq:vdwdfq} was modified
using the known behavior in the limit of large number of
electrons.\cite{elliott2009} Lee et al. report an improvement in the
mean average error obtained in the S22 using vdw-DF2: 0.51
kcal/mol. The improvement is also evident in the calculation of
lattice energies and geometries of molecular
crystals.\cite{oterodelaroza2012b} The vdw-DF family of functionals,
and particularly vdw-DF2, are in widespread use today in the physics
community and are implemented in popular solid-state codes like
Quantum ESPRESSO\cite{espresso} and VASP.\cite{vasp1,vasp2} To our
knowledge, only Q-Chem\cite{qchem} implements these functionals for
gas-phase calculations. The vdw-DF can also be evaluated using the external
program noloco.\cite{noloco} 

Based on the poor performance of the original vdw-DF for
noncovalently bound molecular systems in gas phase, Vydrov and van
Voorhis (VV) proposed a series of modifications, including the
VV09\cite{vydrov2009,langreth2010,vydrov2010reply} and the VV10
functionals.\cite{vydrov2010} VV noted the aforementioned problems in
the vdw-DF functional and its inability to couple with either
Hartree-Fock exchange or with long-range corrected functionals. The VV
family of functionals introduces a reduced number of adjustable
parameters (one or two) and violates some conservation laws enforced
in the functionals by Langreth et
al.\cite{langreth2010,vydrov2010reply} but the results are greatly
improved for molecular systems thanks to the additional
flexibility. In particular, VV09 includes one adjustable parameter
that is fitted to reproduce atomic $C_6$ values, which are known to be
in severe error when calculated using the vdw-DF functionals
(particularly vdw-DF2 with errors slightly over
60\%).\cite{vydrov2010b} In addition, VV is formulated for
spin-polarized (open-shell) systems and the kernel in
equation~\ref{eq:nonlocalec} is analytic rather than numerical. For
the S22, rPW86-VV09 gives a MAE of 1.20
kcal/mol\cite{vydrov2010c} with LDA correlation contributing
appreciably to the binding.\cite{vydrov2010c}

VV10 is the simplest and most accurate functional\cite{vydrov2010} in
the family. In VV10, the exchange functional can either be revised
PW86 (rPW86)\cite{murray2009,lee2010} or LC-$\omega$PBE with $\omega =
0.45$. The former is termed simply VV10 (parameters $C = 0.0093$ and
$b = 5.9$, see below) and the second is LC-VV10 ($C = 0.0089$ and $b =
6.3$). The semilocal correlation functional is PBE correlation in both
of them.  The non-local correlation energy in VV10 is written as in
equation~\ref{eq:nonlocalec}. The correlation kernel is proposed
\emph{ad hoc} based on the authors' experience:
\begin{equation}
\phi(\bm{r},\bm{r}^{'}) = -\frac{3}{2gg^{'}(g+g^{'})}
\end{equation}
with 
\begin{equation}
g = \omega_0(\bm{r})|\bm{r}-\bm{r}^{'}|^2 + \kappa(\bm{r})\quad ; \quad
g^{'} = \omega_0(\bm{r}^{'})|\bm{r}-\bm{r}^{'}|^2 + \kappa(\bm{r}^{'})
\end{equation}
The related quantities are:
\begin{equation}
\omega_0(\bm{r}) = \sqrt{\omega_g^2(\bm{r}) + \frac{\omega_p^2(\bm{r})}{3}}
\end{equation}
with $\omega_p$ the local plasma frequency defined above and
$\omega_g$ the local band gap:\cite{vydrov2010b}
\begin{equation}
\omega_g(\bm{r})^2 = C\left|\frac{\nabla\rho(\bm{r})}{\rho(\bm{r})}\right|
\end{equation}
where $C$ is an adjustable parameter. The other component in $g$ is:
\begin{equation}
\kappa(\bm{r}) = b\frac{3\pi\rho(\bm{r})^{1/3}}{\omega_p(\bm{r})}
\end{equation}
with $b$ another adjustable parameter (VV10 introduces this new
parameter in addition to the $C$ already present in VV09). The
long-range correlation energy is defined as the non-local part plus a
constant times the number of electrons in the system so that it
vanishes in the uniform electron gas limit. The functional provides
correct asymptotics in the infinite separation limit, is easier to
implement, and provides improved statistics thanks to the flexibility
provided by its adjustable parameters. 

The results for the S66 database are reported in
ref.~\cite{vydrov2012} and can be compared with the results in
table~\ref{tab:S66ALL} (described later). Vdw-DF2 tends to show a
systematic underbinding of all molecular dimers, in particular those
involving $\pi$ systems. Semilocal VV10 overbinds hydrogen-bonds but
these effects are fixed when using the long-range corrected version,
LC-VV10, which achieves an outstanding 0.15 kcal/mol error on average.
Both VV functionals have been implemented
self-consistently,\cite{vydrov2010c} including the analytic gradients
for geometry optimizations, in Q-Chem.\cite{qchem}

\section*{Performance of Density-Functionals for Non-Covalent Interactions}
\label{s:performance}

\subsection*{Description of Non-Covalent Interactions Benchmarks}
\label{s:bench}
Developers of new DFT methods for noncovalent interactions have
 focussed largely on the prediction of accurate binding energies. To
achieve this end, researchers make use of sets of benchmark data containing
 structures and binding energies calculated
using reliable {\it ab initio} wavefunction theory methods. A number
of benchmark data sets for a wide range of small, noncovalently-bonded
dimer systems have been developed. The availability of accurate
benchmark data for larger systems is less common. These benchmarks
provide the first steps toward the development of comprehensive DFT
methods that are capable of accurately including noncovalent
interactions into the simulation of materials of all kinds, from small
molecular dimers in vacuum to larger molecules immersed in solvents
and solids.

A preponderance of benchmark data exists for the binding energies
and corresponding structures of noncovalently-interacting dimers in
vacuum owing to the ease with which reference data can be
calculated. One of the most commonly accessed sources for benchmark
data for these systems is the ``Benchmark Energy and Geometry
Database'' of \v{R}ez\'{a}\v{c} et al.~\cite{rezac2008,begdbweb} This
on-line resource contains several sets of benchmarks, including the
A24 set of small molecule dimers,~\cite{rezac2011} water clusters
containing up to 10 monomers obtained from the work of Shields's
group,~\cite{shields2011} the X-40 set of halogen-containing molecular
dimers, the S66 set of molecular dimers containing interactions found
in organic and biomolecular interactions,~\cite{rezac2011,rezac2011a}
the S22 set~\cite{jurecka2006,marshall2011} which is similar to but
smaller than the S66 set, and a few other benchmark sets.  Other
groups have generated or compiled benchmark data sets, including those
of Sherrill,~\cite{marshall2011} Johnson,~\cite{gatsby}
Truhlar,~\cite{m06,minnesotaweb} and Grimme.~\cite{goerigk2011}

\begin{figure}
\centering
{\large Hydrogen-bonded dimers}\\
\hfill
\includegraphics[width=0.15\textwidth]{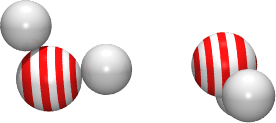}
\hfill
\includegraphics[width=0.15\textwidth]{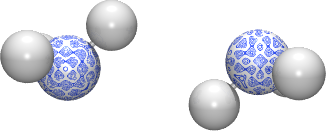}
\hfill
\includegraphics[width=0.15\textwidth]{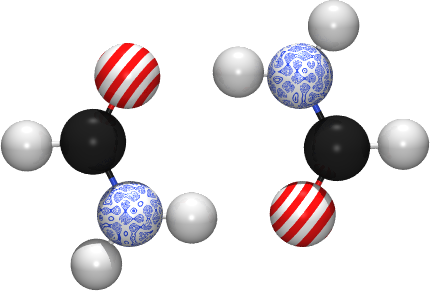}
\hfill
\includegraphics[width=0.15\textwidth]{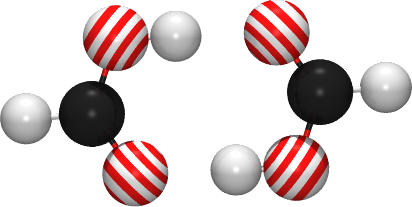}
\hfill\\[0.4\baselineskip]
\includegraphics[width=0.22\textwidth]{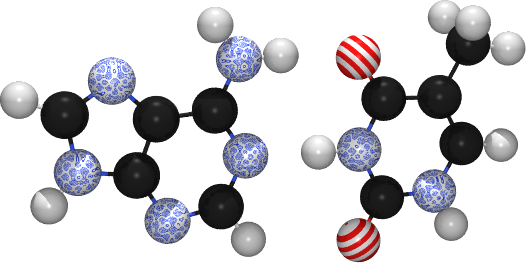}
\hspace{1cm}
\includegraphics[height=0.22\textwidth,angle=90]{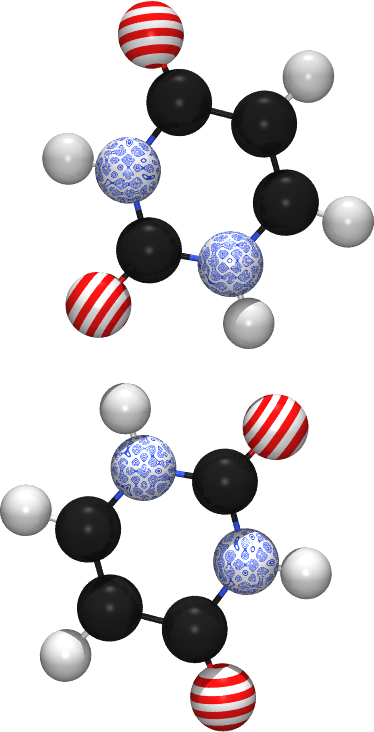}
\hspace{1cm}
\includegraphics[width=0.22\textwidth]{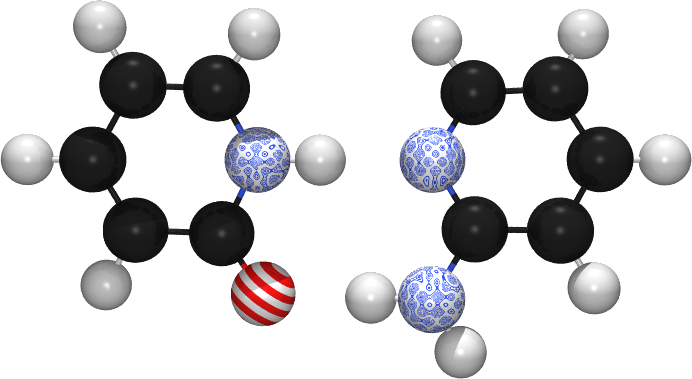}
\hfill
\rule{0.80\textwidth}{1pt}
\vspace{0.5\baselineskip}
{\large Dispersion-bound dimers}\\
\hspace{1cm}
\includegraphics[height=0.12\textwidth]{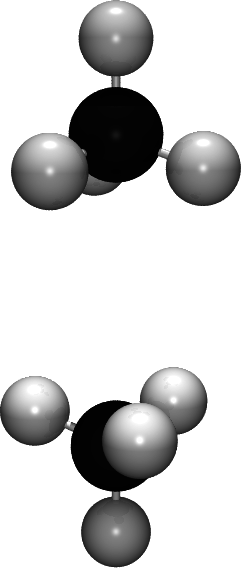}
\hfill
\includegraphics[height=0.12\textwidth]{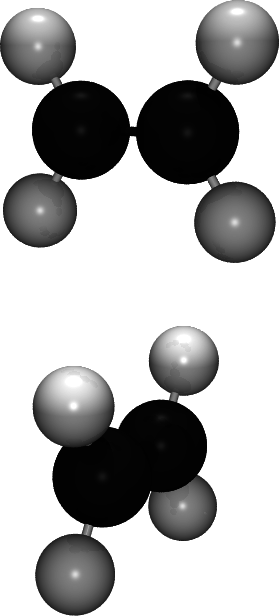}
\hfill
\includegraphics[width=0.12\textwidth]{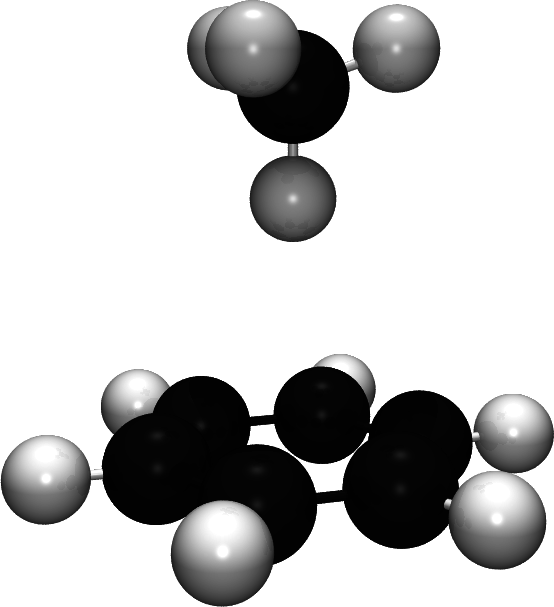}
\hfill
\includegraphics[width=0.15\textwidth]{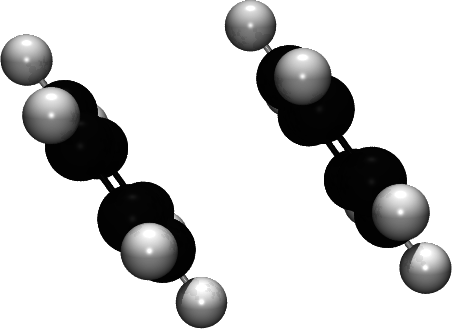}
\hfill\\[0.4\baselineskip]
\includegraphics[width=0.14\textwidth]{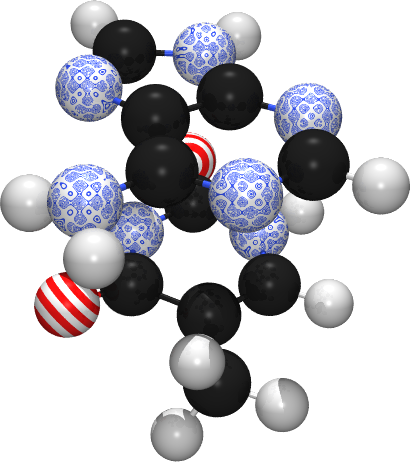}
\hfill
\includegraphics[width=0.15\textwidth]{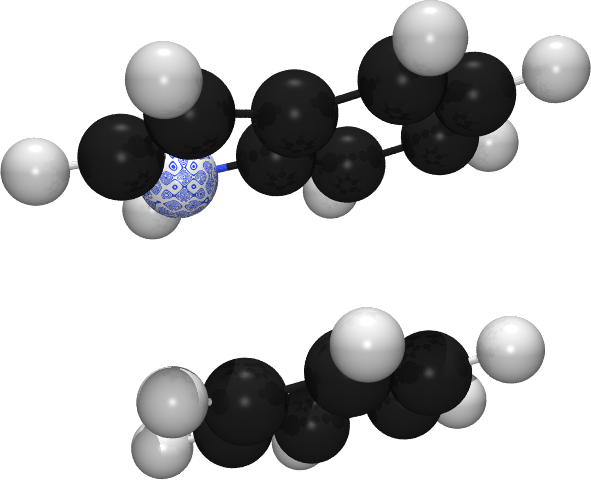}
\hfill
\includegraphics[width=0.13\textwidth]{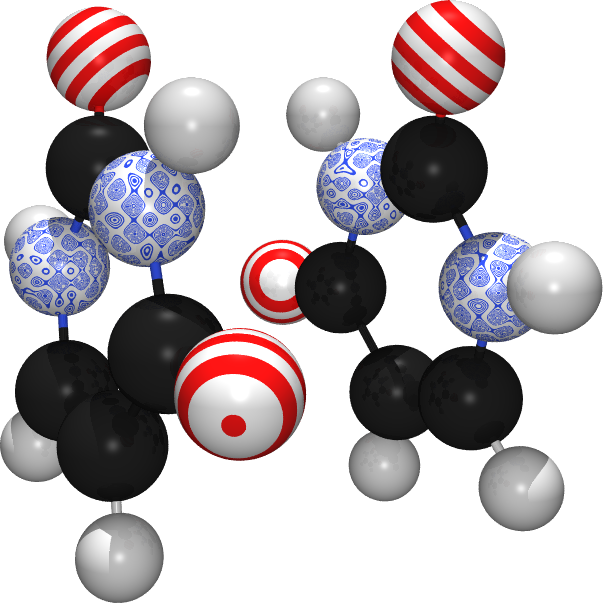}
\hfill
\includegraphics[width=0.12\textwidth]{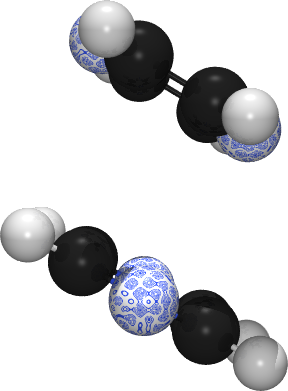}
\hfill
\rule{0.80\textwidth}{1pt}
\vspace{0.5\baselineskip}
{\large Mixed-interaction dimers}\\
\hspace{2cm}
\includegraphics[width=0.07\textwidth]{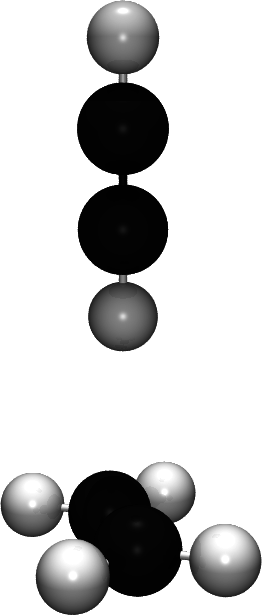}
\hspace{3cm}
\includegraphics[width=0.15\textwidth]{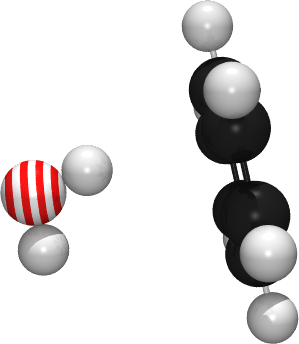}
\hspace{3cm}
\includegraphics[width=0.15\textwidth]{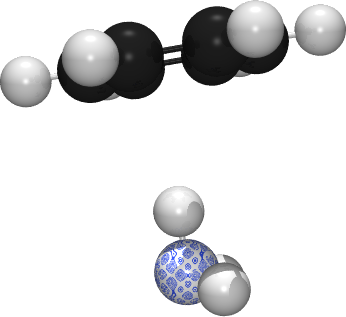}
\hfill\\
\hfill
\includegraphics[width=0.10\textwidth]{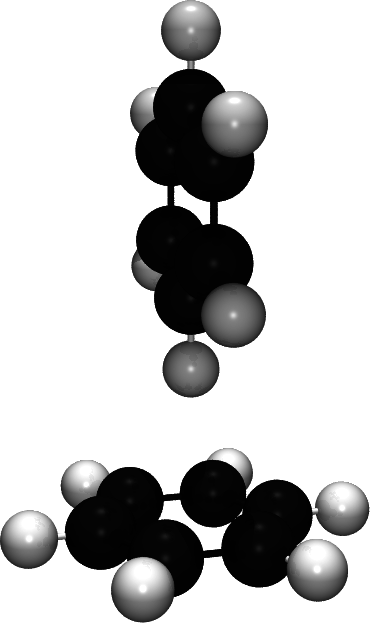}
\hfill
\includegraphics[width=0.15\textwidth]{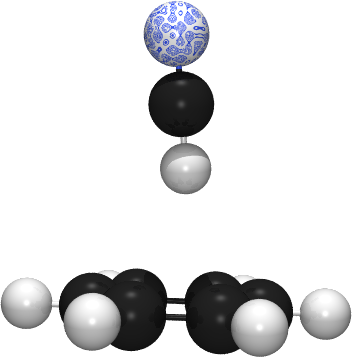}
\hfill
\includegraphics[width=0.20\textwidth]{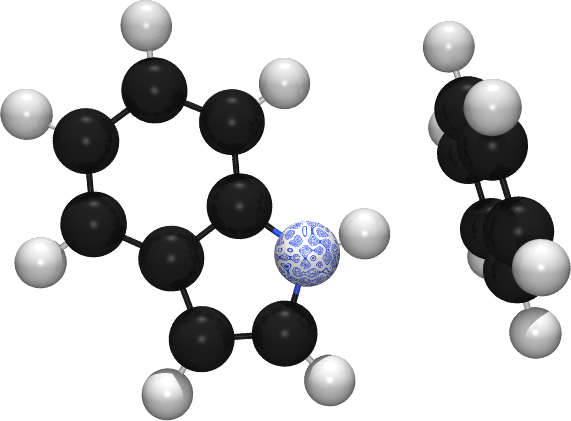}
\hfill
\includegraphics[width=0.20\textwidth]{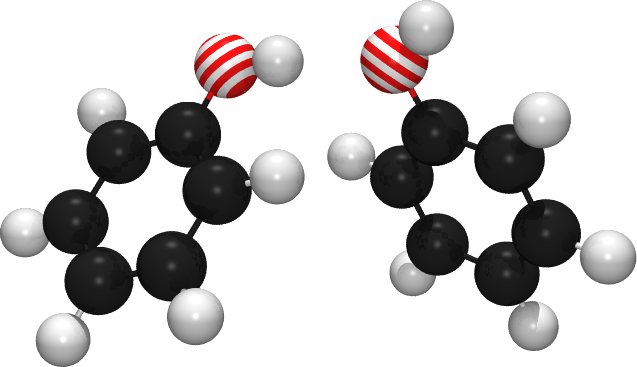}
\hfill
\caption{The dimers in the S22 set, grouped by dominant interaction
type. The atoms are C (black), hydrogen (light gray), oxygen (red
stripes), and nitrogen (blue dots).
\label{fig:S22set}}
\end{figure}

The S22 set contains the atomic coordinates for dimer structures whose
geometry was optimized using an {\it ab initio} wavefunction method,
mostly MP2/cc-pVTZ with counterpoise (CP)
corrections.~\cite{boys1970} Amongst the dimers are seven complexes
predominantly interacting via hydrogen bonding (dimers of ammonia,
water, formic acid, formamide, uracil, and complexes of 2-pyroxidine
with 2-aminopyridine and adenine with thymine), eight complexes bound
mostly by dispersion (dimers of methane, ethene, benzene, pyrazine,
stacked uracil, and complexes of methane-benzene, stacked
indole-benzene and stacked adenine-thymine), and seven complexes
interacting via mixed forces (ethene-ethyne, benzene-water,
benzene-ammonia, benzene-hydrogen cyanide, T-shaped benzene dimer,
T-shaped indole-benzene and the phenol dimer). The structures of these
complexes are illustrated in Figure~\ref{fig:S22set}. The S66
benchmark set is modeled after the S22 set to some degree. It contains
the structures and binding energies of 66 molecular dimers in vacuum:
23 dimers interacting predominately by hydrogen bonding, 23 dimers in
which the dominant interaction can be considered to be dispersion, and
20 dimers in which the interactions are mixed. The benchmark binding
energies were computed using CCSD(T) with complete-basis-set
extrapolation (CCSD(T)/CBS), which is an approach capable of providing
high-quality reference data for noncovalent interactions. Recently,
Marshall et al.\cite{marshall2011} revised the binding energies of the
S22 set using a higher level of theory than was originally used for the
S22.\cite{jurecka2006} The revised S22 set is often referred to as the
S22B set.

As an aside, it is important to keep in mind that the term ``CBS'' is
generic and could apply to any procedure involving basis set
extrapolations regardless of the quality of the basis sets employed as
part of the extrapolation. It is generally understood that certain
extrapolations that make use of energies obtained from double-$\zeta$
basis sets (e.g. aug-cc-pVDZ) are not always able to provide accurate
results when it comes to the binding energies in noncovalently bonded
systems.~\cite{mackie2011,burns2014}  

When making use of the databases to assess the predictions made by a
particular method, the structures are employed as presented
with no adjustments or optimizations. The binding energies are then
computed simply by calculating the electronic energy of the dimer
system and subtracting from that the energy of the two monomers. Users
of the structures in the S22 and S66 sets (as well 
as many other benchmark sets available) should be aware that the
binding energies do not include monomer deformation. When the monomers
are brought together to form the dimer, they undergo a small amount of
structural distortion in order to accommodate their new
environment. The degree of the structural distortion depends on the
strength of the interaction, and is particularly important in
hydrogen-bonded dimers. The reference binding energies reported
for the S22 and S66 benchmark sets were calculated as the difference
in binding energy of the dimer structure and the monomers in their
distorted forms. There is nothing wrong with using these structures to
obtain the benchmark binding energies, so long as users of the data
are aware of their origin and use the database accordingly. Later in this section, we describe the performance of many
dispersion-corrected DFT methods using the S22B and S66 benchmark
sets. 

\begin{figure}
\includegraphics[width=0.70\textwidth]{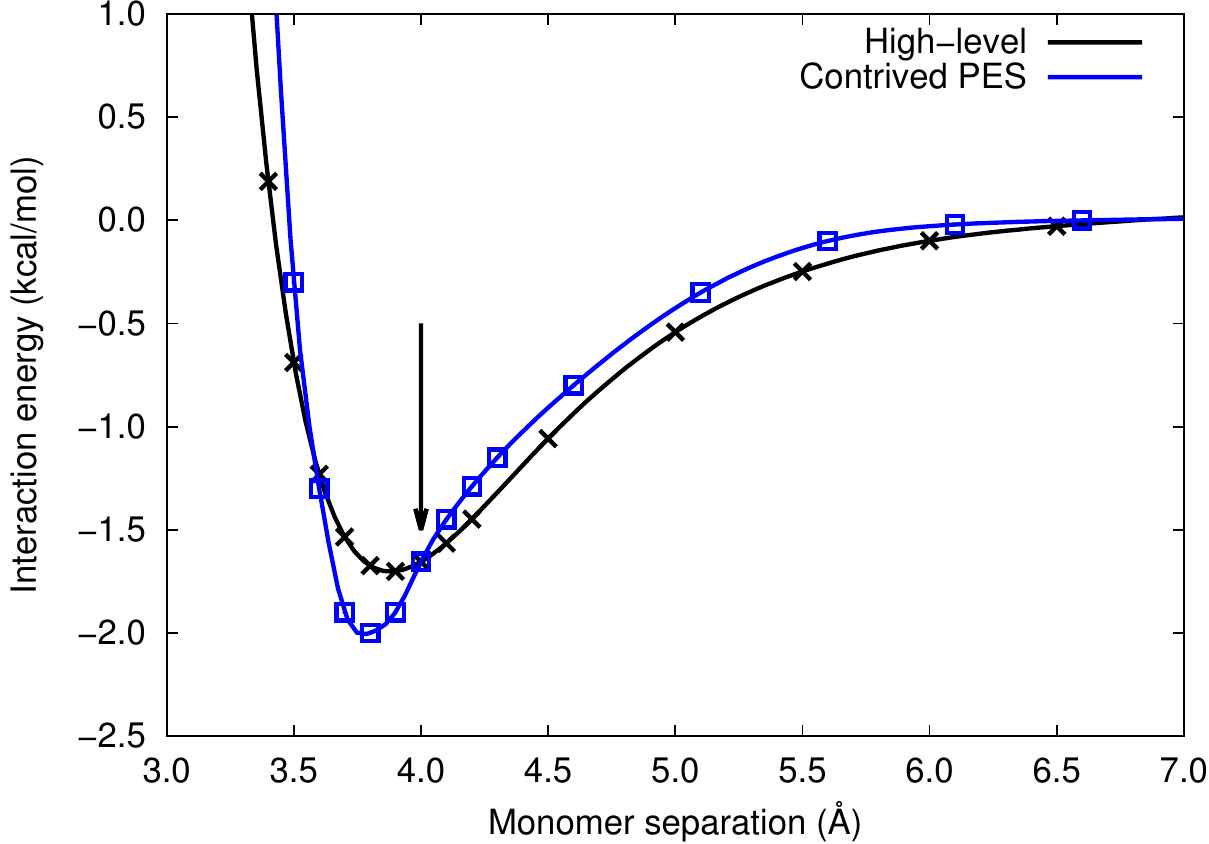}
\caption{Contrived potential energy surfaces representing the
stretching of a noncovalently-interacting dimer. The arrow represents
the point at which the high-quality reference energy is provided, and
which is used in the training of the hypothetical dispersion-corrected
functional.
\label{fig:ContrivedPES}}
\end{figure}

The S22 and S66 databases described above, along with a number of the
other generally-accessible small-molecule databases, focus exclusively
on the binding energies of noncovalently interacting dimer systems at
a single dimer structure. This dimer geometry is close to the minimum
of the potential energy surface (PES) associated with the
dimer. However, it is important to keep in mind that achieving
agreement with the database value of a single binding energy near the
dimer minimum is no guarantee that the DFT method will 
reproduce the features of the whole potential energy surface accurately. To
illustrate this point, consider the two contrived one dimensional PESs
shown in Figure~\ref{fig:ContrivedPES}. The PES defined by the sqaures
represents the benchmark data associated with a particular dimer
structure. The minimum of this PES occurs around 4.0\AA\ and is
highlighted by an arrow. The structure and binding energy at this
geometry is representative of an entry in a database like the S66
set. It is possible for a dispersion-corrected DFT method applied to
the minimum energy structure at 4\AA\ to reproduce the binding energy
exactly, thus giving the impression that the DFT method performs well
in that particular structure. However, it is also possible that the
very same DFT method actually produces a PES that looks like the one
defined by the squares in Figure \ref{fig:ContrivedPES}, rather than
reproducing the entire high-level PES. Taking this wider view of the
PES, even in one dimension, reveals that the DFT method has some
serious deficiencies and performs much worse than is indicated by the
results obtained at the minimum.

To a significant extent, the risks of being misled by an approximate
computational methodology in the fashion suggested in
Figure~\ref{fig:ContrivedPES} is quite high if only a small number of
dimers are used for performance tests. However, using a large
benchmark database like the S66 significantly mitigates these risks
because the set contains a small number of distinct atoms that have a
large number of different spatial arrangements. In this sense, large
benchmark sets offer a means of broadly sampling the different
bonding environments of the atoms contained in the set. 

In any case, the shortcomings associated with databases that contain
information only about noncovalently-interacting dimers at their
minima is beginning to be recognized and efforts are being made to
create databases that contain more PES information about these
dimers. Two prominent examples are the S22x5~\cite{grafova2010} and
S66x8~\cite{rezac2011a} databases, and it seems to be a trend that new
databases are developed containing detailed PES information (see, for
example, reference~\citenum{rezac2012}). One small difficulty, in
particular with the S66x8 database, is that the level of theory
applied to obtain its reference values is slightly lower
that that used for the S66 database. This makes the simultaneous use
of the S66x8 and S66 databases for benchmarking approximate methods
somewhat confusing and there has not yet been the widespread
application of these extended databases to assess dispersion-corrected
DFT methods. Our expectation is that databases that contain more
information about the potential energy surfaces will become more
important in the future. As demonstrated by the generally good
performance of most dispersion-corrected DFT methods
(Tables~\ref{tab:S22ALL} and~\ref{tab:S66ALL}) in the next section,
more stringent and detailed tests will be needed in order to
differentiate them. 

\begin{figure}
\centering
\includegraphics[width=0.25\textwidth]{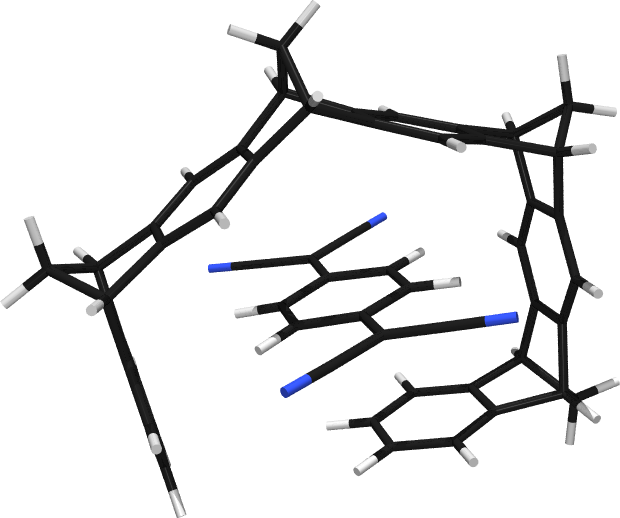}
\hspace{1cm}
\includegraphics[width=0.25\textwidth]{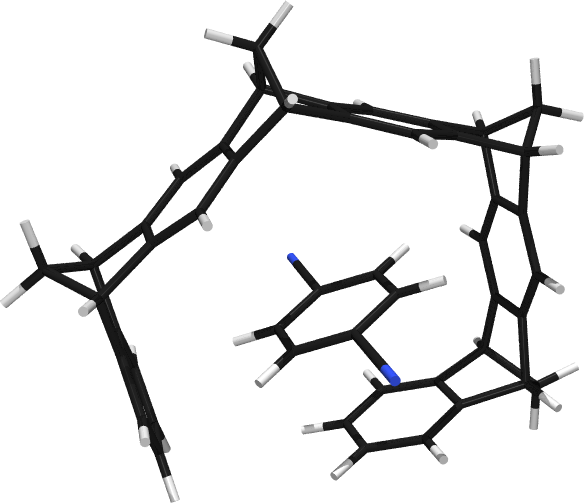}\\
\includegraphics[width=0.25\textwidth]{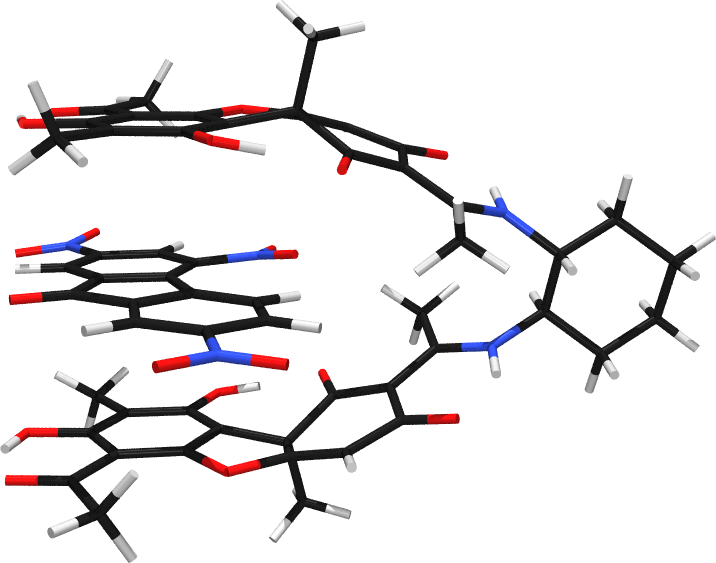}
\hspace{1cm}
\includegraphics[width=0.25\textwidth]{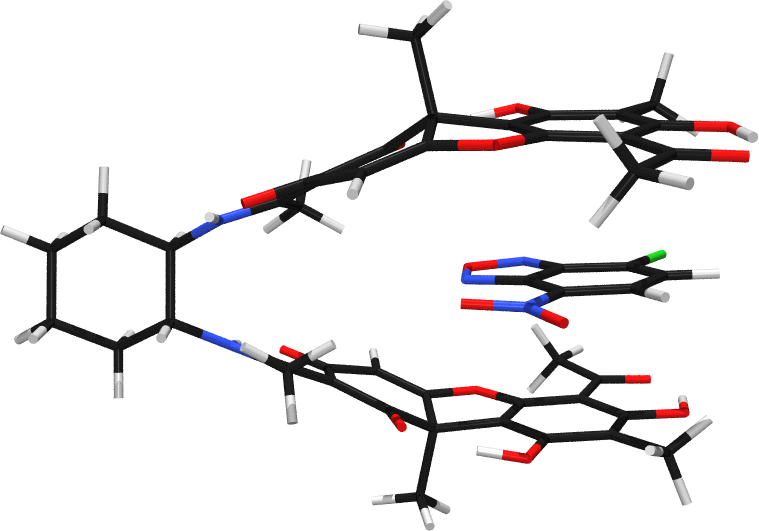}\\
\includegraphics[width=0.25\textwidth]{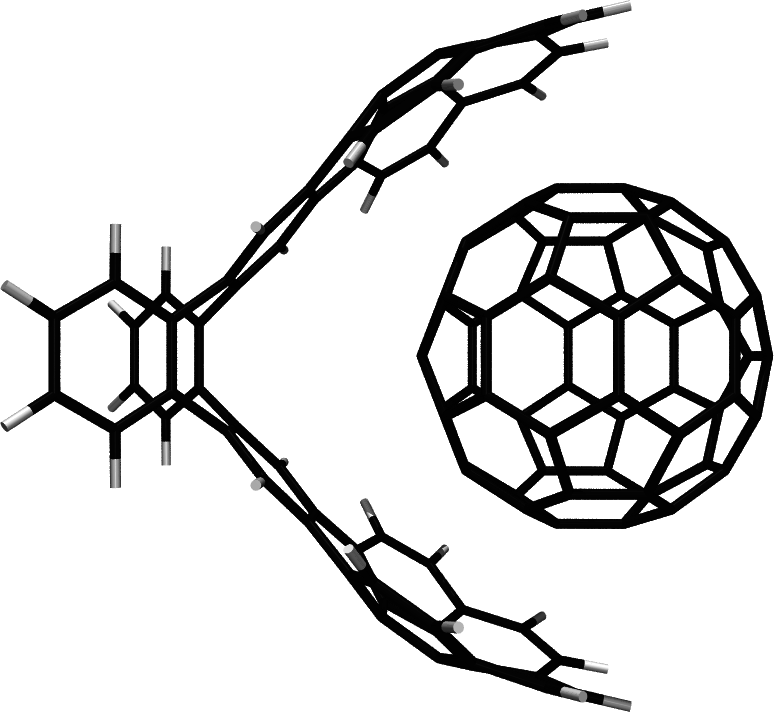}
\hspace{1cm}
\includegraphics[width=0.25\textwidth]{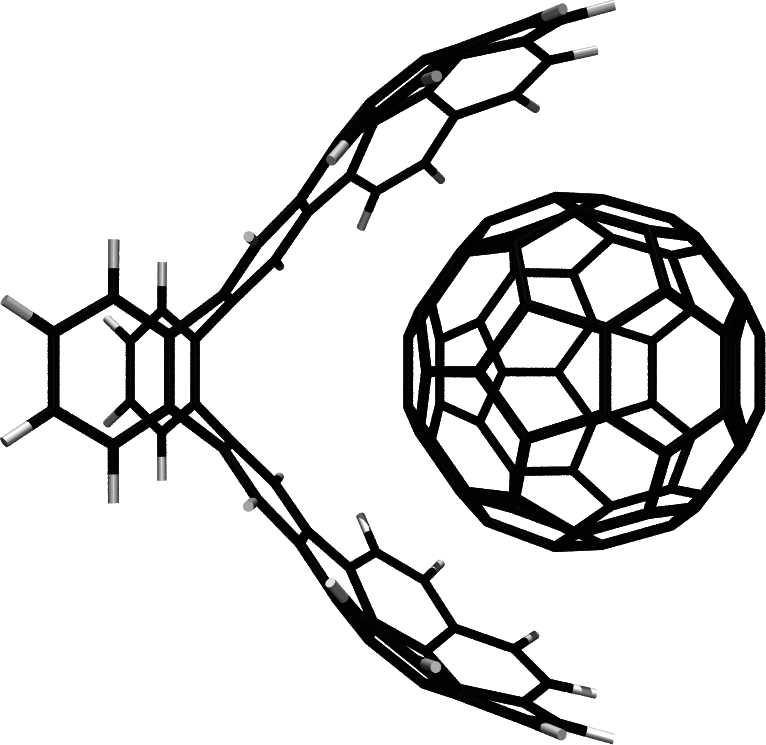}\\
\hspace{1cm}
\includegraphics[width=0.25\textwidth]{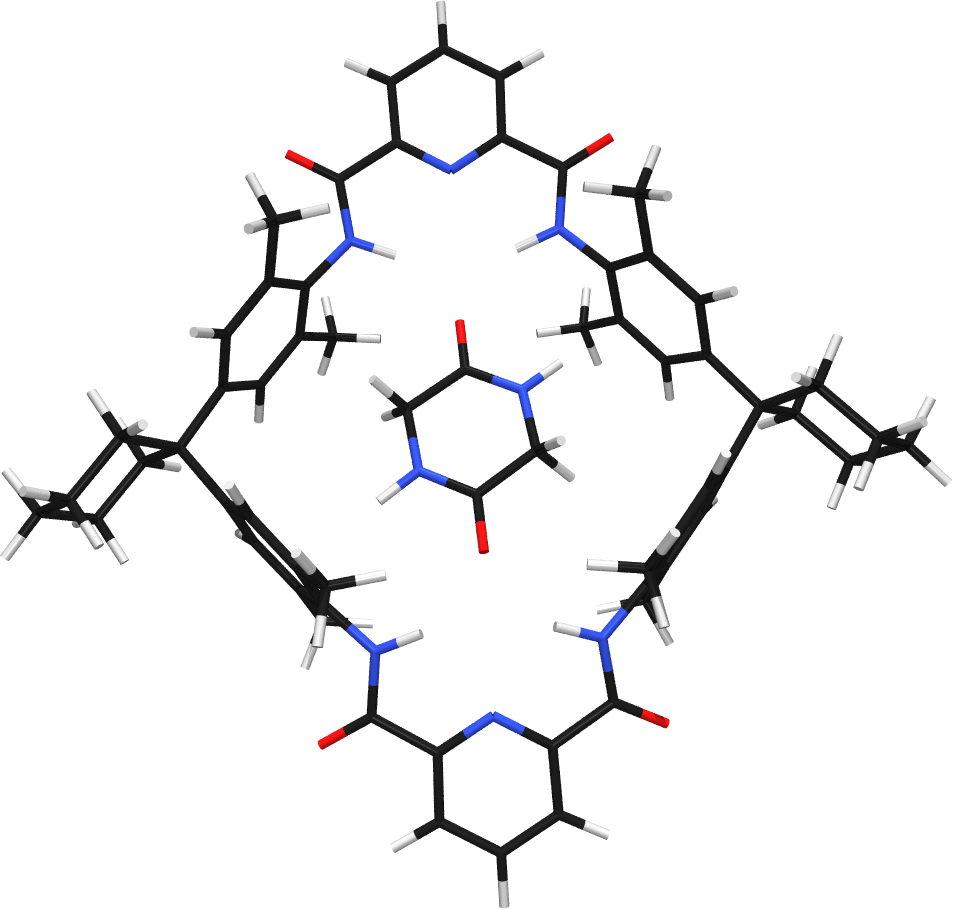}
\hspace{0.5cm}
\includegraphics[width=0.25\textwidth]{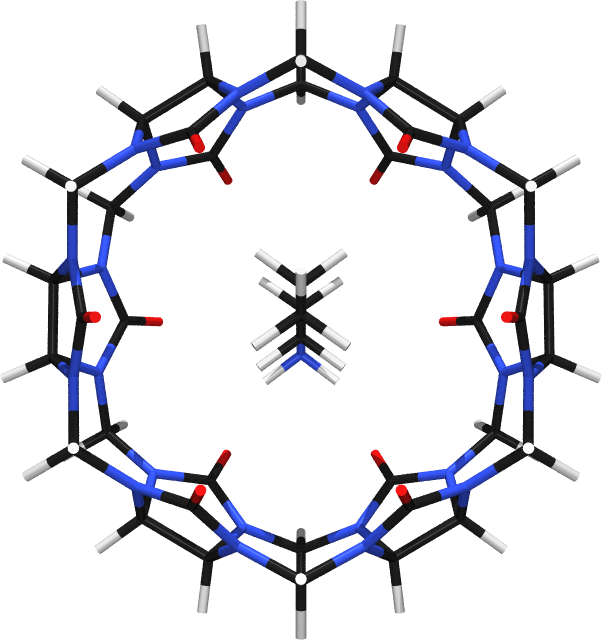}
\hspace{0.5cm}
\includegraphics[width=0.25\textwidth]{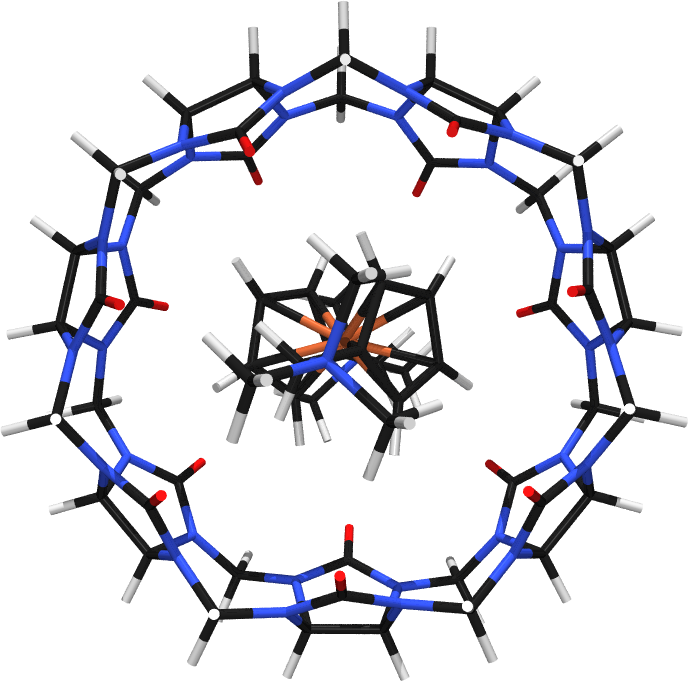}\\
\includegraphics[width=0.25\textwidth]{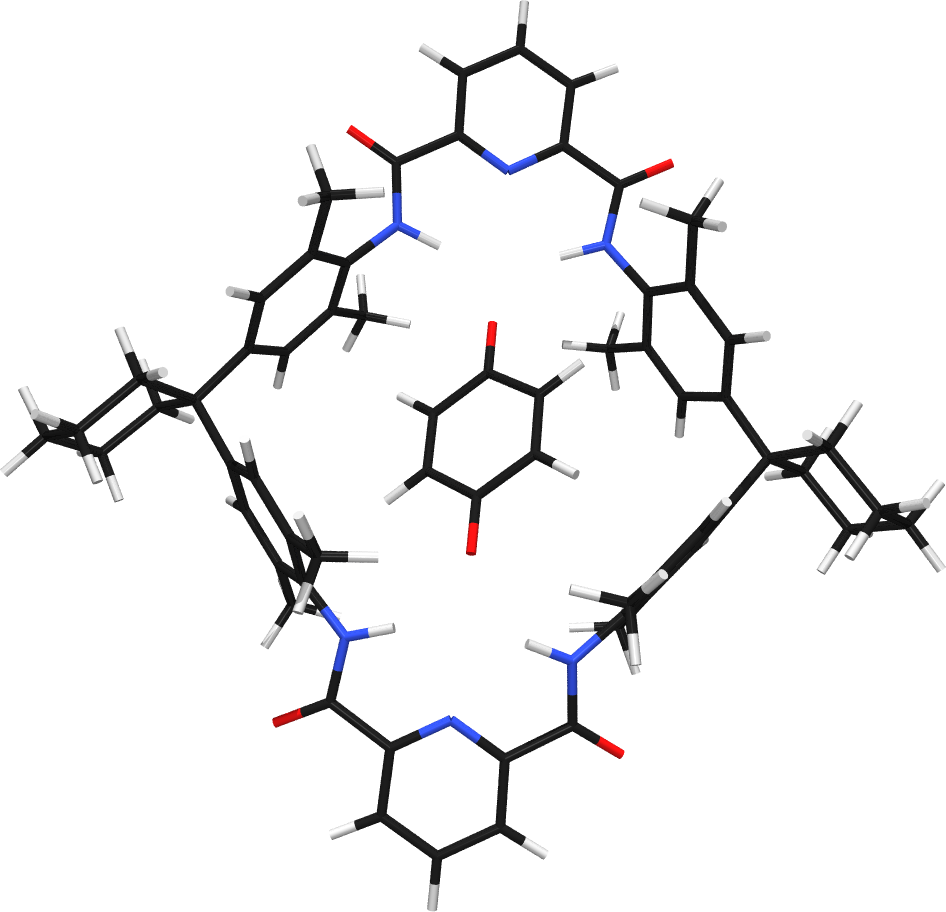}
\hspace{0.5cm}
\includegraphics[width=0.25\textwidth]{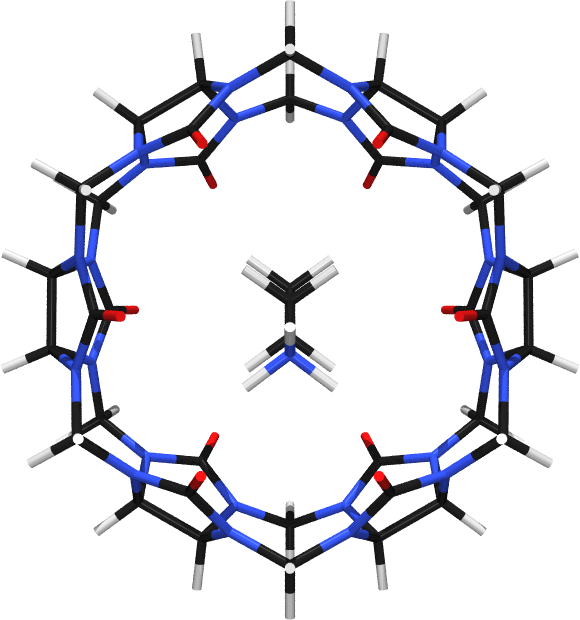}
\hspace{0.5cm}
\includegraphics[width=0.25\textwidth]{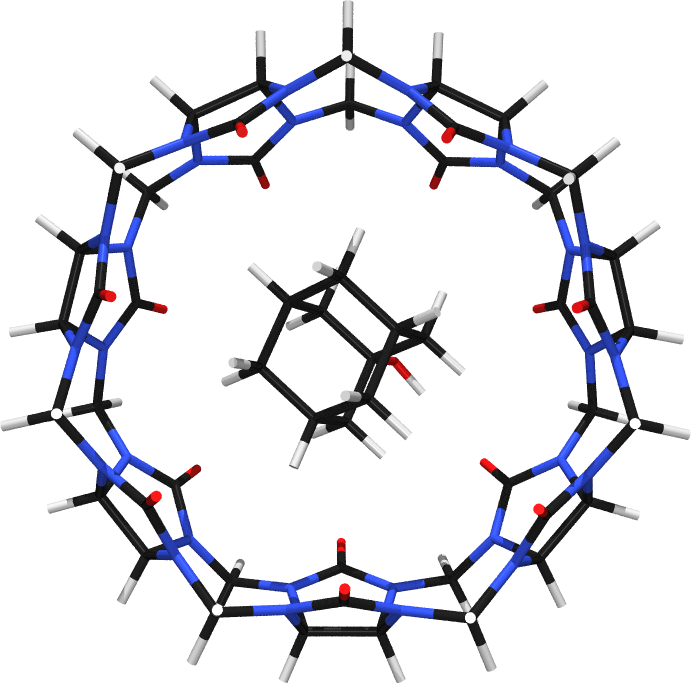}
\caption{The molecular complexes in the S12L benchmark set.\cite{grimme2012}
\label{fig:S12L}}
\end{figure}

One of the important practical strengths of dispersion-corrected DFT is
that it can be used to model large systems containing up to several
hundred atoms, depending on the implementation (plane wave or
gas-phase), the complexity of the DFT method and the basis set
sizes. However, benchmark sets containing large molecular systems
were, until very recently, rare. One set containing 7 large
systems is available online\cite{begdbweb} but is not yet commonly
used. A more popular set due to Grimme is the S12L
set.~\cite{grimme2012} The entries in this benchmark set are displayed
in Figure~\ref{fig:S12L}.  The set includes two ``tweezer'' complexes
with tetracyanoquinone (TCNQ) and 1,4-dicyanobenzene, two ``pincer''
species complexed with heteroatom-substituted $\pi$-delocalized
molecules, a ``buckycatcher'' complexed with C$_{60}$ and C$_{70}$
fullerenes, an amide macrocycle coupled with benzoquinone and glycine
glycine anhydride, complexes of cucurbit[6]uril cation with
butylammonium and propylammonium, and finally complexes of
cucurbit[7]uril bis(trimethylammoniomethyl) ferrocene with neutral
1-hydroxyadamantane. The binding energies of the dimers in the set
range from ca. 20 to 132 kcal/mol and arise from a variety of
noncovalent forces. Like the S22 and S66 sets, the binding energies of
the entries in the S12L set are determined by computing the
differences between the single-point energies of the dimers and their
constituent monomers without modifications to the structures. Grimme
used experimental association data, back-corrected for the effects of
solvent and molecular vibration in order to compute the noncovalent
binding energies. Most recently, Tkachenko's group computed the
noncovalent binding energies of a subset of S12L using a quantum Monte
Carlo (QMC) technique in order to update the reference binding
energies presented by Grimme.\cite{ambrosetti2014} QMC methods can, in
principle, provide CCSD(T)/CBS quality binding energy but results will
be dependent upon the nodal structure of the wavefunction on which the
QMC simulations are based.

Our discussion thus far has focussed largely on molecule-molecule
interactions and on the problems associated with modeling these
systems using modern DFT methods. In condensed matter systems,
noncovalent interactions play a central role in determining
solvent-solute interactions, physisorption at gas/solid interfaces,
and the properties of molecular crystals. Needless to say, in the
development of dispersion-corrected DFT methods, there must be a
convergence between the properties of the molecular and the bulk
regimes.  

In terms of benchmarking DFT based methods, few
reference sets exist for condensed matter systems. Like the
S12L benchmark set described above, computing high-level wavefunction
properties like structures and binding energies are out of the
question for most condensed matter systems. As such, the development
of benchmarking data necessitates the use of experimental data, and this should be viewed poditively because it offers a
direct comparison between theory and reality. However, it does mean
that care must be taken to ensure that all of the effects that are
present in the experiment are properly accounted for. These may
include the effects of solvent, zero-point/lattice vibration and thermal and
entropic effects that come into play at non-zero temperatures, to
offer just a few examples.  

One of the first convenient benchmark sets described for solids is the
C21 set of Otero-de-la-Roza and
Johnson.~\cite{oterodelaroza2012b} It contains 21 crystals of
small molecules, mostly of organic and biomolecular origin, and includes:
1,4-cyclohexanedione, acetic acid, adamantane, ammonia, anthracene,
benzene, CO$_2$, cyanamide, cytosine, ethylcarbamate, formamide,
imidazole, naphthalene, $\alpha$-oxalic acid, $\beta$-oxalic acid,
pyrazine, pyrazole, triazine, trioxane, uracil, and urea. For these
molecular crystals, the experimental sublimation enthalpies, $\Delta
H_{sub}^{0}$, are available\cite{acree2010} and are back-corrected for
zero-point vibration and thermal effects to give values for $\Delta
E^{exp}_{el}$. Examples of the entries in the set include CO$_2$,
having the smallest $\Delta H_{sub}^{0}$ of 5.9 kcal/mol, to
cytosine, which has the largest $\Delta H_{sub}^{0}$ of 39.1
kcal/mol. The C21 set also comprises a set for the molecular
geometries determined experimentally using X-ray and neutron
diffraction, and back-corrected to remove the effects of the crystal
vibrations. The C21 set provides a good bridge from the molecular
to the condensed regime and offers a convenient and somewhat more
comprehensive way of assessing dispersion-corrected DFT methods. That
is, while many of the dimer test sets like the S22 and S66 sets
include the binding energy of only a single geometry for each dimer,
and the S22x5 and S66x8 benchmark sets include one-dimensional PESs,
the sublimation enthalpy data of the C21 set reflect the
three-dimensional environment experienced by a molecule within a
solid. In other words, this environment includes the many-body
interactions and the long-range forces that arise from the crystal
field.

Being able to assess the interactions amongst molecules in
very different orientations accurately will ultimately allow for accurate crystal
structure prediction, and this will lead to the ability of engaging in
meaningful materials design work. However, it should be kept in mind
that there is more to structure prediction than calculating accurate
interaction energies: Other challenges are associated with the
development of algorithms for seeking the local and global
minima associated with the arrangement of atoms within a crystal and,
for example, differentiating crystal polymorphs.~\cite{woodley2008}  
The ability to predict {\it a priori} molecular crystal structure has
implications in the phamaceutical industry from the standpoint of drug
bioavailability and stability.~\cite{cabri2007} Also important are the
legal consequences associated with the protection of intellectual
property related to pharmaceuticals and the possibility of the
existence of multiple drug polymorphs. The interesting case of
polymorphism in the antibiotic Cefdinir is described in
reference~\citenum{cabri2007} and points to the potential benefit of
molecular crystal structure prediction to this industry.  

Before proceding to the next section where the performance of various
dispersion-corrected DFT techniques are compared, we
underscore that conventional DFT methods do, in general, a poor job
in predicting the binding energies of noncovalently-bonded systems by
using one of the benchmark sets described above. Table~\ref{tab:S22DFT} compiles the mean absolute errors in binding
energies for a variety of DFT methods that cover the range of GGA,
hybrid GGA, meta-GGA and range-separated functionals, as reported by
Goerigk and Grimme.~\cite{goerigk2011} Although there are distinct
differences in the predictions made by various base functionals for
the binding energies of the S22B set, there can be no doubt upon
examination of Table~\ref{tab:S22DFT} that conventional DFT methods
fail to model noncovalent systems accurately.

\begin{table}
\caption{Mean Absolute Errors (MAE) of the
Binding Energies in the S22B Benchmark Set\cite{jurecka2006,marshall2011}
Predicted by Various DFT Methods using def2-QZVP Basis Sets
(in kcal/mol). 
\label{tab:S22DFT}}
\begin{tabular}{cc|cc|cc|cc}
\hline \hline 
Functional & MAE  & Functional & MAE & Functional & MAE & Functional & MAE \\
\hline 
\multicolumn{2}{c|}{Semilocal}& OPBE     & 7.73            & BHLYP    & 2.85 & MPWB1K         & 1.81 \\ \cline{1-2}
      B97    & 5.25           & BPBE     & 5.19            & PBE0     & 2.36 & B1B95          & 3.26 \\
      B986   & 4.00           & rPW86PBE & 2.82            & PBE38    & 2.27 & BMK            & 2.61 \\ \cline{7-8}
      BOP    & 6.67           & SSB      & 2.98            & revPBE0  & 4.30 & \multicolumn{2}{c}{Range-separated} \\ \cline{7-8}
      BLYP   & 4.77           & revSSB   & 2.45            & revPBE38 & 3.83 & CAM-B3LYP      & 2.52 \\
      MPWLYP & 3.37           & TPSS     & 3.45            & TPSSh    & 3.28 & LC-$\omega$PBE & 2.81 \\
      OLYP   & 7.33           & oTPSS    & 4.48            & TPSS0    & 3.04 &                &      \\ \cline{3-4}
      PBE    & 2.57           & \multicolumn{2}{c|}{Hybrid}& PW6B95   & 1.95 &                &      \\ \cline{3-4}
      PBEsol & 1.81           & B3LYP    & 3.77            & MPW1B95  & 2.12 &                &      \\
      revPBE & 5.21           & B3PW91   & 4.13            & PWB6K    & 1.20 &                &      \\
\hline \hline 
\end{tabular}
\begin{minipage}{0.7\textwidth}
\raggedright
The functionals are grouped by classes, indicated by horizontal labels
on the table (semilocal functionals, hybrids, and range-separated
hybrids). 
\end{minipage}
\end{table}

\subsection*{Performance of Dispersion-Corrected Methods}
\label{s:dispcorrected}
The popularity of the S22 and S66 sets is fortuitous because it makes
it somewhat straightforward to compare the performance of different
DFT methods for noncovalent interactions. Table~\ref{tab:S22ALL}
contains the mean absolute errors (MAEs) of calculated binding
energies (in kcal/mol) obtained using a variety of density-functional
theory methods and dispersion correction schemes, along with the
reference to the work from which the quoted results were taken. The
Table is arranged so that GGA, hybrid-GGA, meta-hybrid GGA and
range-separated DFT methods are presented from top-to-bottom,
left-to-right. It is important to point out that some groups use the
term mean absolute deviation (MAD) in reference to MAE. However, these
two quantities are very different according to their definitions in
statistics. Generally speaking, when groups report data for the S22
benchmark set as MADs, they are really referring to MAEs.

The variety of basis sets employed in the 
benchmarking studies makes the direct comparison of the results
challenging at first sight. However, most works utilize very large and
nearly complete basis sets (e.g. def2-QZVP, aug-cc-pVTZ, or plane
wave), which permits a nearly direct comparison. The
dispersion-correcting potential approach was designed with the goal
 being efficiently applied to large systems and so it makes
use of smaller 6-31+G(2d,2p) basis sets. Recall that DCPs mitigate to
some extent the effects of basis set incompleteness and so the
expectation is that the performance of DFT-DCP/6-31+G(2d,2p) can be
compared to other approaches that use very large basis sets.

In some cases it is useful to compare the results obtained with
smaller basis sets in order to understand how performance can vary
with basis set size. However, users should be aware of the methods
(DFT approach, basis sets, etc.) under which the dispersion correction
approach of choice was developed and are advised to apply the same
parameters in their calculations, or study the variations caused by
basis set incompleteness in their chosen calculation method
appropriately. 

{
\renewcommand*{\arraystretch}{1.0}
\renewcommand*{\baselinestretch}{1.0}
\begin{table}
\caption{Mean absolute errors (MAEs) in the
Binding Energies for the S22B Benchmark
Set\cite{jurecka2006,marshall2011} of Noncovalently-Bonded Dimers
Predicted by Different Dispersion-Correcting Approaches and
Functionals, in kcal/mol. 
\label{tab:S22ALL}} 
\begin{tabular}{ccl|ccl|ccl|ccl}
\hline \hline
 Func.     &  Disp.   &       MAE  &  Func.      &  Disp.  &   MAE  &  Func.     &  Disp.  &   MAE  &  Func.                   &  Disp.  &       MAE  \\
\hline
 B97            &  D3          &      0.38      &  SSB        &  D3     &  0.63  &  PBE0      &  XDM    &  0.53  &  M05-2X                  &  MN     &      0.79  \\
 B986          &  D3          &      0.66      &  revSSB     &  D3     &  0.49  &  PBE38     &  D3     &  0.63  &  M06                     &  D3     &      0.26  \\
 BOP           &  D3          &      0.52      &  TPSS       &  D3     &  0.32  &  PBEh      &  TS     &  0.30  &  M06                     &  TS     &      0.42  \\
 BLYP          &  D3          &      0.24      &  TPSS       &  TS     &   0.2  &  HSE       &  TS     &  0.39  &  M06                     &  MN     &      1.06  \\
 BLYP          &  XDM       &      0.22       &  oTPSS      &  D3     &  0.31  &  revPBE0   &  D3     &  0.32  &  M06-2X                  &  D3     &      0.36  \\
 BLYP          &  DCACP   &  0.33$^a$  &  M06-L      &  D3     &  0.44  &  revPBE38  &  D3     &  0.39  &  M06-2X                  &  MN     &  0.40$^b$  \\
 MPWLYP    &  D3          &      0.55      &  M06-L      &  TS     &  0.37  &  TPSSh     &  D3     &  0.38  &  M06-HF                  &  D3     &      0.84  \\
 OLYP         &  D3          &      0.71      &  B97-1      &  D3     &  0.36  &  TPSS0     &  D3     &  0.44  &  M06-HF                  &  MN     &      0.62  \\
 PBE           &  D3          &      0.48      &  B97-1      &  XDM    &  0.62  &  PW6B95    &  D3     &  0.34  &  CAM-B3LYP               &  D3     &      0.67  \\
 PBE           &  TS          &      0.28      &  B3LYP      &  D3     &  0.36  &  MPW1B95   &  D3     &  0.29  &  CAM-B3LYP               &  XDM    &      0.50  \\
 PBE           &  XDM       &      0.57      &  B3LYP      &  TS     &  0.23  &  PWB6K     &  D3     &  0.44  &  LC-$\omega$PBE          &  D3     &      0.28  \\
 PBEsol       &  D3         &      1.01      &  B3LYP      &  XDM    &  0.31  &  MPwB1K    &  D3     &  0.32  &  LC-$\omega$PBE          &  XDM    &      0.31  \\
 revPBE       &  D3        &      0.41      &  B3LYP      &  DCP    &  0.27  &  B1B95     &  D3     &  0.43  &  LC-$\omega$PBE          &  DCP    &      0.27  \\
 OPBE          &  D3        &      0.83     &  B3PW91     &  D3     &  0.45  &  BMK       &  D3     &  0.98  &  $\omega$B97X-D$^c$      &  D2     &  0.23$^d$  \\
 BPBE           &  D3       &      0.49     &  BHandHLYP  &  D3     &  0.66  &  M05       &  D3     &  0.52  &  VV10                    &  NL     &      0.31  \\
 rPW86PBE  &  D3        &      0.35    &  BHandHLYP  &  XDM    &  0.47  &  M05       &  MN     &  2.07  &  vdW-DF2                 &  NL     &      0.94  \\
  PW86PBE  &  XDM     &      0.35    &  PBE0       &  D3     &  0.57  &  M05-2X    &  D3     &  0.35  &                          &         &            \\
\hline\hline
\end{tabular}
\begin{minipage}{1.0\textwidth}
\raggedright
{
D3: DFT-D3/def2 using the original damping
function.\cite{goerigk2011} TS: DFT-TS using ``tier2'' numerical
atom-centered orbital bases.\cite{marom2011} XDM:
DFT-XDM/aug-cc-pVTZ.\cite{oterodelaroza2013b} DCACP: DFT-DCACP2/plane
wave.\cite{karalti2014} DCP:
B3LYP-DCP/6-31+G(2d,2p)\cite{dilabiodcp2012} and
LC-$\omega$PBE/6-31+G(2d,2p).\cite{dilabiodcp2014} MN:
Minnesota/def2-QZVP.\cite{goerigk2011} VV10:
VV10/aug-cc-pVTZ.\cite{vydrov2010} vdW-DF2:
vdW-DF2/aug-cc-pVTZ.\cite{vydrov2010} The S22 benchmark was used as
the fitting set for TS, and as component of the fitting set for D3 and
XDM.
}\\
{$^a$ The DFT-DCACP/plane wave approach of reference~\citenum{vonlilienfeld2004} gave a MAE of 0.65 kcal/mol.\cite{karalti2014}}\\
{$^b$ M06-2X/6-31+G(2d,2p) produces a MAE of 0.43 kcal/mol.}\\
{$^c$ Employing the original ``D'' correction for dispersion.}\\
{$^d$ $\omega$B97X-D/6-31+G(2d,2p) produces a MAE of 0.58 kcal/mol.}
\end{minipage}
\end{table}
}

For the S22B set, the method demonstrating the best performance
 is TPSS-TS, with an MAE value of only 0.2
kcal/mol. Ten other approaches give MAE values between 0.2
and 0.3 kcal/mol, namely, BLYP-D3, BLYP-XDM, MPW1B95-D3 , M06-D3,
LC-$\omega$PBE-D3, PBE-TS, B3LYP-TS, B3LYP-DCP, LC-$\omega$PBE-DCP,
and $\omega$B97X-D. It is interesting to note that all of the
dispersion-correcting methods are capable of predicting noncovalent
binding energies for the S22B set with very low average errors but not
necessarily when used in combination with the same functionals. This
suggests that each of the dispersion-correcting techniques is best
suited to correct for the underlying deficiencies of certain bare
functionals, although the broader development and application of
methods besides the D3 approach will be required before making
definitive statements in this connection. It is also interesting to
note that the best performing methods represent various ``rungs'' of
the DFT ladder,\cite{tpss} with dispersion-corrected GGAs, hybrids,
and range-separated functionals making an appearance.

Of the remaining of the methods listed in Table \ref{tab:S22ALL}, 21 give MAE values in the 0.3--0.4 kcal/mol range, 12 fall
into the 0.4--0.5 kcal/mol range, and the other 23 methods predict
MAE values between 0.5 and 2.1 kcal/mol. The non-local functionals are
not amongst the top performers for this benchmark set. All of the
methods, except for the Minnesota functionals for which there is not a
non-dispersion-corrected version to compare against, represent
significant improvements over the results obtained with the bare
functionals, as provided in Table \ref{tab:S22DFT}. However, the fact
that so many methods still perform poorly when dispersion corrections
are incorporated implies that these functionals have underlying
deficiencies beyond just dispersion.

The successful application of D3 and TS corrections to the Minnesota
functionals demonstrates that dispersion-correction approaches do not
mutually exclude each other. In the particular case of the Minnesota
DFT methods, the parameters of the functionals were optimized to
minimize the errors associated with a number of molecular properties
and not just noncovlaent interactions. It follows that Minnesota
functionals are, in general, systematically underbinding, and that
there is potential to improve them through the use of other
dispersion-correction techniques. This is demonstrated in
Table~\ref{tab:S22ALL} most dramatically for M05, M05-2X, and M06
where the MAEs are reduced by factors of 2--4 through the use of these
functionals with the D3 dispersion correction. However, better results
are not always achieved in this way: The combination of M06-HF and D3
increases the MAE from 0.62 to 0.84 kcal/mol.

It is worth mentioning the relative computational costs (i.e. run
time) associated with the different dispersion-correction methods. The
costs are not very relevant for the S22 set because most of the dimers
contained therein are fairly small. However, computational cost
becomes a concern when large systems, or a large number of systems,
are simulated. The dispersion-correction method with lowest cost of
those listed in Table~\ref{tab:S22ALL} is the DCP approach owing to
the fact that it is designed for use with small basis sets. Despite
the small basis sets, the DCP method produces small MAEs for the
S22B set of 0.27 and 0.23 kcal/mol, depending on the underlying DFT
method.

On the point of the basis set dependence of dispersion-corrected
methods, the M06-2X functional demonstrates remarkably consistent MAEs
for the S22 set as obtained using def2-QZVP and 6-31+G(2d,2p)
bases. The differences in the MAEs so calculated is only 0.03
kcal/mol, or about 8\%. Conversely, the $\omega$B97X-D method display
significantly greater basis set dependence, giving differences in the
def2-QZVP and 6-31+G(2d,2p) MAEs of 0.35 kcal/mol, which represents a
more than two-fold increase in average error.  This underscores an
important issue that should not be overlooked by users of these
methods: There may be significant basis-set dependencies associated
with a particular dispersion correction approach that should be
determined if basis sets other than the recommended ones are
utilized. For DFT methods that are coupled with pair-wise
dispersion-correction (D3 and XDM) methods, users can expect basis
sets dependencies similar or identical to those of the
underlying functional. DFT approaches that are coupled with DCPs are
expected to have smaller basis set dependencies than the bare
functional.

\begin{figure}
\includegraphics[width=0.99\textwidth]{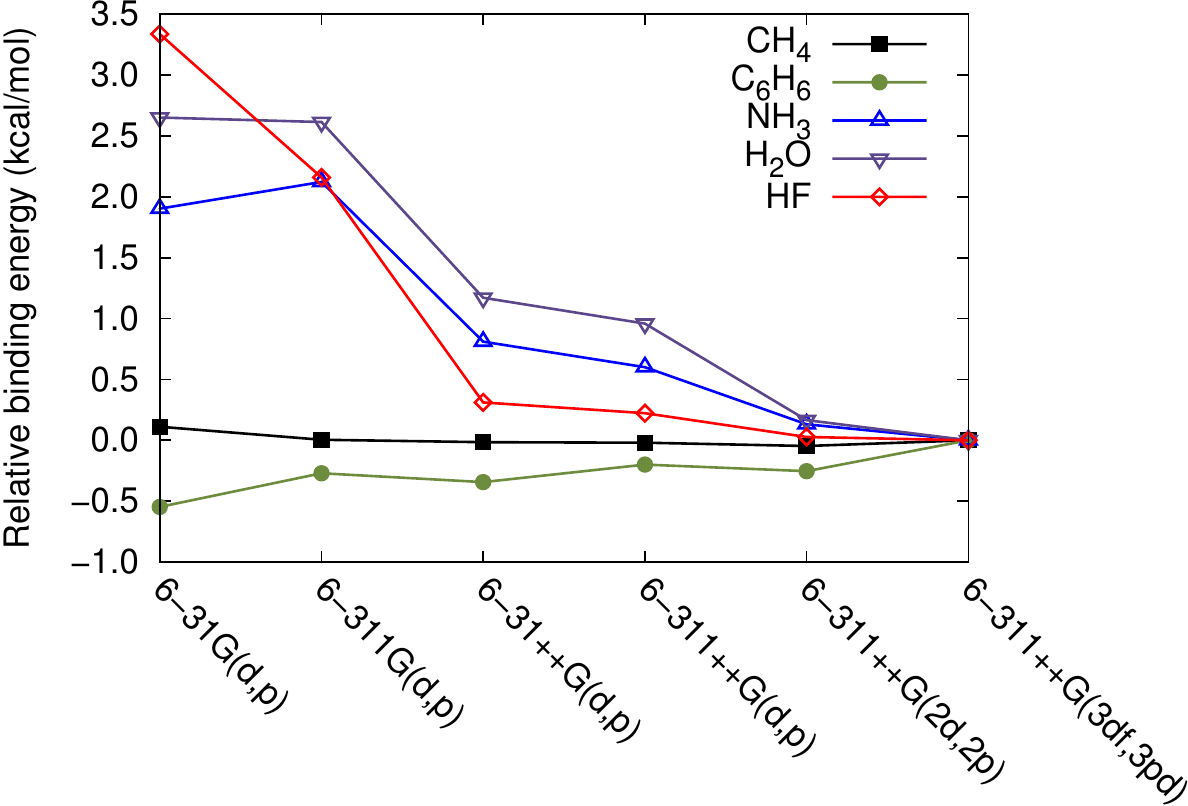}
\caption{Basis set dependence of the LC-$\omega$-PBE-XDM binding
energies of various noncovalent interacting dimers (from
reference~\citenum{johnson2013}). 
\label{fig:BasisSetDep}}
\end{figure}

Figure~\ref{fig:BasisSetDep} demonstrates the basis-set dependence of
the LC-$\omega$-PBE functional with XDM. Because XDM is a post-SCF
energy correction, the basis set dependence for the bare functional is
roughly the same as that illustrated in the Figure. The plot shows
that the performance of LC-$\omega$PBE-XDM degrades substantially when
basis sets smaller than 6-311++G(2d,2p) are used. Interestingly, the
basis set dependence increases with the strength of the interaction
type. That is, dispersion-dominated dimers present a weak basis set
dependence whereas hydrogen-bonded systems are much more affected by
basis-set incompleteness errors. This makes sense in view that
hydrogen bonds involve significant intramolecular charge transfer and
orbital interactions, whereas the former does not.

With respect to DCACPs, it is noted that the BLYP-DCACP approach of
von Lilienfeld et al.~\cite{vonlilienfeld2004} produces rather poor
results (see footnote a in Table \ref{tab:S22ALL}). However, keeping
in mind that only one DCACP function was used for each atom and that
the fitting data used to generate the functions were very small, the
performance is reasonable. The recent work of Karalti et
al.~\cite{karalti2014} convincingly demonstrates that using two
functions per atom can offer much improved performance in the
treatment of noncovalent interactions, with the MAE reduced by almost
a factor of 2 over that obtained with the single function DCACPs.

{
\renewcommand*{\arraystretch}{1.0}
\renewcommand*{\baselinestretch}{1.0}
\begin{table}
\caption{Mean Absolute Errors (MAE) in the Binding Energies for the
S66 Benchmark Set of Noncovalently-Bonded Dimers Predicted by DFT
Methods Employing Various Dispersion-Correcting Approaches in
kcal/mol.
\label{tab:S66ALL}}
\begin{tabular}{ccl|ccl|ccl}
\hline \hline 
 Func.    &  Disp.  &   MAE  &  Func.      &  Disp.  &       MAE     &  Func.                   &  Disp.    &        MAE     \\
\hline 
 B97      &  D3     &  0.29  &  M06-L      &  D3     &  0.34$^\dag$  &  M06-2X                  &  D3       &   0.24$^\dag$  \\
 BLYP     &  D3     &  0.19  &  M06-L      &  MN     &  0.60         &  M06-2X                  &  MN       &   0.28$^a$     \\
 BLYP     &  XDM    &  0.19  &  B3LYP      &  D3     &  0.28         &  CAM-B3LYP               &  XDM      &       0.35     \\
 PBE      &  D3     &  0.40  &  B3LYP      &  XDM    &  0.22         &  LC-$\omega$PBE          &  D3       &       0.19     \\
 PBE      &  TS     &  0.44  &  B3LYP      &  DCP    &  0.19         &  LC-$\omega$PBE          &  XDM      &       0.21     \\
 PBE      &  XDM    &  0.39  &  B97-1      &  XDM    &  0.38         &  LC-$\omega$PBE          &  DCP      &       0.21     \\
 PBEh     &  TS     &  0.39  &  BHandHLYP  &  XDM    &  0.31         &  $\omega$B97X-D$^b$      &  D2       &   0.29$^c$     \\
 PW86PBE  &  TS     &  0.39  &  PBE0       &  XDM    &  0.36         &  VV10                    &  VV10     &       0.35     \\
 PW86PBE  &  XDM    &  0.26  &  PW6B95     &  D3     &  0.18         &  LC-VV10                 &  VV10     &       0.15     \\
 revPBE   &  D3     &  0.29  &  MPW1B95    &  D3     &  0.20         &  vdW-DF2$^k$             &  vdw-DF2  &       0.48     \\
 TPSS     &  D3     &  0.30  &  M05-2X     &  D3     &  0.30$^\dag$  &                          &           &                \\
 oTPSS    &  D3     &  0.30  &  M05-2X     &  MN     &  0.58         &                          &           &                \\
\hline \hline 
\end{tabular}
\begin{minipage}{0.8\textwidth}
\raggedright
{D3: DFT-D3/def2-QZVP\cite{goerigk2011b}  using the
Becke-Johnson damping function\cite{becke2005a} except where indicated
by $\dag$. TS: DFT-TS using ``tier2'' numerical atom-centered
orbital bases.\cite{marom2011} XDM:
DFT-XDM/aug-cc-pVTZ.\cite{oterodelaroza2013b} DCP:
B3LYP-DCP/6-31+G(2d,2p)\cite{dilabiodcp2012} and
LC-$\omega$PBE/6-31+G(2d,2p).\cite{dilabiodcp2014} MN:
Minnesota/def2-QZVP.\cite{goerigk2011} VV10:
VV10/aug-cc-pVTZ.\cite{vydrov2010} vdW-DF2:
vdW-DF2/aug-cc-pVTZ.\cite{vydrov2010}
}\\
{$^a$ M06-2X/6-31+G(2d,2p) produces a MAE of 0.25 kcal/mol.}\\
{$^b$ Employing the original ``D'' correction for dispersion.}\\
{$^c$ $\omega$B97X-D/6-31+G(2d,2p) produces a MAE of 0.30
kcal/mol.\cite{dilabiodcp2012}}
\end{minipage}
\end{table}
}

Table~\ref{tab:S66ALL} summarizes the performance of various
dispersion-corrected DFT methods on the larger S66 set. As a point of
reference, B3LYP without corrections for dispersion gives an MAE of
3.8 kcal/mol for this set. Inclusion of dispersion corrections by any
means is expected to reduce the MAE and this is realized in the data
provided in Table~\ref{tab:S66ALL}. The best performing method in the
table is the long-range corrected, non-local version of the VV10
functional, giving an MAE of only 0.15 kcal/mol. D3, XDM
and DCP-corrected functionals round-out the top ten list giving MAEs
of 0.18--0.21 kcal/mol, with all classes of functionals
represented. 

As happened in the S22 set, the combination of dispersion-corrections
with different functionals are capable of performing very well for the
prediction of noncovalent binding energies of the S66 set. This offers
potential users of these methods some choice in terms of balancing the
desired accuracy in noncovalent properties with those of other
properties (see, for example, Table \ref{tab:thermo}). Alternatively,
users may wish to consider the use of more computationally efficient
methods, such as DCP-based approaches or functionals that allow for
the use of smaller basis sets owing to the small impact of basis set
incompleteness in their performance. With respect to molecular
properties, it should also be understood that the application of
dispersion corrections may also alter the performance of the DFT
method for properties other than simple noncovalent interactions. This
issue is discussed to some extent in the final section of this chapter.

Other results of note presented in Table~\ref{tab:S66ALL} are those of
the M06-2X and M05-2X functionals, which give MAEs of 0.28 and 0.58
kcal/mol, respectively. Both functionals perform better on the S66 set than they do
on the S22 set, but the performance of M05-2X may be considered by
most to be too poor to be useful for noncovalent interactions. The
inclusion of D3 corrections to these two functionals improves their
performance - marginally for M06-2X but by almost a factor of 2 for
M05-2X. Again this underscores the notion that different dispersion
approaches can be combined in order improve their performance for
noncovalent interactions.

{
\renewcommand*{\arraystretch}{1.0}
\renewcommand*{\baselinestretch}{1.0}
\begin{table}
\caption{Mean Absolute Error (MAE) in the Binding Energies Predicted by
Selected Density Functionals, Basis Sets and Dispersion Correction
Schemes on the S12L set of Noncovalently Bonded
Dimers\cite{risthaus2013} (in kcal/mol). 
\label{tab:S12L}}
\begin{tabular}{ccccc}
\hline \hline
Density functional & Basis set & Disp. Correction & MAE & Ref.\\
\hline 
\hline 
PBE 				& def2-QZVP	 	& NL 	+ $\Delta$E$^{\text{ABC}}$				& 2.1		& \citenum{risthaus2013}\\
PBE 				& def2-QZVP$^a$ 	& D2 + $\Delta$E$^{\text{ABC}}$				& 2.3		& \citenum{risthaus2013}\\
PBE 				& def2-QZVP$^a$ 	& D3 + $\Delta$E$^{\text{ABC}}$				& 2.4		& \citenum{risthaus2013}\\
M06-L			& def2-QZVP		& MN + $\Delta$E$^{\text{ABC}}$ 				& 4.1		& \citenum{risthaus2013}\\
PBE 				& def2-TZVP$^b$ 	& D2 + $\Delta$E$^{\text{ABC}}$				&  1.6	& \citenum{risthaus2013}\\
PBE 				& def2-TZVP$^b$ 	& NL	+ $\Delta$E$^{\text{ABC}}$				&  2.3	& \citenum{risthaus2013}\\
PBE 				& def2-TZVP$^b$ 	& D3 + $\Delta$E$^{\text{ABC}}$				&  2.3		& \citenum{risthaus2013}\\
M06-2X			& def2-TZVP$^b$ 	& MN + $\Delta$E$^{\text{ABC}}$ 				&  3.3		& \citenum{risthaus2013}\\
M06-L			& def2-TZVP$^b$ 	& --- + $\Delta$E$^{\text{ABC}}$				&  4.6		& \citenum{risthaus2013}\\
\hline 
PBE 				& pc-2(spd)$^{c}$ 	& XDM 				& 2.3$^{d}$ 	& \citenum{oterodelaroza2014b}\\
BLYP 				& pc-2(spd)$^{c}$ 	& XDM 				& 3.9$^{d}$ 	& \citenum{oterodelaroza2014b}\\
B3LYP 			& pc-2(spd)$^{c}$ 	& XDM 				& 3.7$^{d}$ 	& \citenum{oterodelaroza2014b}\\
LC-$\omega$~PBE 	& pc-2(spd)$^{c}$ 	& XDM 				& 6.5$^{d}$ 	& \citenum{oterodelaroza2014b}\\
\hline 
B3LYP 			& 6-31+G(2d,2p)	& DCP 				& 2.6		& \citenum{dilabiodcp2013}\\
LC-$\omega$~PBE 	& 6-31+G(2d,2p) 	& DCP 				& 3.4 	& \citenum{dilabiodcp2014}\\
\hline \hline
\end{tabular}\\
\centering
\begin{minipage}{0.8\textwidth}
\raggedright
{NL: Hujo et al. tweak\cite{hujo2011} to the original VV10 by Vydrov
and van Voorhis.\cite{vydrov2010} D2: DFT-D2. D3: DFT-D3. $\Delta
E^{\text{ABC}}$: a three-body Axilrod-Teller-Muto term has been added
(Eq.~\ref{eq:atm}).}\\
{$^{a}$These basis sets were employed without g-functions.}\\
{$^{b}$With counterpoise corrections.~\cite{boys1970}}\\
{$^{c}$Using the pc-2 basis set of
Jensen,~\cite{jensen2004,jensen2002,jensen2001} with the heavy atom
f-basis functions removed, as described in
reference~\citenum{johnson2013}. The pc-2(spd) basis set has the same
number of contracted functions as the 6-31+G(2d,2p) basis set.}\\ 
{$^{d}$These MAEs decrease by 0.2 to 0.3 kcal/mol when the Quantum
Monte-Carlo (QMC) reference data of Ambrosetti et
al.~\cite{ambrosetti2014} are used.}\\  
\end{minipage}
\end{table}
}

As described in previous sections, the power of DFT becomes very clear
when it is applied to very large systems that are well outside of the
range of accurate wavefunction theory methods. The S12L set allows
these limits to be explored. The results presented in
Table~\ref{tab:S12L} show the MAEs predicted by a relatively small
number of dispersion-corrected DFT methods. To put the data into some
context, the LC-$\omega$PBE functional with 6-31+G(2d,2p) basis sets
predicts an MAE of 17.9 kcal/mol for the set. The binding energies
predicted using this approach are consistently underbinding despite
benefiting from some stabilization through basis set incompleteness
effects. The MAE of 25.4 kcal/mol (underbinding) obtained for the bare
PBE functional used with def2-QZVP (without g
functions)\cite{risthaus2013} confirms the expectation that
underbinding by bare functionals will become more pronounced as the
size of the basis set increases.

The MAEs listed in Table~\ref{tab:S12L} obtained by
dispersion-corrected DFT methods are much larger than those found for
the S22 and S66 sets. This in part relates to the fact that the
underlying DFT methods perform very poorly on the S12L set and also
because the binding energies of some members of the set are very
large.

Risthaus and Grimme's recent work explored the utility of some
dispersion-corrected DFT methods in treating the large systems of the
S12L set. In several cases, they found it important to include
corrections to the binding energies for 3-body terms, and estimated
the magnitude of this contribution using C$_9$ coefficients estimated
from C$_6$ coefficients. The 3-body corrections range from about 0.7
to 4.6 kcal/mol and reduce the strength with which the complexes are
bound. The authors paired the D2 and D3 dispersion corrections with the PBE
functional and basis sets of QZ quality and found MAEs of 2.3 and 2.4
kcal/mol, respectively. PBE-NL provided a slightly smaller MAE while
M06-L gave an MAE that was nearly twice as large.

Recognizing that the large QZ basis sets (with which the D3
corrections were developed and are normally used) are impractical for
very big systems, Risthaus and Grimme also explored the use of
smaller, def2-TZVP basis sets and counterpoise corrections on the S12L
set. These smaller basis sets with the PBE-D2 approach gave the lowest
MAE in binding energy for the set---only 1.6 kcal/mol---whereas the
performance of PBE-NL and PBE-D3 were unchanged. The MAEs for M06-2X
and M06-L are factors of ca. 2 and 3 larger than the best performing
method in the Table.

Considering both low MAEs and basis set size, some of the best
performing methods on the S12L set appear to be those based on the XDM
approach. These used pc-2(spd) basis sets, the
polarization-consistent-2 set of
Jensen~\cite{jensen2004,jensen2002,jensen2001} with the heavy atom
f-functions removed. The pc-2(spd) basis sets were found to offer an
excellent compromise between quality of results for noncovalent
interactions and computational cost.~\cite{johnson2013} The PBE
approach coupled with XDM gives an MAE for the S12L set of 2.3
kcal/mol. If the QMC reference binding energies are used, the MAE
drops to 2.1 kcal/mol. The average of the 3-body correction terms
computed by Grimme for the S12L set is 2.2 kcal/mol, and so the good
agreement provided by the PBE-XDM approach raises some interesting
questions about the role that 3-body corrections actually play in
dispersion-corrected DFT. The BLYP and B3LYP were also coupled with
XDM and the small pc-2(spd) basis sets and these approaches gave MAEs
that are on par with the M06-2X results.

The DCP approach also produces reasonable results for the S12L
set. B3LYP-DCP/6-31+G(2d,2p) gives an MAE of 2.6 kcal/mol, which is
competitive with many of the other methods listed in Table
~\ref{tab:S12L}. The MAE for LC-$\omega$PBE-DCP is 3.6 kcal/mol,
fairly close to that of M06-2X/def2-TZVP with counterpoise
corrections.

\begin{table}
\caption{Mean Absolute Errors (MAE, kcal/mol) and
Mean Absolute Percent Errors (MAPE) in the Experimental Lattice
Energies Predicted by Selected Functionals and Dispersion Corrections
on the C21 Set of Molecular Crystals.\cite{oterodelaroza2012b} 
\label{tab:C21}} 
\begin{tabular}{cccc}
\hline \hline
Functional & Disp. Correction & MAE & MAPE \\
\hline 
PBE & --- & 8.6 & 47.2 \\
PBE & D2 & 2.2 &  11.9 \\
PBE & TS & 4.1 & 22.1 \\
PBE & XDM & 1.3 &6.7  \\
vdw-DF1 & NL & 2.4 &13.5 \\
vdw-DF2 & NL & 2.4 &13.1 \\
B86b & XDM & 1.1 &6.2 \\
\hline 
PBE0 & MBD & 0.9 &5.7 \\
\hline \hline
\end{tabular}
\begin{minipage}{0.8\textwidth}
\raggedright
D2: DFT-D2 as implemented by Barone et al.\cite{barone2009} TS:
Tkatchenko-Scheffler.\cite{tkatchenko2009} vdw-DF1: Dion et
al. version of the vdw-DF functional.\cite{dion2004} vdw-DF2: Lee et
al. version of vdw-DF.\cite{lee2010} PBE0-MBD: TS functional with
many-body dispersion corrections.\cite{reilly2013} All calculations
were carried out in a plane wave/pseudopotentials approach.
\end{minipage}
\end{table}

The last benchmark set for which we discuss the performance of
dispersion-corrected DFT method is the heats of sublimation molecular
crystal database (C21) of Otero-de-la-Roza and
Johnson.~\cite{oterodelaroza2012b} Table \ref{tab:C21} summarizes the
performance of XDM and TS pair-wise dispersion corrections along with
non-local DFT methods, applied with plane wave basis sets. As an aside,
DCACPs could be applied to this set because they were designed for use with
plane wave codes but, to our knowledge, this has not been done
yet. DCPs cannot be applied easily to the C21 set as they are
readily applicable only in quantum chemistry (``cluster'') codes. The
C21 set also contains a test set using the same crystals in which the
X-ray or neutron diffraction structures are compared against the
calculated crystal geometries after relaxation under a ``thermal
pressure'', that encapsulates the effects of crystal vibrations.

For comparison, the PBE functional without corrections for dispersion
does poorly in predicting the experimental lattice energies for
the set giving a MAE of 8.6 kcal/mol, or nearly 50\%\ average
error. Including dispersion corrections  improves the
predictions greatly. The ``TS'' parameters for pair-wise dispersion perform
the most poorly, giving an MAPE of about 22\%; however, recent
modifications to the ``TS'' treatment that include of a
many-body correction and incorporate Hartree-Fock exchange
via the PBE0 functional, reduce the MAE to below 1
kcal/mol\cite{reilly2013} (5.7\%). Nevertheless, the PBE0-MBD results
are only slightly better than those derived from the PBE-XDM approach,
which has neither an explicit correction for many body terms nor does
the functional have Hartree-Fock exchange.

The results that are achieved by PBE-XDM and PBE0-MBD point to an
important aspect of calculations that should be considered by users
who are interested in applying these methods to their own
problems. In periodic calculations, the inclusion of Hartree-Fock
exchange is far more computationally expensive (up to two orders of
magnitude longer running times) than the cost of doing so in cluster
calculations (usually less than a factor of two). With any
computational approach, users must balance the costs associated with a
given approach against the accuracy of the methods. In the case of
the C21 set, for example, users should assess whether they are willing to wait
10 to 100 times longer to
achieve an average improvement in performance of 0.4 kcal/mol (or 1\%). In addition, the cost of the
MBD calculation has, to our knowledge, not been reported.

The results of Table~\ref{tab:C21} show that dispersion-corrected DFT
methods are becoming advanced enough such that accurate relative
stability of molecular crystals (particularly in the context of {\it a
priori} molecular crystal structure prediction) should be possible.

\section*{Non-Covalent Interactions in Perspective}
\label{s:perspective}
In this chapter, we demonstrated that dispersion-corrected DFT methods
are now capable of providing excellent agreement with benchmark
binding energies for a variety of dimer systems.  There is no doubt
that the methods have advanced to the point where many of them are
robust and can be generally applied beyond noncovalently-bonded
systems to good effect. Coupled with the demonstration that the XDM
and TS approaches are capable of predicting heats of sublimation,
dispersion-corrected DFT methods are now promising tools for the
{\it a priori} design of new molecular materials.

It is worthwhile at this point to consider dispersion-corrected DFT
from a broader chemical perspective. Thus, we may ask: Does the
inclusion of dispersion corrections in DFT methods influence
properties other than binding energies? The answer to this question is
partially answered by recognizing that many implementations of
pair-wise dispersion correction schemes are not self-consistent, which
means that the electron density is not affected by them. It follows
that, for a given molecular structure, properties not directly taken
from the molecular energy will be exactly the same with and without
the pair-wise dispersion correction. From this perspective, pair-wise
dispersion corrections cannot \emph{directly} influence properties
that depend on the electronic structure. The same is not true for the
non-local and Minnesota functionals that are formulated
self-consistently, although examining ``with and without'' dispersion
scenarios with these families of functionals is not possible. DCPs
also operate self-consistently in that they are incorporated into the
Hamiltonian of the system and thereby alter electron distributions and
thus have the potential to alter properties, which may include among
others NMR chemical shifts, dipole moments, and hyperfine coupling
constants.

All dispersion corrections, regardless of the type, have the potential
to \emph{indirectly} affect molecular properties by altering molecular
structure. The relationship between structure and properties is well
known in chemistry. As a simple example, the catalytic properties of
an enzyme depend crucially on its structure~\cite{solomon2000} and
denaturing the enzyme (i.e. changing its structure through, for
instance, heating) reduces or destroys its catalytic properties. In
general, by providing a realistic description of the energy landscape
that determines the structure by introducing 
dispersion corrections into DFT methods, the properties depending on that 
structure will also be better described.

Understanding the ability of dispersion-corrected DFT
methods to predict accurate structures can be important for making
decisions about how such methods should be used for solving problems 
 in chemistry and physics. A simple and effective example of
the impact of noncovalent interactions in the area of organic electronic materials
was recently published .~\cite{mcclure2010} In cases where electron transport within
these materials is dominated by a hopping mechanism, the process can
be modeled as an electron transfer reaction between adjacent molecules
in a molecular film. The rate constant for the electron transfer
process between molecules can be approximated using Marcus theory
according to:
\begin{equation}
\label{eq:marcustheory}
k_{\text{ET}} = \frac{1}{\sqrt{4{\pi\lambda}k_{\text{B}}T}} V^2 exp(-\frac{\lambda}{4k_{\text{B}}T})
\end{equation} 
where $k_{\text{B}}$ is the Boltzmann's constant, and $\lambda$ is the
reorganization energy associated with the geometry change that occurs
during the electron transfer process (atomic units are used).

The variable $V$ is the electron coupling matrix element that relates
orbital overlap between adjacent molecules; that ultimately dictates
the carrier transport efficacy of a material. Thus, the
structure-activity relationship is defined: the spatial arrangement of
adjacent molecules in a material dictates the ability of the material
to transport charge. In the context of modeling charge transport in
organic electronic materials, the ability to predict reasonably
accurate structures may be considered to be more important that the
ability to predict accurate binding energies. We refer back to
Figure \ref{fig:ContrivedPES} to illustrate this point: although the
fictitious approximate method predicts the correct binding energy at
the accepted dimer minimum indicated at 4 {\AA}, a full structure
optimization would result in a intermonomer minimum that is too short
by 0.2 {\AA} and this would give an overly-large electron transfer
rate constant if applied to organic electronic material. 

Structure is also a critical aspect in all problems related to
reaction chemistry and covalent bonding. However, the deficiencies
with respect to dispersion treatment in many DFT methods become overwhelmed
by other shortcomings in the base functionals. This becomes clear
when modeling simple thermochemistry, like the C-H bond
dissociation enthalpy (BDE) in methane. Here, base DFT
methods may predict BDEs that are in error by 3 or 4 kcal/mol. These
errors are more than an order of magnitude larger than those 
obtained without dispersion.
Always keep in mind that including accurate dispersion-corrections into a
chosen method is usually insufficient for alleviating all of the
problems of a DFT method. 

In some cases, including dispersion-corrections in a DFT treatment can
produce much worse results than might be achieved without them. This is 
illustrated by considering one of the early
studies on the ability of some common DFT based methods to predict
accurate barrier heights in simple hydrogen atom exchange
reactions. Lynch and Truhlar~\cite{lynch2001} presented results for a
benchmark set of 21 simple forward and reverse bimolecular hydrogen
atom transfer reactions (e.g. OH $+$ H$_2 \rightarrow$ H $+$ H$_2$O,
C$_2$H$_6 +$ NH$_2 \rightarrow$ C$_2$H$_5 +$ NH$_3$) and one
intramolecular hydrogen atom transfer (namely, \emph{s-trans
cis}-C$_5$H$_8 \rightarrow $ \emph{s-trans cis}-C$_5$H$_8$) for a
total of 43 barrier heights. B3LYP with reasonably large basis sets
predicts all but two of the 43 reaction barriers to be too low. How
would the results be affected if pair-wise dispersion corrections were
applied with B3LYP to these barrier heights? Consider that 
the transition state for any
bimolecular reaction is composed of more atoms
than the reactant and product states. Because the energy contribution of
any pair-wise dispersion scheme is always stabilizing, it follows that
the energy stabilization of the larger transition states will be
preferentially stabilized relative to the smaller reactants and
products. In this case, applying of pair-wise dispersion
schemes will result in predicted barrier heights that are in worse
agreement with the benchmark values as compared to the results for the base
functional. Of course, in cases where a DFT method predicts barrier
heights to be too high, pair-wise dispersion
corrections will improve the results.

\begin{figure}
\includegraphics[width=0.60\textwidth]{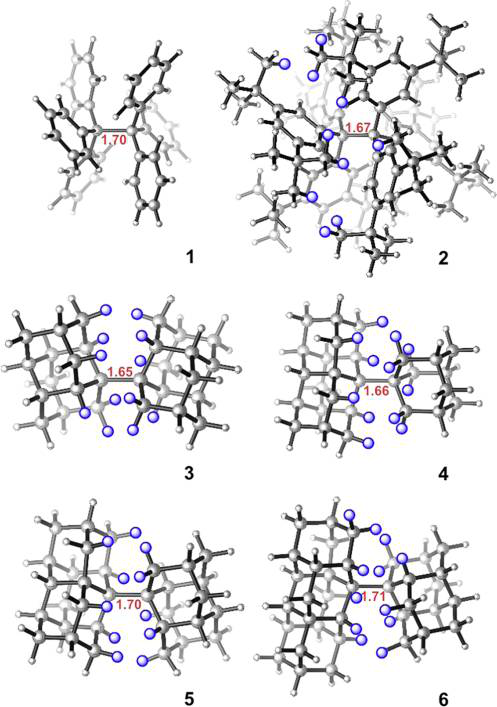}
\caption[Diamondoid structures]{Structures of coupled diamondoid
molecules showing unusually long C-C bonds (taken with permission from
reference~\citenum{schreiner2012}).
\label{fig:Xschreiner2012}} 
\end{figure}

Recent work by Schreiner's team provides an interesting counter
example to the barrier height problem described above. This group
prepared and studied (experimentally and computationally)
sterically-crowded molecules based on coupled diamondoid
species.~\cite{schreiner2011,schreiner2012} Diamondoids are small,
hydrogen-terminated three-dimensional carbon clusters consisting of
adamantane subunits. The coupled diamondoid species, a sampling of
which are shown in Figure~\ref{fig:Xschreiner2012}, have one or more
very long C-C bonds as a result of the steric strain between
juxtaposed diamondoid groups. Some of the C-C bonds are elongated by
up to 0.17 {\AA} compared to a ''normal'' C-C bond.  These structures
provide a harsh test of dispersion-corrected DFT methods by probing
their ability to capture the effects of very close contacts on
covalent bonding. Schreiner et al.\ found that dispersion-corrections
were required in order for DFT methods to attain reasonable agreement
with the experimentally determined, long C-C bond lengths in the
coupled diamondoid systems. Interestingly, most of the DFT methods
applied to the rotation barrier about the elongated C-C bond,
including those without dispersion-corrections, predicted values in
reasonable agreement with the experimentally derived values,
most likely because of error cancellation.

The foregoing discussion reminds us that dispersion-corrections are
important in the modeling of some, but not all, physical
systems. Until underlying functionals themselves are improved, the
value derived from dispersion-correction DFT methods will be
limited. In this connection, we remind the reader that the Minnesota
functionals considers this by fitting a highly parameterized
functional to a broad set of benchmark data that includes
thermochemical and kinetic information in addition to noncovalent
binding energies. There also seems to be some promise for the
application of potential-based correction methods to offer
improvements to underlying problems with functionals beyond noncovalent
interactions. Since the DCP and DCACP approaches alter electron
distributions, it follows that they may be utilized to provide better
descriptions of more than just noncovalent binding energies. von
Lilienfeld recently demonstrated how DCACPs
can be used with the BLYP functional to give molecular vibrational
frequencies that are of B3LYP quality, and closer to experimental
values.~\cite{vonlilienfeld2013} DiLabio and Koleini developed DCPs
for use with LC--$\omega$PBE that, in addition to very good binding
energies in noncovalently bonded systems (see
Tables~\ref{tab:S22ALL},~\ref{tab:S66ALL}, and~\ref{tab:S12L}), are
capable of providing excellent bond dissociation enthalpies for X-H
and X-Y covalent bonds (X,Y = C, N, O).~\cite{dilabiodcp2014}

We expect that the future development of approximate DFT methods will
continue to include the accurate treatment of noncovalent
interactions, with near-term focus on many-body effects and on regions
intermediate to the covalent and noncovalent bonding
regimes. Furthermore, we expect the development of less-empirical and
more efficient approaches to dispersion-corrected DFT with the
evolution of more detailed understanding of how exchange and
correlation functionals interplay in the treatment of noncovalent
interactions. Ultimately, considerations of the broader performance of
DFT methods for thermochemical, kinetics and other properties should
remain the priority.


\end{document}